\documentclass[preprint,aps,prd,amsmath,amssymb]{revtex4}
\pdfoutput=1
\usepackage{graphicx}
\usepackage{color}
\newcommand{\be}{\begin{equation}}
\newcommand{\ee}{\end{equation}}
\newcommand{\bea}{\begin{eqnarray}}
\newcommand{\eea}{\end{eqnarray}}
\newcommand{\beq}{\begin{eqnarray}}
\newcommand{\eeq}{\end{eqnarray}}

\newcommand{\bmp}{\noindent\begin{minipage}{16cm}}
\newcommand{\emp}{\end{minipage}\vskip 7mm} 

\usepackage{graphicx}
\usepackage{dcolumn}
\usepackage{bm}
\usepackage{amsmath}
\usepackage{amsfonts}
\usepackage{bbm}
\usepackage{subfigure}
\usepackage{pxfonts}
\usepackage{slashed}

\usepackage{ulem}

\def\drawbox#1#2{\hrule height#2pt
        \hbox{\vrule width#2pt height#1pt \kern#1pt
              \vrule width#2pt}
              \hrule height#2pt}
\def\Fund#1#2{\vcenter{\vbox{\drawbox{#1}{#2}}}}
\def\Asym#1#2{\vcenter{\vbox{\drawbox{#1}{#2}
              \kern-#2pt 
              \drawbox{#1}{#2}}}}

\def\fund{\Fund{6.4}{0.3}}

\begin{document}

\title{ 
{\LARGE {\color{magenta} The Electroweak Phase Transition \\in\\ Ultra Minimal Technicolor }} }
\author{{\color{blue}Matti {\sc J\"arvinen}}$^{\color{blue}{\varheartsuit}}$}
\email{mjarvine@ifk.sdu.dk}
\author{{\color{blue}Thomas A. {\sc Ryttov}}$^{\color{blue}{\spadesuit}}$}
\email{ryttov@nbi.dk}
\author{{\color{blue}Francesco {\sc Sannino}}$^{\color{blue}{\varheartsuit}}$}
\email{sannino@ifk.sdu.dk} \affiliation{$^{\color{blue}{\varheartsuit}}${\color{magenta}University of Southern Denmark, Campusvej 55, DK-5230 Odense M} \\ $^{\color{blue}{\spadesuit}}${\color{magenta}Niels Bohr Institute, Blegdamsvej 17, DK-2100 Copenhagen \O, Denmark}}

\begin{abstract}
We unveil the temperature-dependent electroweak
phase transition in new extensions of the Standard Model in which the
electroweak symmetry is spontaneously broken via strongly coupled,
nearly-conformal dynamics achieved by the means of multiple matter representations. In particular, we focus on the low energy
effective theory introduced to describe Ultra Minimal Walking Technicolor at the
phase transition. Using the one-loop effective potential with ring
improvement, we identify regions of parameter space which
yield a  strong first order transition. 
A striking feature of the model is the existence of a second phase
 transition associated to the electroweak-singlet sector. The interplay between these two transitions leads to an extremely rich phase diagram.

\end{abstract}

\maketitle

\section{Introduction}
Models of dynamical electroweak symmetry breaking constitute some of the best motivated extensions of the Standard Model (SM) of particle interactions.  Within these models one can simultaneously address, in a natural way, the breaking of the electroweak symmetry and the origin of dark matter.

An intriguing research topic to explore within these extensions of the SM is the one associated to the possible generation of  the experimentally observed baryon asymmetry of the universe at the electroweak phase transition (EWPT)\cite{Shaposhnikov:1986jp,Shaposhnikov:1987tw,Shaposhnikov:1987pf,
Farrar:1993sp}.  {}For the
mechanism to be applicable it requires the presence of new physics
beyond the SM
\cite{Nelson:1991ab,Joyce:1994bi,Joyce:1994fu,Cline:1995dg,cjk,wkb}. {An essential condition for electroweak
baryogenesis is that the  baryon-violating interactions induced by
electroweak sphalerons are sufficiently slow immediately after the
phase transition to avoid the destruction of the baryons that have
just been created.  This is achieved when the thermal average of the
Higgs field evaluated on the ground state,  in the broken phase of
the electroweak symmetry, is large enough compared to the critical
temperature at the time of the transition (see for example
ref.~\cite{Cline:2006ts} and references therein),
\beq
        \phi_c/ T_c   > 1.
\label{cond}
\eeq
In the SM, the bound (\ref{cond}) was believed to be satisfied
only for very light Higgs
bosons \cite{Carrington:1991hz,Arnold:1992fb,Arnold:1992rz,
Anderson:1991zb,Dine:1992wr}.  However, this was before the mass
of the top quark was known.  With $m_t=175$ GeV, nonperturbative studies
of the phase transition \cite{Kajantie:1995kf,Kajantie:1996mn,Rummukainen:1998as} show that
the bound (\ref{cond}) cannot be satisfied for {\it any} value of the
Higgs mass (see also \cite{Gynther:2005av,Gynther:2005dj}).
In addition to the
difficulties with producing a large enough initial baryon asymmetry,
the impossibility of satisfying the sphaleron constraint (\ref{cond})
in the SM provides an incentive for seeing whether the situation
improves in various extensions of the SM \cite{CQW,DR,CM,LR}.

In this paper we explore the electroweak phase transition in the model put forward in \cite{Ryttov:2008xe}.
This is an explicit example of (near) conformal (NC) technicolor \cite{Holdom:1984sk,Holdom:1983kw,Eichten:1979ah,Holdom:1981rm,Yamawaki:1985zg,Appelquist:an,Appelquist:1999dq} with two types of technifermions transforming according to two different representations of the underlying technicolor gauge group \cite{Dietrich:2006cm,Lane:1989ej}. The model possesses a number of interesting properties to recommend it over the earlier models of dynamical electroweak symmetry breaking:
\begin{itemize}
\item
 Features the lowest possible value of the naive $S$ parameter \cite{Peskin:1990zt,{Peskin:1991sw}} while possessing a dynamics which is NC.

 \item Contains, overall, the lowest possible number of fermions.

\item Yields natural dark matter (DM) candidates.
\end{itemize}
Due to the above properties we termed this model {\it Ultra Minimal near conformal Technicolor} (UMT). It is constituted by an $SU(2)$ technicolor gauge group with two Dirac flavors in the fundamental representation also carrying electroweak charges, as well as, two additional Weyl fermions in the adjoint representation but singlets under the SM gauge groups.

We arrived at the UMT model while investigating, in a systematic way, the parameter space of the possible strongly coupled theories one can use to construct models of dynamical electroweak symmetry breaking \cite{Sannino:2004qp,Dietrich:2006cm,Ryttov:2007sr,Ryttov:2007cx,Sannino:2008ha,Sannino:2009aw}. More specifically we uncovered, using various analytic methods, the phase diagrams as function of colors and flavors for $SU(N)$ gauge theories in  \cite{Sannino:2004qp,Dietrich:2006cm,Ryttov:2007sr,Ryttov:2007cx,Sannino:2008ha,Sannino:2008pz} and for $SO(N)$ and $Sp(2N)$ theories in  \cite{Sannino:2009aw}. By direct comparison of the various phase diagrams an intriguing universal picture emerges \cite{Sannino:2009aw}. The results of these investigations led to Minimal Walking Technicolor (MWT)\cite{Sannino:2004qp,Dietrich:2005jn,Dietrich:2006cm,Foadi:2007ue} and also to UMT \cite{Ryttov:2008xe}. These models make use of higher dimensional representations which have already been used in the past. Time-honored examples are grand unified models. Theories with fermions transforming according to higher dimensional representations develop an infrared fixed point (IRFP) for a very small number of flavors and colors  \cite{Sannino:2004qp,Dietrich:2006cm,Ryttov:2007cx}. This was considered unlikely to occur for nonsupersymmetric gauge theories with fermionic matter \cite{Hill:2002ap}.  The relevance of this discovery  \cite{Sannino:2004qp} is that it allows for the construction of several explicit UV-complete models able to break the electroweak symmetry dynamically while the models naturally feature small contributions to the electroweak precision parameters \cite{Appelquist:1998xf,Kurachi:2006mu,Foadi:2007ue}. It also helps alleviating the Flavor Changing Neutral Currents problem while featuring explicit candidates of asymmetric dark matter \cite{Foadi:2007ue,Ryttov:2008xe}. The models are also economical since they require the introduction of a very small number of underlying elementary fields and can feature a light composite Higgs \footnote{Note that even the traditional technicolor model, which is  a scaled up copy of QCD (i.e. three technicolors), contains at low energies a spin zero composite state the $\sigma$, with the same quantum numbers of the Higgs, lighter than the technirho. In fact it is impossible to fit the pion-pion scattering data in QCD without including such a spin zero state lighter than the associated vector meson $\rho$. The fact that the $\sigma$ state is broad is inconsequential to the argument that to unitarize pion-pion scattering one needs such a state \cite{Sannino:1995ik,Harada:1995dc,Harada:1996wr,Black:1998zc,Black:1998wt,Harada:2003em,Sannino:2007yp}. Recently other analysis also confirm our  original findings\cite{Maiani:2004uc,Caprini:2005zr}. Another important point is that in a field theoretical framework the only way, independent from perturbation theory, that one can classify states is as poles in the scattering amplitudes. In the Appendix F of \cite{Sannino:2008ha} we have showed in which limits one expects a truly Higgless strongly interacting theory. {}For example, if the number of (techni)colors is larger than six with (techni)flavors in the fundamental representation then one can unitarize pion-pion scattering only with a vector meson.  } \cite{Dietrich:2005jn,Dietrich:2006cm,Hong:2004td}.  Recent analysis further support this latter observation \cite{Doff:2009nk,Doff:2008xx} for walking models.

On the astrophysical side, technicolor models are capable of
providing interesting dark matter candidates, since the  new strong
interactions confine techniquarks in technimeson and technibaryon
bound states. The spin of the technibaryons depends on the
representation according to which the technifermions transform, and
the numbers of flavors and colors. The lightest technimeson is
short-lived, thus evading BBN constraints, but the lightest
technibaryon typically  has \footnote{there may be situations in which
the technibaryon is a goldstone boson of an enhanced flavor
symmetry} a mass of the order
\begin{equation}
 m_{TB}  \sim  1-2\ {\rm TeV} \ .
\end{equation}
However, in UMT the technibaryon is a pseudo Goldstone boson and hence can be substantially lighter then a TeV making it possible to observe it at the Large Hadron Collider experiment \cite{Foadi:2008qv}.
 Technibaryons are therefore natural dark matter candidates
\cite{Nussinov:1985xr,Barr:1990ca,Gudnason:2006yj}. In fact it is
possible to {naturally} understand the observed ratio of the dark to
luminous matter mass fraction of the universe if the technibaryon
possesses an asymmetry
\cite{Nussinov:1985xr,Barr:1990ca,Gudnason:2006yj}. If the latter is
due to a net $B-L$ generated at some high energy scale, then this
would be subsequently distributed among all electroweak
doublets by fermion-number violating processes in the SM at
temperatures above the electroweak scale
\cite{Shaposhnikov:1991cu,Kuzmin:1991ft,Shaposhnikov:1991wi}, thus
naturally generating a technibaryon asymmetry as well. To avoid
experimental constraints the technibaryon should be constructed in
such a way as to be a complete singlet under the electroweak
interactions\cite{Foadi:2008qv}

Interesting applications have been envisioned for the LHC phenomenology \cite{Sannino:2004qp,Foadi:2007ue,Belyaev:2008yj,Christensen:2005cb,Gudnason:2006mk,Dietrich:2009ix} and for Cosmology \cite{Nussinov:1985xr,Barr:1990ca,Foadi:2008qv,Ryttov:2008xe,Nardi:2008ix, Gudnason:2006yj,Kainulainen:2006wq,Kouvaris:2007iq,Cline:2008hr,Kouvaris:2008hc,Kikukawa:2007zk,Jarvinen:2009wr,Antipin:2009ch}.

The nonperturbative dynamics of these models is being investigated via first principles lattice computations by several groups \cite{Catterall:2007yx,Catterall:2008qk,
Shamir:2008pb,DelDebbio:2008wb,DelDebbio:2008zf, Hietanen:2008vc,Hietanen:2008mr,Appelquist:2007hu,Deuzeman:2008sc,Fodor:2008hn}. In the literature the reader can also find attempts to gain information using  gauge-gravity type duality \cite{Hirn:2008tc,Dietrich:2008ni,Nunez:2008wi}.

The order of the electroweak
phase transition (EWPT) depends on the underlying type of strong dynamics
and  plays an important role for baryogenesis
\cite{Cline:2002aa,Cline:2006ts}. The technicolor chiral phase
transition at finite temperature is mapped onto the electroweak one.
Attention must be paid to the way in which the electroweak symmetry
is embedded into the global symmetries of the underlying technicolor
theory.  We analyzed the EWPT, at the effective Lagrangian level, for MWT in \cite{Cline:2008hr} while an interesting analysis dedicated to earlier
models of technicolor has been performed in \cite{Kikukawa:2007zk}.

In this work, we investigate the EWPT in a new class of realistic
and viable technicolor models such as the UMT model. We will uncover an extremely rich finite temperature phase diagram.

We will use as a template the low energy
effective theory developed in  \cite{Ryttov:2008xe}. We start in
section \ref{sect2} by summarizing the UMT model. In section \ref{sect3} we highlight the
degrees of freedom relevant near the phase transition and the zero temperature effective Lagrangian.  In section
\ref{sect4} we present the finite-temperature effective potential
computed at the one-loop order, including the resummation of ring
diagrams.  We set up the analysis in section \ref{sect5} and analyzes the results in \ref{sect6}.  We briefly summarize the main point in the last section. Several appendices are provided, which give
details concerning our analytical results.

As a preliminary investigation we adopt the high-temperature expansion
results for the effective potential.  We then explore the regions of
the effective theory parameters yielding first-order phase
transitions and study their strength and interplay. The ratios of the composite Higgses
thermal expectation values at the critical temperatures divided by the
corresponding temperatures are determined as function of the parameters
of the low energy effective theory. We identify a region
of parameter space where this ratio is sufficiently large to induce
electroweak baryogenesis.  An extremely rich structure emerges.  {}For example, we observe (depending on the parameters) the existence of extra electroweak phase transitions \cite{Jarvinen:2009wr} emerging  at different values of the temperature. We also consider the possibility of having simultaneous phase transitions of the various global symmetries related to the new strong dynamics. We find regions of the effective Lagranian parameters allowing for a sufficiently strong first-order electroweak phase transition able to support electroweak baryogenesis.

\section{The Model}
\label{sect2}

The model proposed in \cite{Ryttov:2008xe} consists of an $SU(2)$ gauge group with two Dirac fermions belonging to the fundamental representation and two Weyl fermions belonging to the adjoint representation. In order not to be in conflict with the Electroweak Precision Tests only the fundamental fermions are charged under the electroweak symmetry. They are arranged into electroweak doublets in the standard way and may be written as
\begin{eqnarray}
T_L = \left( \begin{array}{c} U \\ D  \end{array}  \right)_L \ , \qquad U_R \ , \ \  D_R \ .
\end{eqnarray}
The adjoint fermions needed to render the theory near conformal are denoted as $\lambda^f$ with $f=1,2$. The global symmetries of the theory are most appropriately handled by first arranging the fundamental fermions into a quadruplet of $SU(4)$

\begin{eqnarray}
Q &=& \left( \begin{array}{c}
U_L \\
D_L \\
-i\sigma^2U_R^* \\
-i\sigma^2D_R^*
\end{array} \right) \ .
\end{eqnarray}

Since the fermions belong to pseudo-real and real representations of the gauge group the global symmetry of the theory is enhanced and can be summarized as
\begin{eqnarray}
\begin{array}{c||ccc}\label{symmetries}
 &\qquad SU(4) &\qquad SU(2) &\qquad U(1) \\ \hline\hline
Q &\qquad \fund &\qquad 1 &\qquad -1  \\
\lambda &\qquad 1 &\qquad \fund &\qquad \frac{1}{2}
\end{array}
\end{eqnarray}
The abelian symmetry is anomaly free. The global symmetry group $G=SU(4)\times  SU(2) \times U(1)$ breaks to $H= Sp(4) \times SO(2)\times Z_2$. The stability group $H$ is dictated by the (pseudo)reality of the fermion representations and the breaking is triggered by the formation of the following two condensates
\begin{eqnarray}\label{condensate1}
\langle Q_F^{\alpha, c} Q_{F'}^{\beta, c'} \epsilon_{\alpha\beta} \epsilon_{cc'} E_4^{FF'} \rangle &=& -2 \langle \overline{U}_R U_L + \overline{D}_R D_L \rangle \\
\langle \lambda_{f}^{\alpha, k}\lambda_{f'}^{\beta, k'} \epsilon_{\alpha\beta} \delta_{kk'} E_2^{ff'} \rangle &=& -2 \langle \lambda^1 \lambda^2 \rangle \label{condensate2}
\end{eqnarray}
where
\begin{eqnarray}
E_4 = \left( \begin{array}{cc}
0_{2\times2} & \mathbf{1}_{2\times2} \\
-\mathbf{1}_{2\times2} & 0_{2\times2}
\end{array} \right) \ , \qquad E_2 = \left( \begin{array}{cc}
0 & 1 \\
1 & 0
\end{array} \right)
\end{eqnarray}
The flavor indices are denoted with $F,F'=1,\ldots,4$ and $f,f'=1,2$, the spinor indices as $\alpha,\beta=1,2$ and the color indices as $c,c'=1,2$ and $k,k'=1,\ldots,3$. Also the notation is such that $U_L^{\alpha} U_R^{*\beta} \epsilon_{\alpha\beta} = -\overline{U}_RU_L$ and $\lambda^{1,\alpha} \lambda^{2, \beta} \epsilon_{\alpha\beta} = \lambda^1 \lambda^2$. Under the $U(1)$ symmetry $Q$ and $\lambda$ transform as
\begin{equation}
Q \rightarrow e^{-i\alpha}Q \ , \qquad {\rm and } \qquad \lambda \rightarrow e^{-i\frac{\alpha}{2}} \lambda \ ,
\end{equation}
and the two condensates are simultaneously invariant if
\begin{equation}
\alpha = 2 k \pi \ , \qquad {\rm with}~ k ~{\rm an~integer}\ .
\end{equation}
Only the $\lambda$ fields will transform nontrivially under the remaining $Z_2$, i.e.  $\lambda \rightarrow - \lambda$.

In principle this extension suffers a vacuum alignment problem \cite{Peskin:1980gc} due to the fact that the electroweak sector would tend to destabilize the chosen vacuum direction. This would be true in absence of new interactions needed to, for example, provide mass to some of the unwanted Goldstone bosons. In fact, the first operator in Eq.~(\ref{LETC}) needed to give mass to one of these Goldstones naturally re-aligns the vacuum. A recent analysis of the vacuum alignment effects in walking technicolor theories such as MWT can be found in \cite{Dietrich:2009ix}. 

\section{Low Energy Spectrum}
\label{sect3}

We shall only consider the effect of the composite scalar mesons which are expected to be the lightest particles in the theory. Their masses have the strongest dependence on the vacuum expectations values of the Higgs fields.

The relevant degrees of freedom are efficiently collected in two distinct matrices, $M_4$ and $M_2$, which transform as $M_4 \rightarrow g_4M_4g_4^T$ and $M_2 \rightarrow g_2M_2g_2^T$ with $g_4 \in SU(4)$ and $g_2 \in SU(2)$. Both $M_4$ and $M_2$ consist of a composite iso-scalar and its pseudoscalar partner together with the Goldstone bosons and their scalar partners
\begin{eqnarray}
M_4 &=& \left[ \frac{\sigma_4 + i \Theta_4}{2} + \sqrt{2}\left( i \Pi_4^i+ \tilde{\Pi}_4^i \right) X_4^i \right] E_4 \ , \qquad i=1,\ldots,5 \ , \\
M_2 &=& \left[ \frac{\sigma_2 + i \Theta_2}{\sqrt{2}} + \sqrt{2} \left( i \Pi_2^i+ \tilde{\Pi}_2^i \right) X_2^i \right] E_2 \ , \qquad i=1,2 \ .
\end{eqnarray}

The notation is such that $X_{4}$ and $X_{2}$ are the broken generators of $SU(4)$ and $SU(2)$ respectively. An explicit realization can be found in Appendix \ref{generators}. Also $\sigma_{4}$ and $\Theta_{4}$ are the composite Higgs and its pseudoscalar partner while $\Pi_{4}^i$ and $\tilde{\Pi}_{4}^i$ are the Goldstone bosons and their associated scalar partners. For $SU(2)$ one simply substitutes the index $4$ with the index $2$.

Having included the $\Theta$ and $\tilde{\Pi}^i$ states one should note that the matrices are actually form invariant under $U(4)$ and $U(2)$ with the abelian parts being broken by anomalies. With the above normalization of the $M$ matrices the kinetic term of each component field is canonically normalized.

The relation -- which can be found in Appendix \ref{AppA} -- between the composite scalars and the underlying degrees of freedom can be found by noting that $M_4$ and $M_2$ transform as
\begin{eqnarray}
M_4^{FF'} \sim Q^FQ^{F'} \ , \qquad M_2^{ff'} \sim \lambda^f \lambda^{f'}
\end{eqnarray}
where both color and spin indices have been contracted.

To describe the interaction with the weak gauge bosons we embed the electroweak gauge group in $SU(4)$ as done in \cite{Appelquist:1999dq}.  First we note that the following generators
\begin{eqnarray}
L^a = \frac{S^a_4+X^a_4}{\sqrt{2}} = \left( \begin{array}{cc}
\frac{\tau^a}{2} &  \\
 & 0
\end{array} \right) \ , \qquad
R^a = \frac{X^{aT}_4-S^{aT}_4}{\sqrt{2}} = \left( \begin{array}{cc}
0 & \\
 & \frac{\tau^a}{2}
\end{array} \right)
\end{eqnarray}
with $a=1,2,3$ span an $SU(2)_L\times SU(2)_R$ subalgebra. By gauging $SU(2)_L$ and the third generator of $SU(2)_R$ we obtain the electroweak gauge group where the hypercharge is $Y=-R^3$. Then as $SU(4)$ breaks to $Sp(4)$ the electroweak gauge group breaks to the electromagnetic one with the electric charge given by $Q=\sqrt{2}S^3$.

Due to the choice of the electroweak embedding the weak interactions explicitly reduce the $SU(4)$ symmetry to $SU(2)_L \times U(1)_Y \times U(1)_{TB}$ which is further broken to $U(1)_{\rm em} \times U(1)_{TB}$ via the technicolor interactions. $U(1)_{TB}$ is the technibaryon number related to the fundamental fermions and its generator corresponds to the $S_4^4$ diagonal generator (Appendix \ref{generators}). The remaining $SU(2)\times U(1)$ spontaneously breaks, via the extra technifermion condensates, to $SO(2)\times Z_2$. Here $SO(2)\cong U(1)$ is the technibaryon number related to the adjoint fermions. 

With the above discussion of the electroweak embedding the covariant derivative for $M_4$ is
\begin{eqnarray}
D_{\mu} M_4 &=& \partial_{\mu} M_4 - i \left[ G_{\mu} M_4 + M_4 G_{\mu}^T  \right] \ , \qquad G_{\mu} =
\left( \begin{array}{cc}
g W_{\mu}^a \frac{\tau^a}{2} & 0 \\
0 & -g' B_{\mu} \frac{\tau^3}{2}
\end{array} \right)  \ .
\end{eqnarray}

We are now in a position to write down the effective Lagrangian. It contains the kinetic terms and a potential term
\begin{eqnarray} \label{Leff}
\mathcal{L} &=& \frac{1}{2} \text{Tr} \left[ D_{\mu} M_4 D^{\mu} M_4^{\dagger} \right] + \frac{1}{2} \text{Tr} \left[ \partial_{\mu} M_2 \partial^{\mu} M_2^{\dagger} \right] - \mathcal{V} \left( M_4, M_2 \right)  \ ,
\end{eqnarray}
where the potential is
\begin{eqnarray}
\mathcal{V} \left( M_4, M_2 \right) &=& -\frac{m_4^2}{2} \text{Tr}\left[ M_4M_4^{\dagger} \right] + \frac{\lambda_4}{4}\text{Tr}\left[ M_4M_4^{\dagger} \right]^2 + \lambda_4' \text{Tr} \left[ M_4M_4^{\dagger}M_4M_4^{\dagger} \right] \nonumber \\
&&
-\frac{m_2^2}{2} \text{Tr}\left[ M_2M_2^{\dagger} \right] + \frac{\lambda_2}{4}\text{Tr}\left[ M_2M_2^{\dagger} \right]^2 + \lambda_2' \text{Tr} \left[ M_2M_2^{\dagger}M_2M_2^{\dagger} \right] \\
&&
+ \frac{\delta}{2}\text{Tr}\left[ M_4M_4^{\dagger} \right] \text{Tr}\left[ M_2M_2^{\dagger} \right] +4\delta' \left[ \left( \det M_2 \right)^2 \text{Pf}\ M_4 + \text{h.c.} \right] \ . \nonumber
\end{eqnarray}

Once $M_4$ and $M_2$ each develop a vacuum expectation value the electroweak symmetry breaks and three of the eight Goldstone bosons - $\Pi^0,\ \Pi^+$ and $\Pi^-$ - will be eaten by the massive gauge bosons. In terms of the parameters of the theory the vacuum states $\langle \sigma_4 \rangle = v_4$ and $\langle \sigma_2 \rangle = v_2$ which minimize the potential are a solution of the two coupled equations
\begin{eqnarray}\label{vacua1}
0 &=& -m_4^2 - \left( \delta' v_2^2 - \delta \right) v_2^2 + \left( \lambda_4 + \lambda_4' \right) v_4^2 \ , \\
0 &=& -m_2^2 - \left(   2\delta' v_2^2 - \delta \right) v_4^2 + \left( \lambda_2 + 2\lambda_2' \right) v_2^2 \ . \label{vacua2}
\end{eqnarray}

Expanding around the symmetry breaking vacua all of the Goldstone bosons scalar partners are seen to be mass eigenstates with masses
\begin{eqnarray}
M^2_{\tilde{\Pi}^0} = M^2_{\tilde{\Pi}^{\pm}} = M^2_{\tilde{\Pi}_{UD}} = 2 \left( \lambda_4' v_4^2 + \delta' v_2^4 \right) \ , \qquad M^2_{\tilde{\Pi}_{\lambda\lambda}} = 4v_2^2 \left( \lambda_2'+ \delta' v_4^2 \right) \ ,
\end{eqnarray}
while the Goldstone bosons which are not eaten by the massive gauge bosons of course have vanishing mass $M^2_{\Pi_{UD}} = M^2_{\Pi_{\lambda\lambda}} = 0$. Here the vacuum expectation values $v_4$ and $v_2$ are solutions to Eq.~(\ref{vacua1}) and (\ref{vacua2}). Due to the presence of the determinant/Pfaffian term in the potential the remaining states are not mass eigenstates. Specifically $H_4$ and $H_2$ and their associated pseudoscalar partners will mix. In the diagonal basis we find the following mass eigenstates
\begin{equation}
\begin{array}{rclcrcl}\label{mesons}
\Theta  & \equiv &  \sin (\alpha)\ \Theta_4 + \cos (\alpha)\ \Theta_2 & ,~~~~ & M_{\Theta}^2 & = & 0 \ , \\
\tilde{\Theta}  & \equiv & \cos (\alpha)\ \Theta_4 - \sin (\alpha)\ \Theta_2  & ,~~~~ & M_{\tilde{\Theta}}^2 & = & 2 \delta' v_2^2 \left( v_2^2 + 4 v_4^2 \right) \ , \\
H_-  & \equiv & \sin (\beta)\ H_4 + \cos (\beta)\ H_2 & ,~~~~ & M^2_{H_-} & = &  m_2^2 +m_4^2 + k_-  \ , \\
H_+  & \equiv & \cos (\beta)\ H_4 -  \sin(\beta) \ H_2 & ,~~~~ & M^2_{H_+} & = & m_2^2 +m_4^2 + k_+ \ , \\
\end{array}
\end{equation}
with
\begin{eqnarray}
\tan (2\alpha) = \frac{4v_4v_2}{v_2^2-4v_4^2} \ , \qquad \tan (2\beta) = \frac{2v_2v_4\left(   2 \delta' v_2^2 - \delta \right)}{m_2^2 - m_4^2 - \delta v_4^2 - \left(   \delta' v_2^2 - \delta \right) v_2^2}  \ ,
\end{eqnarray}
\begin{eqnarray}
k_{\pm}= \left(   \delta'v_2^2 - \delta \right) v_2^2 - \delta v_4^2 \pm \left[ \left( m_4^2 - m_2^2 + \left(   \delta' v_2^2 - \delta \right)v_2^2 + \delta v_4^2 \right)^2 + \left( 2v_2v_4 \left(   \delta' v_2^2 - \delta \right) \right)^2 \right]^{\frac{1}{2}}
\end{eqnarray}
Note that we have one massless state $\Theta$ which we identify with the original $U(1)$ Goldstone boson while $\tilde{\Theta}$ is massive. In the limit $\delta' \rightarrow 0$ both states are massless and at the classical level the global symmetry is enhanced to $U(4)\times U(2)$.

For the model to be phenomenologically viable some of the Goldstones must acquire a mass. Here we parameterize the ETC interactions by adding at the effective Lagrangian level the operators needed to give the unwanted Goldstone bosons an explicit mass term.

The effective ETC Lagrangian breaks the global $SU(4)\times SU(2) \times U(1)$ symmetry. The $SU(4)$ generator commuting with the $SU(2)_L \times SU(2)_R$ generators is $B_4 = 2\sqrt{2} S_4^4$. To construct, at the effective Lagrangian level, the required ETC terms we find it useful to split
$M_4$ ($M_2$) -- form invariant under $U(4)$ ($U(2)$) --  as follows:
\begin{equation}
M_4 = \tilde{M}_4  + i P_4 \ , \qquad  {\rm and} \qquad M_2 = \tilde{M}_2  + i P_2 \ ,
\end{equation}
with
\begin{eqnarray}
\tilde{M}_4 &=& \left[ \frac{\sigma_4}{2} + i\sqrt{2}  \Pi_4^i   X_4^i \right]   E_4 \ , \quad P_4 = \left[ \frac{ \Theta_4}{2}  -i\sqrt{2} \tilde{\Pi}_4^i X_4^i \right] E_4 \ , \qquad i=1,\ldots,5 \ , \\
\tilde{M}_2 &=& \left[ \frac{\sigma_2 }{\sqrt{2}} + i\sqrt{2}  \Pi_2^i  X_2^i \right] E_2 \ , \quad P_2 = \left[ \frac{ \Theta_2}{\sqrt{2}}  -i\sqrt{2} \tilde{\Pi}_2^i  X_2^i \right] E_2 \ , \qquad i=1,2 \ .\end{eqnarray}
$\tilde{M}_4 $ ($\tilde{M}_2$) as well as $P_4$ ($P_2$) are separately $SU(4)$ ($SU(2)$) form invariant. A set of operators able to give masses to the electroweak neutral Goldstone bosons is:
\begin{eqnarray} \label{LETC}
\mathcal{L}_{ETC} &=&  \frac{m_{4,ETC}^2}{4} \text{Tr} \left[ \tilde{M}_4B_4 \tilde{M}_4^{\dagger} B_4 + \tilde{M}_4 \tilde{M}_4^{\dagger} \right] + \frac{m_{2,ETC}^2}{4} \text{Tr} \left[ \tilde{M}_2B_2 \tilde{M}_2^{\dagger} B_2 + \tilde{M}_2 \tilde{M}_2^{\dagger} \right] \nonumber \\
&& - m_{1,ETC}^2 \left[ \text{Pf}\ P_4 + \text{Pf}\ P_4^{\dagger} \right] - \frac{m_{1,ETC}^2}{2} \left[ \det(P_2) + \det(P_2^{\dagger}) \right] \ ,
\end{eqnarray}
where $B_2=2S_2^1$. The spectrum is:
\begin{eqnarray}
M^2_{\Pi_{UD}} = m_{4,ETC}^2 \ , \qquad M^2_{\Pi_{\lambda\lambda}} = m_{2,ETC}^2 \ , \qquad M^2_{\Theta} = m_{1,ETC}^2 \ ,
\end{eqnarray}
for the Goldstone bosons that are not eaten by the massive vector bosons and:
\begin{eqnarray}
M^2_{\tilde{\Pi}_{UD}} &=& M^2_{\tilde{\Pi}^0} = M^2_{\tilde{\Pi}^{\pm}} = 2\left( \lambda'_4 v_4^2 + \delta' v_2^4 \right) + m^2_{1,ETC} \ , \\
M^2_{\tilde{\Pi}_{\lambda\lambda}} &=& 4v_2^2 \left( \lambda_2' + \delta' v_4^2 \right) + m^2_{1,ETC} \ , \\
M_{\tilde{\Theta}}^2 &=& 2\delta' v_2^2 \left( v_2^2 + 4v_4^2 \right) + m_{1,ETC}^2 \ ,
\end{eqnarray}
for the pseudoscalar and scalar partners. The masses of the two Higgs particles $H_+$ and $H_-$ are unaffected by the addition of the ETC low energy operators.

\section{Effective Potential} 
\label{sect4}

We shall construct the one-loop finite-temperature effective potential for two different cases:
\begin{itemize}
 \item[i)] UMT without EW, only the strong dynamics described by (\ref{Leff}) without coupling to the SM and ETC interactions;
 \item[ii)] UMT with EW, including electroweak gauge bosons, the top quark, and the ETC term (\ref{LETC}) of the effective Lagrangian.
\end{itemize}
Let us start by reviewing the formalism for the effective potential in general. It is obtained by adding to the tree level potential $V^{(0)}$ the one-loop correction $V^{(1)}$
\begin{eqnarray}
 V(\sigma_4,\sigma_2,T) = V^{(0)}(\sigma_4,\sigma_2) + V^{(1)}_{T=0}(\sigma_4,\sigma_2) + V^{(1)}_{T}(\sigma_4,\sigma_2,T) \ .
\end{eqnarray}
For brevity, we denote by $\sigma_i$ the expectation values of the Higgs fields (as well as the fields themselves) that characterize the two techniquark condensates.
The standard zero-temperature one loop contribution to the potential reads:
\begin{eqnarray} \label{VT0}
V^{(1)}_{T=0} = \frac{1}{64\pi^2} \sum_{i} \bar{n}_i\, f_{i}(M_i(\sigma_4,\sigma_2)) \ ,
\end{eqnarray}
where the index $i$ runs over all of the mass eigenstates and $\bar{n}_i$ is the multiplicity
factor for a given scalar particle $n_b$ while for Dirac fermions it is $-4$
times the multiplicity factor of the specific fermion $n_f$. The function $f_i$ is:
\begin{equation} \label{fdef}
f_i = M^4_i(\sigma_4,\sigma_2) \left[\log\frac{M^2_{i}(\sigma_4,\sigma_2)}{M^2_i(v_4,v_2)}  - \frac{3}{2}\right] + 2M^2_i(\sigma_4,\sigma_2) \, M^2_i(v_4,v_2)  \ ,
\end{equation}
where $M^2_i(\sigma_4,\sigma_2)$ is the background dependent mass term of the
$i$-th particle.

The one-loop, ring-improved, correction can be divided into fermionic, scalar and vector contributions,
\begin{eqnarray}
V_T^{(1)} = {V_T^{(1)}}_{\rm f}+ {V_T^{(1)}}_{\rm b}+ {V_T^{(1)}}_{\rm gauge}\ .
\end{eqnarray}
The high temperature expansion of the fermionic contribution reads:
\begin{eqnarray} \label{VTf}
{V_T^{(1)}}_{\rm f} &=& 2\frac{T^2}{24} \sum_{f} n_{f} M_{f}^2(\sigma_4,\sigma_2) +\frac{1}{16 \pi^2}\sum_fn_{f} M_{f}^4(\sigma_4,\sigma_2)\left[\log\frac{M_{f}^2(\sigma_4,\sigma_2)}{T^2}-c_f\right] \nonumber\\
&& - 2\sum_{f} n_{f} M_{f}^2(\sigma_4,\sigma_2)T^2 \sum_{l=2}^\infty\left(\frac{-M_{f}^2(\sigma_4,\sigma_2)}{4\pi^2T^2}\right)^l \frac{(2l-3)!!\zeta(2l-1)}{(2l)!!(l+1)}\left(2^{2l-1}-1\right)
\end{eqnarray}
where $c_f \simeq 2.63505$.
For the scalar part of the thermal potential one must resum the
contribution of the ring diagrams. Following Arnold and Espinosa
\cite{Arnold:1992rz} we write
\begin{eqnarray} \label{VTb}
{V_T^{(1)}}_{\rm b} &=&\frac{T^2}{24} \sum_{b} n_{b} M_{b}^2(\sigma_4,\sigma_2)  - \frac{T}{12\pi} \sum_b\,n_b \,M_{b}^3(\sigma_4,\sigma_2,T) \nonumber\\ &&-\frac{1}{64 \pi^2}\sum_bn_{b} M_{b}^4(\sigma_4,\sigma_2)\left[\log\frac{M_{b}^2(\sigma_4,\sigma_2)}{T^2}-c_b\right] \nonumber\\
 && +\sum_{b} n_{b} \frac{M_{b}^2(\sigma_4,\sigma_2)T^2}{2} \sum_{l=2}^\infty\left(\frac{-M_{b}^2(\sigma_4,\sigma_2)}{4\pi^2T^2}\right)^l \frac{(2l-3)!!\zeta(2l-1)}{(2l)!!(l+1)}
\end{eqnarray}
where $c_b\simeq 5.40762 $ and $M_b(\sigma_4,\sigma_2,T)$ the thermal mass
which follows from the tree-level plus one-loop thermal contribution to the potential.  {}For the gauge bosons,
\begin{eqnarray} \label{VTgb}
{V_T^{(1)}}_{\rm gauge} &=&\frac{T^2}{24} \sum_{gb} 3 M_{gb}^2(\sigma_4,\sigma_2)  - \frac{T}{12\pi} \sum_{gb}\left[ 2 M_{T,gb}^3(\sigma_4,\sigma_2)+ M_{L,gb}^3(\sigma_4,\sigma_2,T)\right] \nonumber \\
&&-\frac{1}{64 \pi^2}\sum_{gb} 
 M_{gb}^4(\sigma_4,\sigma_2)\left[\log\frac{M_{gb}^2(\sigma_4,\sigma_2)}{T^2}-c_b\right] \nonumber \\
&& + \sum_{gb}\frac{M_{gb}^2(\sigma_4,\sigma_2)T^2}{2} \sum_{l=2}^\infty\left(\frac{-M_{gb}^2(\sigma_4,\sigma_2)}{4\pi^2T^2}\right)^l \frac{(2l-3)!!\zeta(2l-1)}{(2l)!!(l+1)}\ .
\end{eqnarray}
Here $M_{T,gb}$ ($M_{L,gb}$ ) is the transverse (longitudinal) mass
of a given gauge boson and we have $M_{T,gb}(\sigma_4,\sigma_2)=
M_{L,gb}(\sigma_4,\sigma_2,T=0) = M_{gb}(\sigma_4,\sigma_2)$.

\subsection{Effective potential in UMT: without EW} 

Let us now discuss how the different terms of the effective potential appear in UMT. We shall start with the case without electroweak interactions. By using (\ref{Leff}) we find for the tree level potential
\begin{eqnarray} \label{Vtree}
V^{(0)} &=& \frac{1}{4}\left( \lambda_4 + \lambda_4' \right) \left( \sigma_4^2-v_4^2 \right)^2 + \frac{1}{4}\left( \lambda_2 +2\lambda_2' - 2v_4^2\delta' \right) \left( \sigma_2^2-v_2^2 \right)^2 \nonumber \\
&& + \frac{1}{2} \left( \delta -2v_2^2\delta' \right) \left( \sigma_4^2-v_4^2 \right) \left( \sigma_2^2-v_2^2 \right) - \frac{1}{2} \delta' \left( \sigma_4^2-v_4^2 \right) \left( \sigma_2^2-v_2^2 \right)^2 \ ,
\end{eqnarray}
where we have added a constant term. It is clear that at very large values of the fields the potential is unstable for a positive value of $\delta^{\prime}$. This is needed to provide a positive  mass squared term for the $\tilde{\Theta}$ particle. However this is not a problem since the potential is valid only below the TeV scale where the minimum is present. In fact, for reasonable values of the $\tilde{\Theta}$ mass the value of $\delta^{\prime}$ is extremely small.

In the temperature-independent one-loop correction (\ref{VT0}) we sum over all scalar states of the effective Lagrangian except the Goldstone bosons and the $\Theta$ particle \footnote{We have also neglected the one-loop zero temperature contribution of the $\tilde{\Theta}$ state which is also a Goldstone boson when $\delta^{\prime}$ vanishes. We have retained these states in the section in which the electroweak sector is taken into account.}. Our prescription would lead to  infrared
divergences in the 't Hooft-Landau gauge for these contributions, when evaluated at the tree-level VEV, due to the
vanishing of the masses.
A simple approach is to neglect the massless
states here, since their effect on the
phase transition is small. The background dependent scalar masses are those of Appendix~\ref{AppA}. Given that we are not discussing the electroweak physics in this subsection we set $m_{i,ETC}$ to zero.

For the temperature dependent scalar contribution ${V_T^{(1)}}_b$ we keep terms up to and including $l=4$, i.e., ${\cal O}\left(1/T^6\right)$. The background dependent masses are given in Appendix~\ref{AppA}, and the temperature-dependent masses are obtained from the expressions of Appendix~\ref{AppB} by setting $m_{\rm top}=g=g^\prime=0$. There are no fermion nor gauge boson contributions when the electroweak effects are excluded.

\subsection{Effective potential in UMT: with EW} 

Here we add the effects of the electroweak gauge bosons, the ETC operators and the top quark mass operator.  The addition of ETC terms and the EW gauge bosons affect the phase transitions weakly. The main difference between this scenario and the previous one is due to the top quark Yukawa coupling.

The tree level potential (\ref{Vtree}) remains unchanged. In the zero-temperature one-loop correction we add the contribution of the gauge bosons, with $n_W=6$ and $n_Z=3$, and the top quark with $n_{\rm top}=3$ (this amounts to  $-12$ for the $\bar{n}_{\rm Top}$) . Since the ETC terms give mass to the $\Theta$ particles, we are now also able to add them without causing infrared divergencies.

In the bosonic contributions in the temperature-dependent term ${V_T^{(1)}}_b$ we now use the full expressions of Appendices~\ref{AppA} and~\ref{AppB} with nonzero ETC masses, top mass and the electroweak couplings. In the fermionic part  ${V_T^{(1)}}_f$ we include the top contribution with $n_{\rm top}=3$ and $M_{\rm top} (\sigma_4) = m_{\rm top}\sigma_4/v_4$ where $m_{\rm top}$ is the physical zero-temperature mass.

We also add the electroweak gauge bosons in ${V_T^{(1)}}_{\rm gauge}$.
Only the longitudinal gauge bosons acquire a thermal mass squared at the leading order,
$O(g^2T^2)$. The transverse bosons acquire instead a
magnetic mass squared  of order $g^4T^2$ which we have neglected.
The explicit form of the transverse and longitudinal gauge boson
mass matrix is given in Appendix~\ref{AppC}.
Similarly as for the scalars, for the top and the gauge bosons we include the terms in the high temperature series up to $l=4$.

\section{Analysis of the phase transition: setup}
\label{sect5}
We will next perform a detailed study of the phase transition structure in UMT by using the one-loop effective potential described above. We start by analyzing the parameter space. 
\subsection{Parameter space}

After fixing the electroweak scale $v_4$ to 246~GeV, the effective scalar Lagrangian (\ref{Leff}) includes eight free parameters. In the ``with EW'' scenario one has in addition three free mass parameters in the effective ETC Lagrangian (\ref{LETC}), and also additional parameters of the electroweak and SM sectors ($g$, $g^\prime$ and the top mass). However, the depedence of the phase transitions on the actual value of the ETC masses is relatively weak whence we fix them at 150~GeV. The values of $g$, $g^\prime$ and the top mass are fixed to their experimental values. The remaining parameters of the with/without EW scenarios are the same.

These parameters can be chosen to be $v_2$, the Higgs mixing angle $\beta$, and the zero temperature scalar masses. 
It is convenient to define:
\bea
 \left(\Delta M_{\Pi_4}\right)^2 &\equiv& M_{\tilde \Pi_{UD}}^2 - m_{1,ETC}^2 = 2 \left(\lambda^\prime_4 v_4^2+\delta^\prime v_2^4\right) \ , \nonumber\\
 \left(\Delta M_{\Pi_2}\right)^2 &\equiv& M_{\tilde \Pi_{\lambda\lambda}}^2 - m_{1,ETC}^2 = 4 \left(\lambda^\prime_2 v_2^2+\delta^\prime v_2^2v_4^2\right) \ , \nonumber\\
 \left(\Delta M_{\Theta}\right)^2 &\equiv& M_{\tilde \Theta}^2 - m_{1,ETC}^2 = 2\delta^\prime v_2^2 \left( v_2^2 + 4 v_4^2\right) \ .
\eea
In this way we factor out the dynamical contribution to the masses $\propto \Delta M$ which is due to $v_2$ and $v_4$. Their effect on the nature of the phase transitions is much larger than that of the ``hard'' ETC masses. Note that we have dropped the tilde on the symbols on the $\Delta$ definition to make the notation lighter, however we stress that these mass differences do not refer to the Goldstones but to their scalar partners. 

Our convention for the Higgs masses $M_{H_\pm}$ and $\beta$ is such that the Higgs mass matrix reads
\be \label{Hmm}
 \left.\left(\frac{\partial^2 V^{(0)}}{\partial \sigma_i \partial \sigma_j}\right)\right|_{\sigma_i=v_i} = \left(\begin{array}{cc}
 M_{H_+}^2 \cos^2\beta + M_{H_-}^2 \sin^2\beta   \ \  & \ \  \cos \beta \sin \beta \left(M_{H_+}^2-M_{H_-}^2\right)\\
 \cos \beta \sin \beta \left(M_{H_+}^2-M_{H_-}^2\right)\ \  & \ \ M_{H_-}^2 \cos^2\beta + M_{H_+}^2 \sin^2\beta
\end{array}\right)
\ee
where the first (second) row and column correspond to derivatives wrt. $\sigma_4$ ($\sigma_2$).  One can restrict $M_{H_+}>M_{H_-}$ and $\beta = -\pi/2 \ldots \pi/2$, or equivalently $\beta=-\pi/4 \ldots \pi/4$ if the order of the Higgs masses is not fixed. For $M_{H_+}>M_{H_-}$ this convention coincides with the formulae given above in (\ref{mesons}). Note that for $\beta=0$ the decoupled Higgs masses $M_{H_{4,2}}$ of the $SU(4)$ and $SU(2)$ sectors are given by $M_{H_4} = M_{H_+}$ and $M_{H_2} = M_{H_-}$, respectively, while for $\beta=\pi/2$ the order is reversed.

The Higgs masses $M_{H_\pm}$, and the Higgs mixing angle $\beta$  are the most relevant parameters. Also the $\tilde \Pi$ masses
$\Delta M_{\Pi_{4,2}}$, the zero-temperature vacuum $v_2$, and  $\Delta M_\Theta$, or the $\delta'$ coefficient
of the term in the Lagrangian which is reponsible for the breaking of the anomalous $U(1)$ symmetry, affect the transitions rather strongly.

Since the parameter space of the theory is vast we have to make choices for reference values of the various parameters. We mainly plot quantities as functions of the Higgs masses.  As we do not fix the order of the Higgs masses $M_{H_+}$ and $M_{H_-}$ the possible values of the Higgs mixing angle $\beta$ can be taken to lie in between $-\pi/4$ and $\pi/4$. We choose as reference values $\beta=-0.6$, 0, and $+0.6$.
The value of $v_2$ is expected to lie close to $v_4$, which is fixed to the electroweak scale 246~GeV, since both arise from the same strongly interacting theory. We will be mostly using reference values from $v_2=250$~GeV to $v_2=350$~GeV. For $\Delta M_\Theta \propto \delta'$ we use  $0$ and $200$~GeV. As mentioned above, the former case ($\delta'=0$) is special (but unnatural) since the chiral symmetry of the theory is enhanced to $U(4)\times U(2)$. Then neither of the $\Theta$ particles receives a dynamical mass.

In the following we mean by a ``$\sigma_i$ transition'' a transition where the chiral condensate $\sigma_i$ (with either $i=4$ or $i=2$) changes from a nonzero value to zero when increasing the temperature.
We assume that a first order phase
transition takes place when the broken and symmetric phase vacua are exactly degenerate, which defines the critical temperature $T_{c,i}$ and the critical (broken phase) value of the condensate $\sigma_{c,i}$. Since the barriers between the vacua will be low wrt. the electroweak scale in all the cases which we shall study, this should be a good approximation.

It is instructive to first consider the (unrealistic) case where the two phase transitions
are 
decoupled, i.e., $\beta=0$ and $\delta'\propto \Delta M_{\Theta} = 0$. 

\subsection{Decoupled transitions}

\begin{figure}
{
 \includegraphics[height=6.5cm,width=5.82cm]{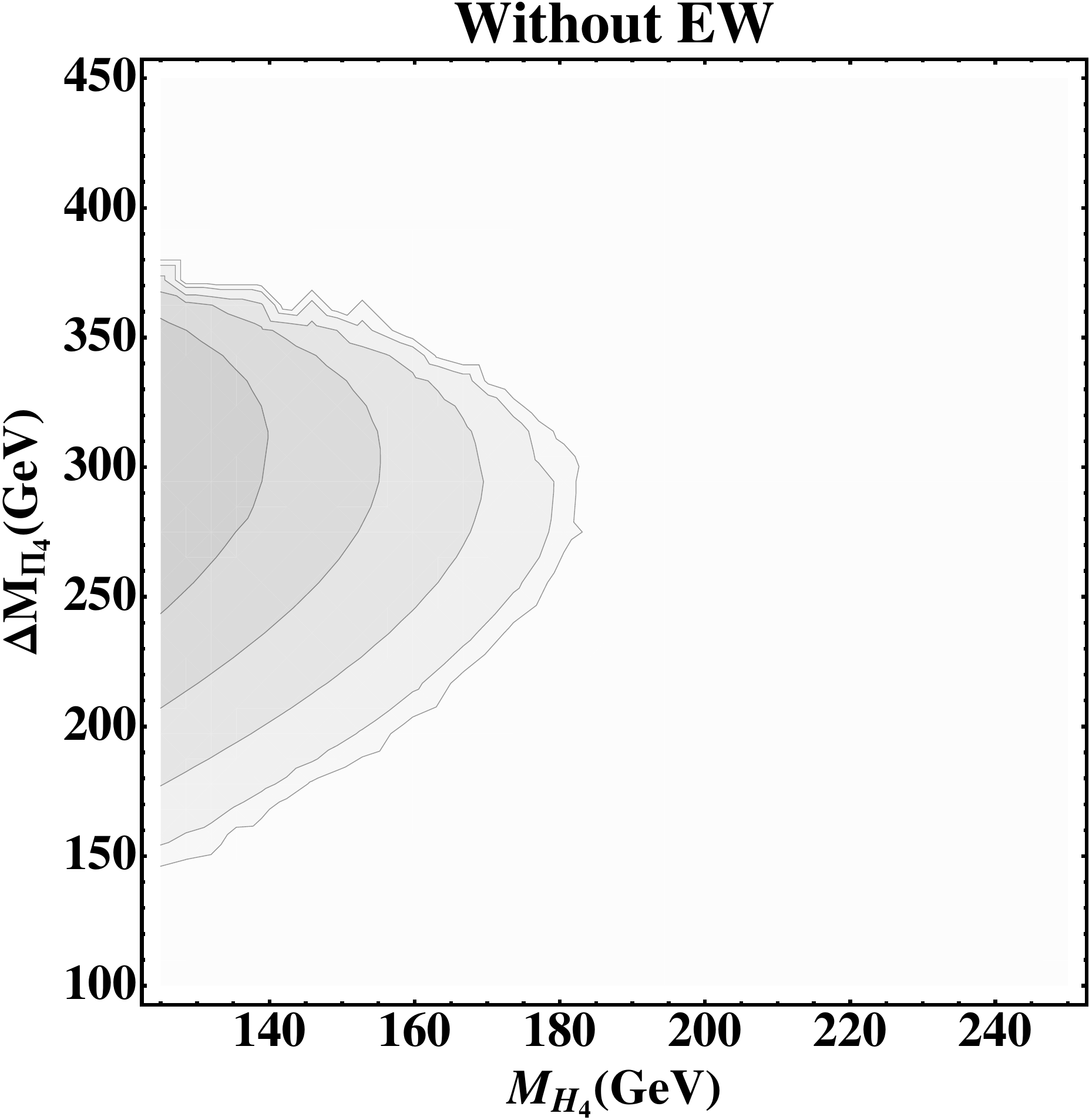}\hspace{0.8cm}\includegraphics[height=6.5cm,width=5.82cm]{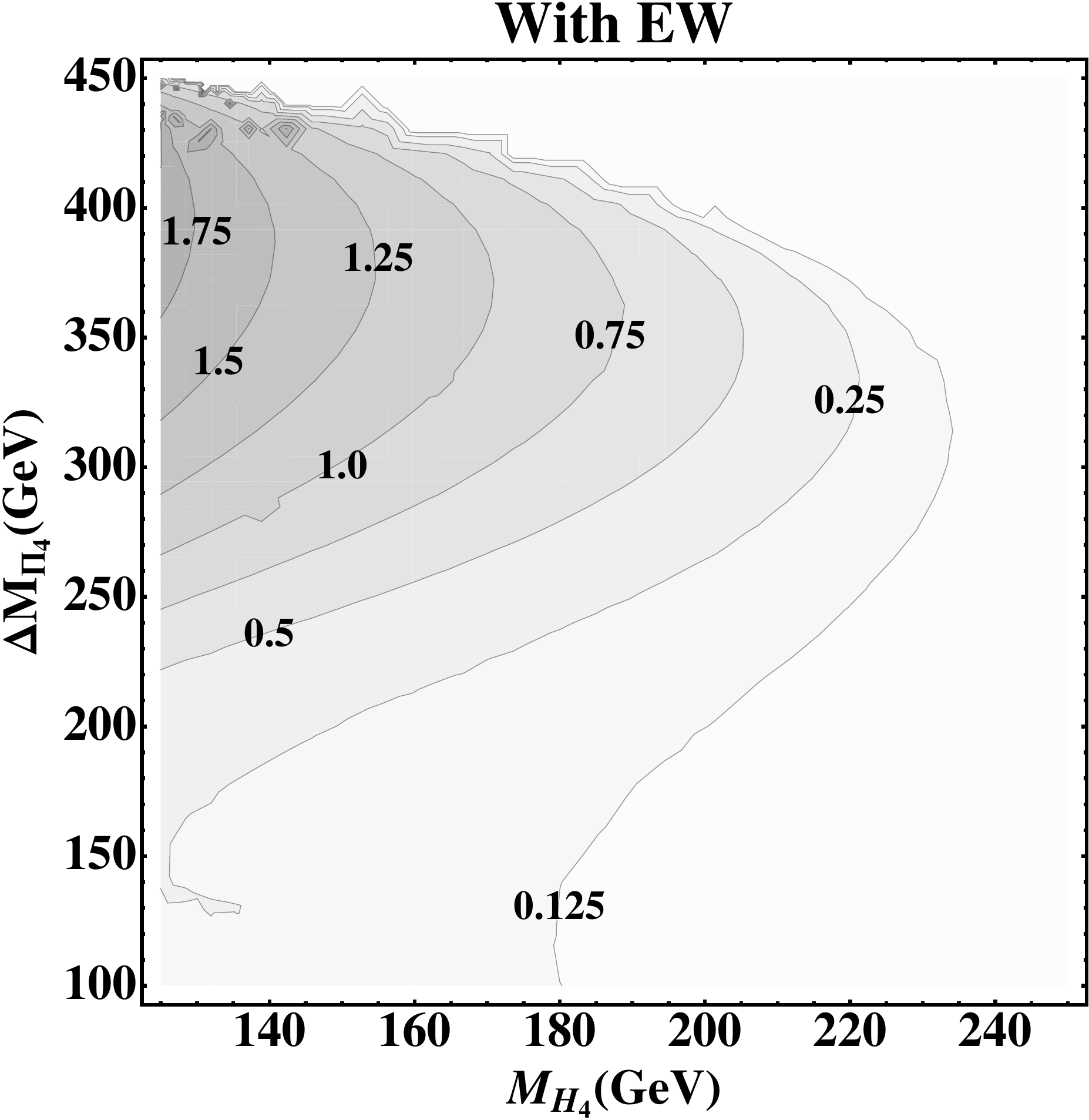}
}
\caption{The strength of the $\sigma_4$ phase transition ($\sigma_{4,c}/T_{4,c}$) in the
$M_{H_{4}}$-$\Delta M_{\Pi_{4}}$ plane, without EW interactions (left) and with EW interactions (right). The contours are at $\sigma/T = 0.125$, 0.25, 0.5, 0.75, 1, 1.25, 1.5, and 1.75 from lighter to darker shades.}\label{fig:decoupled4}
\end{figure}

When $\beta=0$ and $\Delta M_{\Theta} = 0$ the phase transitions of the chiral condensates $\sigma_4$ and $\sigma_2$ 
take place independently.
Then the nature of the $\sigma_4$-transition 
only depends on the Higgs mass $M_{H_4}=M_{H_+}$ and on $\Delta M_{\Pi_4}$. In Fig.~\ref{fig:decoupled4} we plot the strength $\sigma_{c,4}/T_{c,4}$ for the ``without EW'' (left) and ``with EW'' (right) scenarios. The difference between the plots is mainly due to the top Yukawa coupling,
which makes the transition stronger in particular for $\Delta M_{\Pi_4}>300$~GeV. Recall that electroweak baryogenesis requires  $\sigma_{4,c}/T_{4,c} \gtrsim 1$.
We see that first order phase transitions are possible, but the transition is mostly rather weak, i.e., $\sigma/T \lesssim 1$ except for a rather small region of parameter space with light Higgs.

\begin{figure}
{
 \includegraphics[height=5.5cm,width=4.82cm]{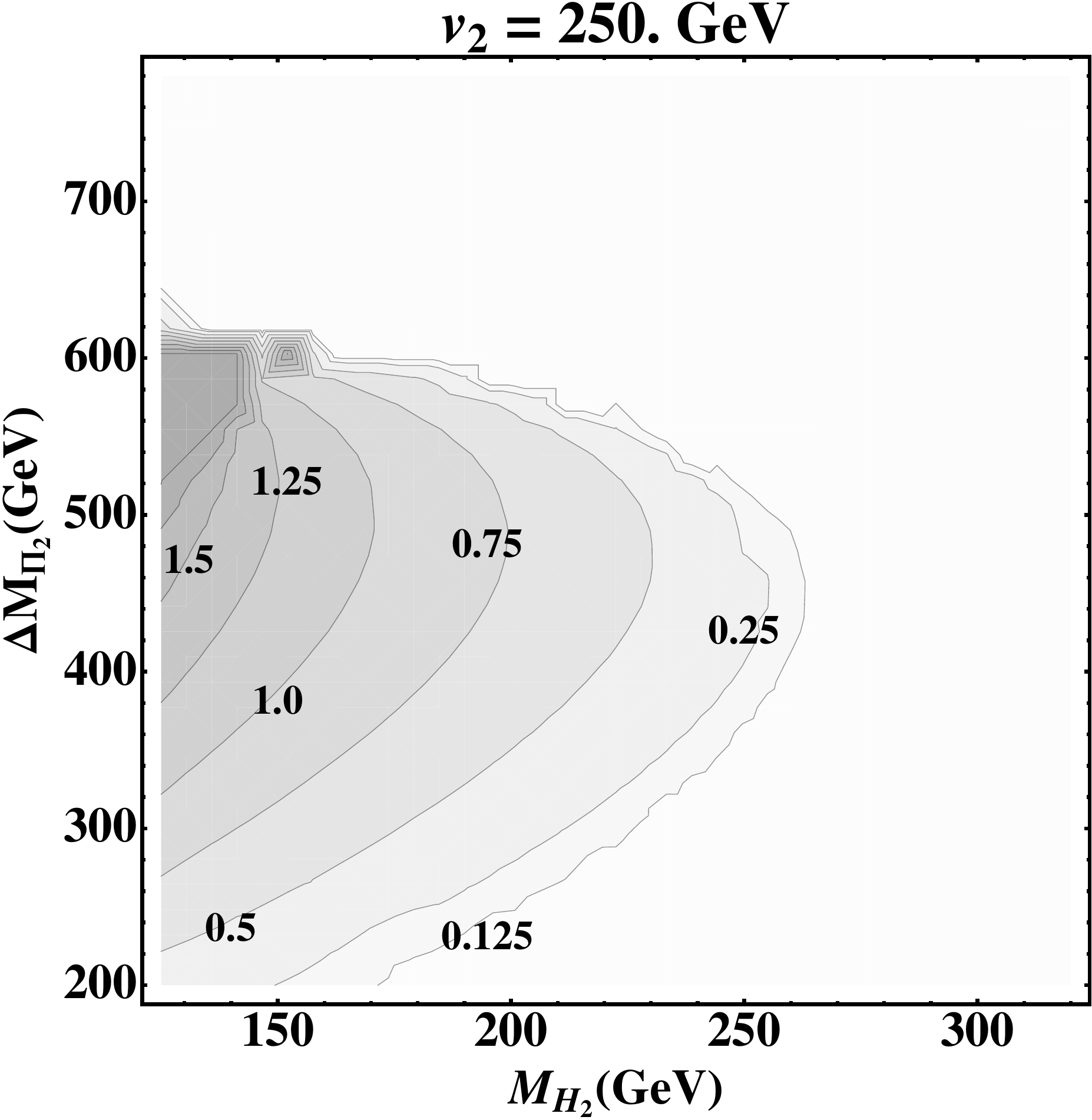}\hspace{0.6cm}\includegraphics[height=5.5cm,width=4.82cm]{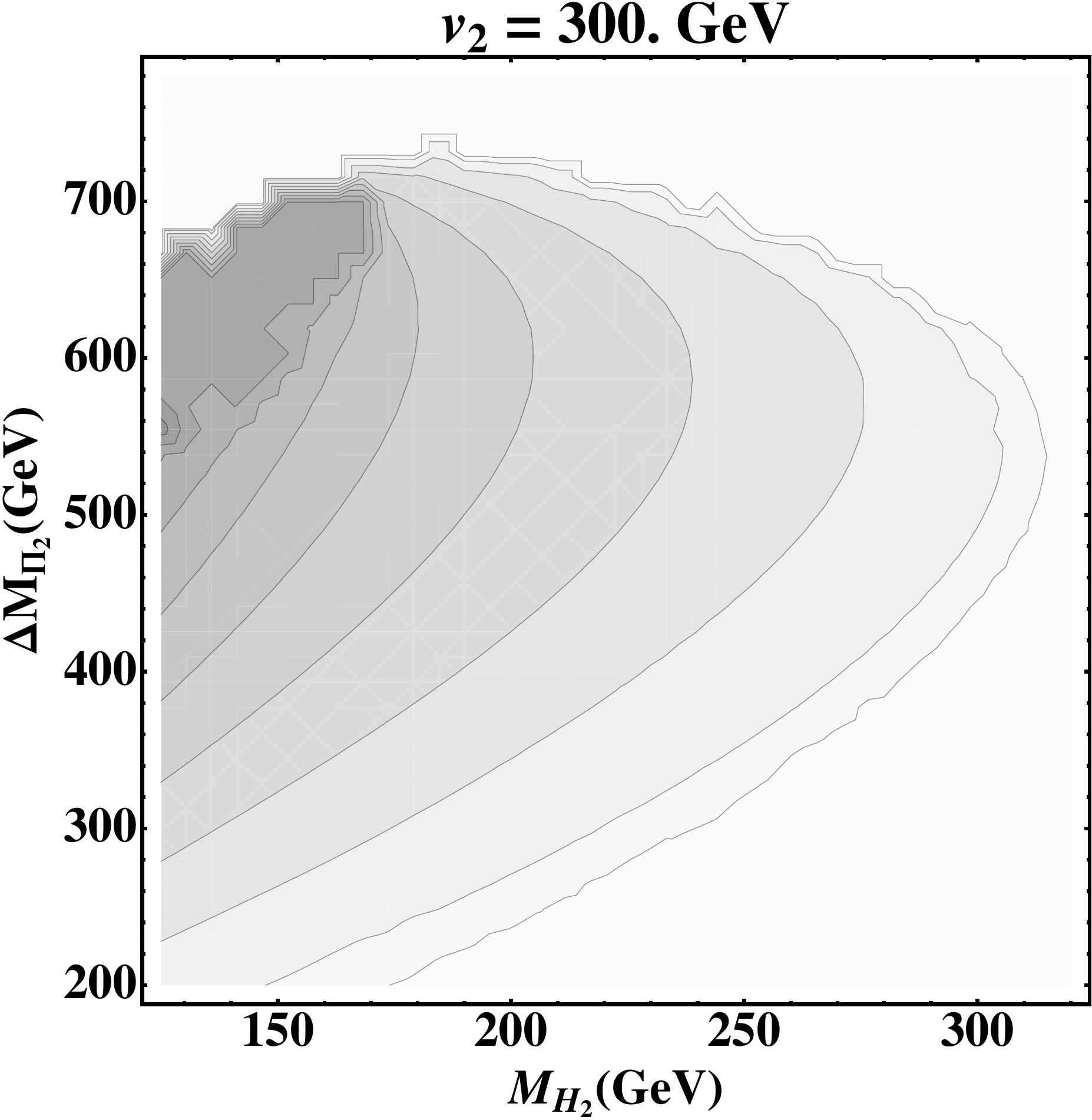}\hspace{0.6cm}\includegraphics[height=5.5cm,width=4.82cm]{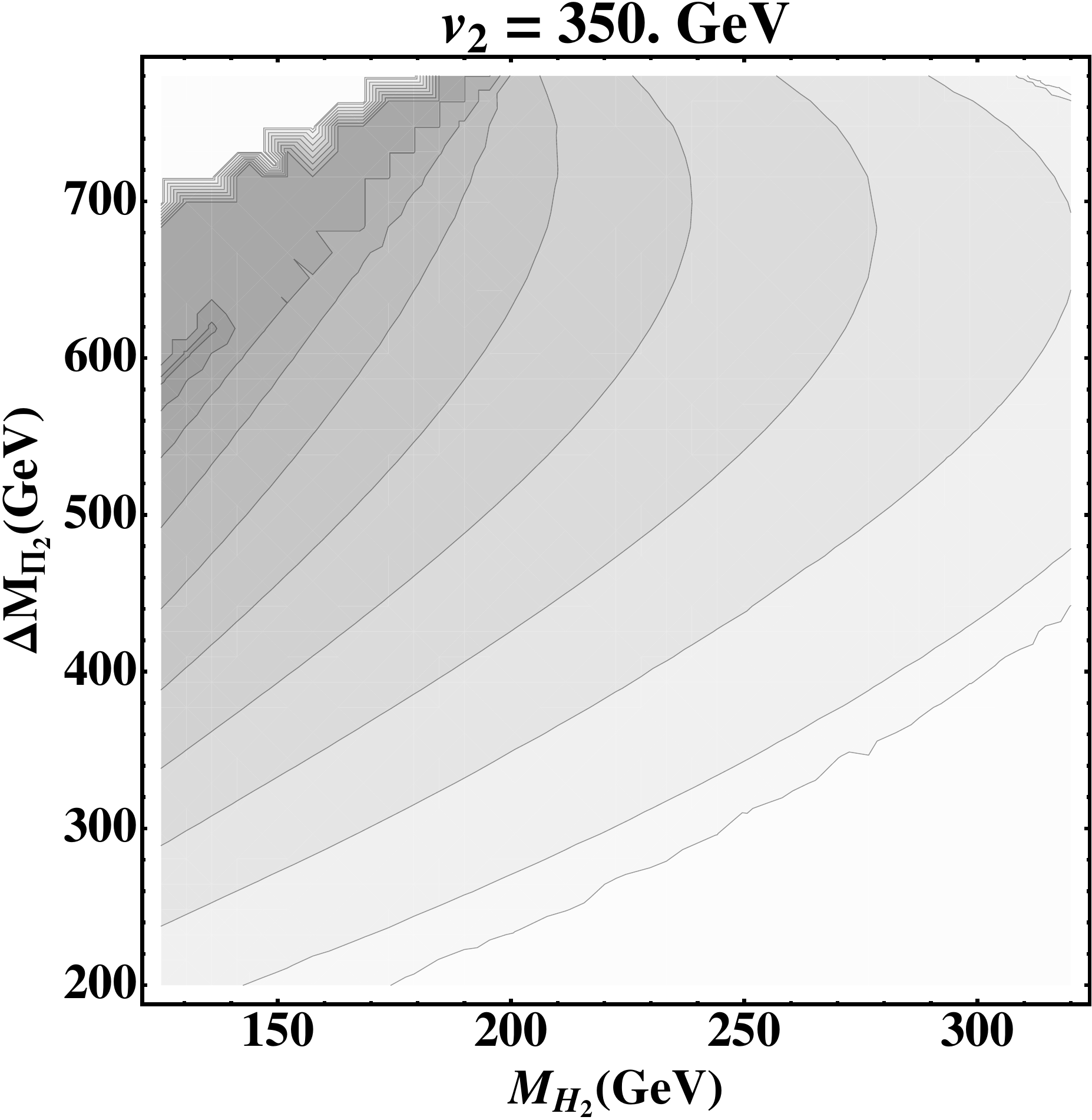}
}
\caption{The strength of the $\sigma_2$ phase transition ($\sigma_{2,c}/T_{2,c}$) in the
$M_{H_{2}}$-$\Delta M_{\Pi_{2}}$ plane, without EW interactions, and for $v_2=250 \ldots 350$~GeV as indicated in the labels. The contours are at $\sigma/T = 0.125$, 0.25, 0.5, 0.75, \ldots, 2.5 from lighter to darker shades.}\label{fig:decoupled2}
\end{figure}

Similarly, the $\sigma_2$-transition depends on $M_{H_2}$ and $\Delta M_{\Pi_2}$, but also on $v_2$. Notice that since the electroweak gauge fields only couple to the sector 4, the transition 2 is independent of the electroweak physics and hence similar in the with/without EW scenarios. In Fig.~\ref{fig:decoupled2} we plot the strengths $\sigma_{2,c}/T_{2,c}$ of the phase transitions in the $M_{H_2}$-$\Delta M_{ \Pi_2}$ plane for various values of $v_2$ without including EW interactions. 
With increasing $v_2$ the phase transition gets stronger, and the region of maximal $\sigma/T$ moves to higher $\Delta M_{ \Pi _2}$. In the dark gray region near the top right corner of the plots the critical temperature is too low to determine $\sigma/T$ reliably when the high temperature expansion is used. In the white region there is either a second order phase transition or no phase transition takes place at all.

We have also checked the dependence of the phase transitions on the ETC mass scale.
With increasing the ETC mass scale, both the transitions are slightly weakened.

The qualitative behavior of the phase transition strengths is very similar to that of MWT \cite{Cline:2008hr} which has the chiral symmetry breaking pattern of $SU(4) \to SO(4)$ and to that of the two Higgs doublet models \cite{Land:1992sm}. This is understandable since the origin of the first order transitions is the same: with  light Higgs the tree level potential is flat, and the one-loop contributions from relatively heavy other scalars (here the $\tilde \Pi$'s) can make the symmetric and broken phase vacua almost degenerate even at $T=0$. Then the (one-loop) temperature-dependent corrections favor the symmetric vacuum already at low temperatures, and the phase transition is strongly first order.

Notice that the phase transition for the  $SU(4) \to Sp(4)$ chiral symmetry breaking pattern of Fig.~\ref{fig:decoupled4} is considerably weaker than what was observed in MWT for large $\Theta$ masses.

A strongly $\sigma_4$ dependent $\Theta_4$ mass that would lead to a stronger phase transition cannot be achieved within UMT. This would require the presence of an ``anomalous'' determinant term $\propto $Pf$\left[M_4\right]^2$ which is forbidden by the anomaly free $U(1)$ symmetry of UMT. 

The analogous term $\propto \delta'$Pf$\left[M_4\right]$ in the UMT Lagrangian does not yield the desired $\sigma_4$ dependence of the $\Theta_4$ mass, but rather a ``hard'' mass term where essentially $M_{\Theta_4}^2 \propto \sigma_2^2$.

\section{Results } 
\label{sect6}

We will now start investigating the effects of the coupling among the two phase transitions.
For nonzero $\beta$ and $\Delta M_\Theta^2$ a variety of different phase transition structures are possible as we have suggested in \cite{Jarvinen:2009wr}. 
Both the transitions can be independently first or second order transitions, and the relative ordering of their critical temperatures must be assessed. In addition, as we shall see, more complicated structures can appear, mostly when both the transitions are first order and their critical temperatures are close to each other.  A similar phenomenon was observed in \cite{Sannino:2002wb,Mocsy:2003tr,Mocsy:2003qw}. 
We shall first consider the case where only the strong technicolor interactions are considered, without coupling to the SM.

\begin{figure}
{
 \includegraphics[height=6.5cm,width=5.82cm]{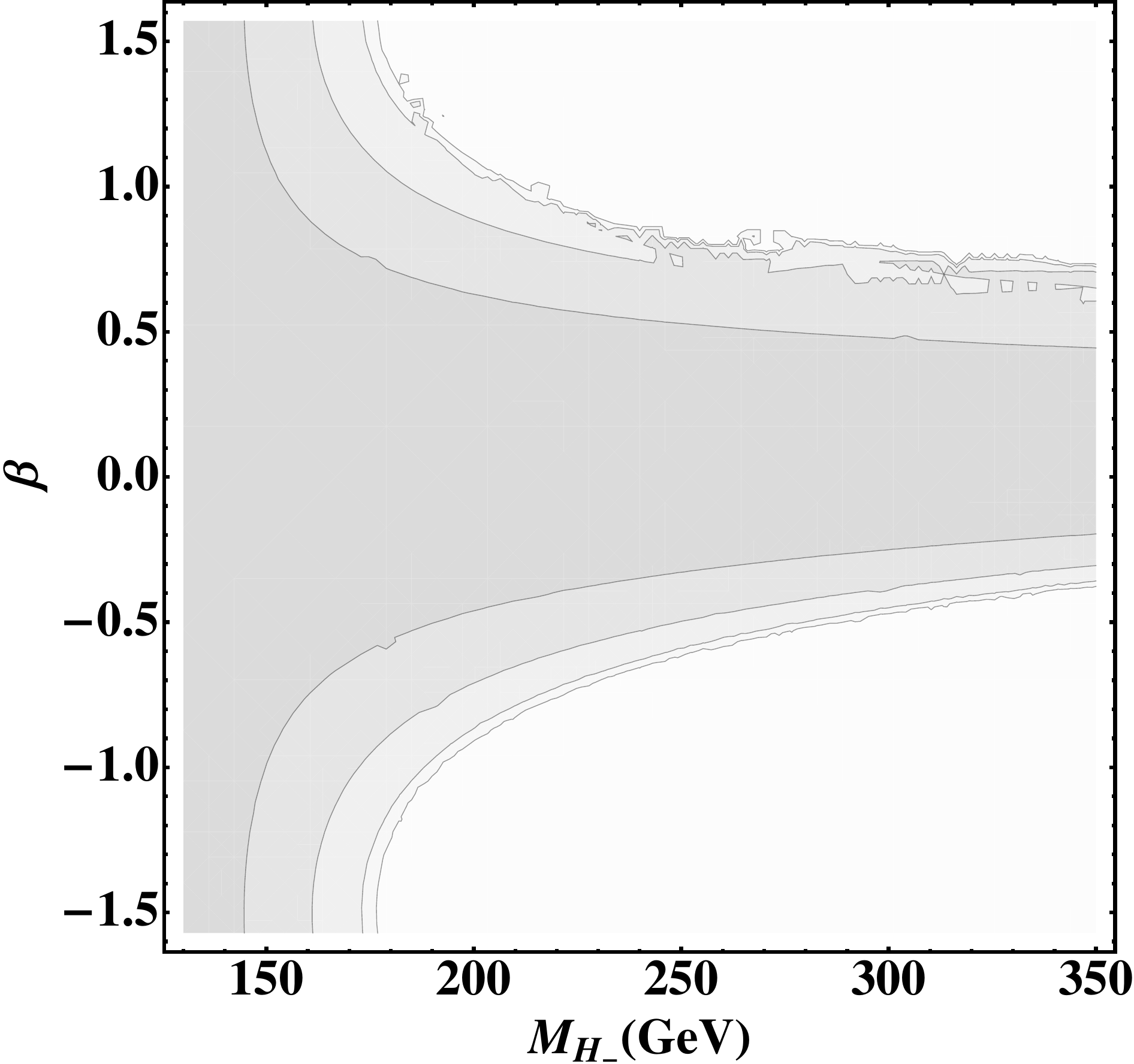}\hspace{0.8cm}\includegraphics[height=6.5cm,width=5.82cm]{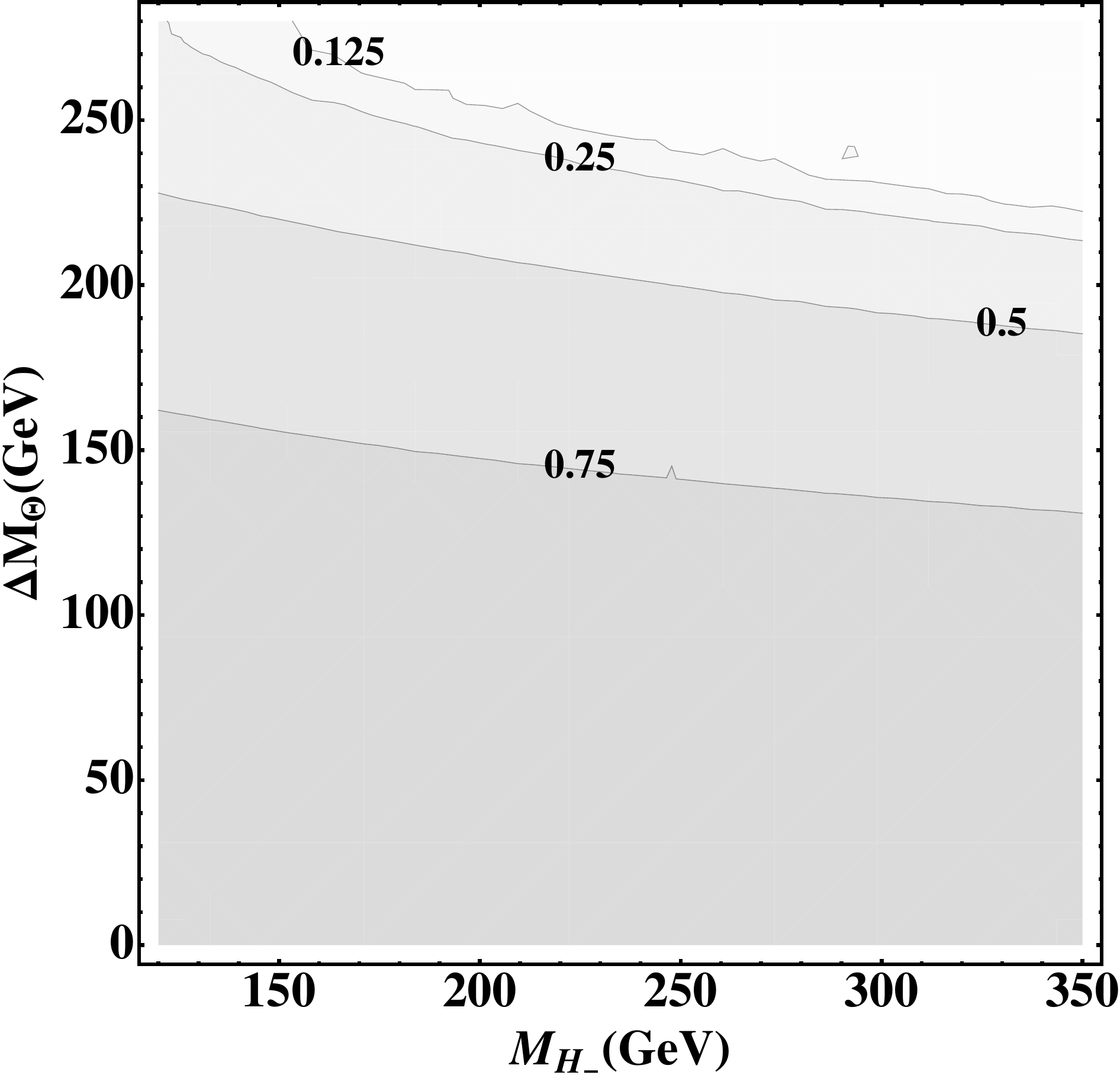}
}
\caption{Left: the strength of the $\sigma_4$ phase transition ($\sigma_{4,c}/T_{4,c}$) in the
$M_{H_{-}}$-$\beta$ plane, with $\Delta M_\Theta = 0$~GeV. Right: $\sigma_{4,c}/T_{4,c}$ in the
$M_{H_{-}}$-$\Delta M_\Theta$ plane, with $\beta=0$. Other parameter values are $M_{H_+}=150$~GeV~$=\Delta M_{\Pi_2}$, $\Delta M_{\Pi_4}=350$~GeV, $v_2=300$~GeV and also $m_{i,ETC}=150$~GeV. The contours are at $\sigma/T = 0.125$, 0.25, 0.5, and 0.75 from lighter to darker shades.}\label{fig:coupl1}
\end{figure}

\subsection{Without EW} 

A relevant question which we wish to address is if the coupling between the two sectors can induce stronger phase transitions.
Let us first consider a simple case where a first order $\sigma_4$ transition is coupled with a mostly second order $\sigma_2$ transition, which is the case for $\Delta M_{\Pi_2} \lesssim $~GeV. By using Fig.~\ref{fig:decoupled4}, we choose a reference point of $M_{H_+}=150$~GeV and $\Delta M_{\Pi_4}=350$~GeV where the $\sigma_4$ transition is strongly first order. We also set $\Delta M_{ \Pi_2}$ and use $v_2=300$~GeV. Fig.~\ref{fig:coupl1} shows the dependence of the phase transition strength $\sigma_{4,c}/T_{4,c}$ on the Higgs mixing angle $\beta$ and on $\Delta M_\Theta$ with varying $M_{H_-}$.
We see that for nonzero $\beta$ and  $\Delta M_\Theta$ the phase transition is weakened at least as long as $\Delta M_\Theta>0$.


\begin{figure}
{
 \includegraphics[height=5.5cm,width=5.2cm]{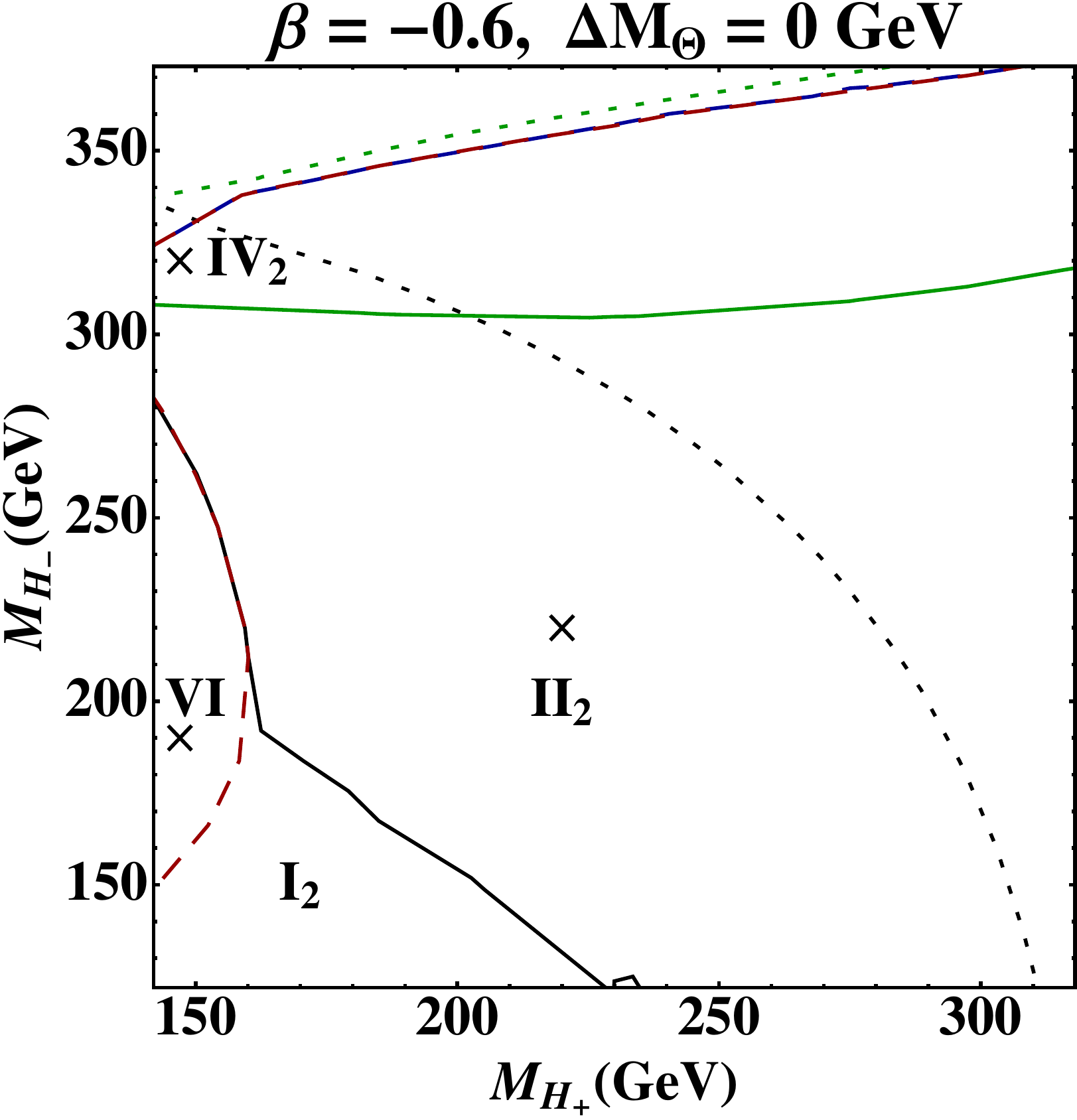}\hspace{0.4cm}\includegraphics[height=5.5cm,width=5.2cm]{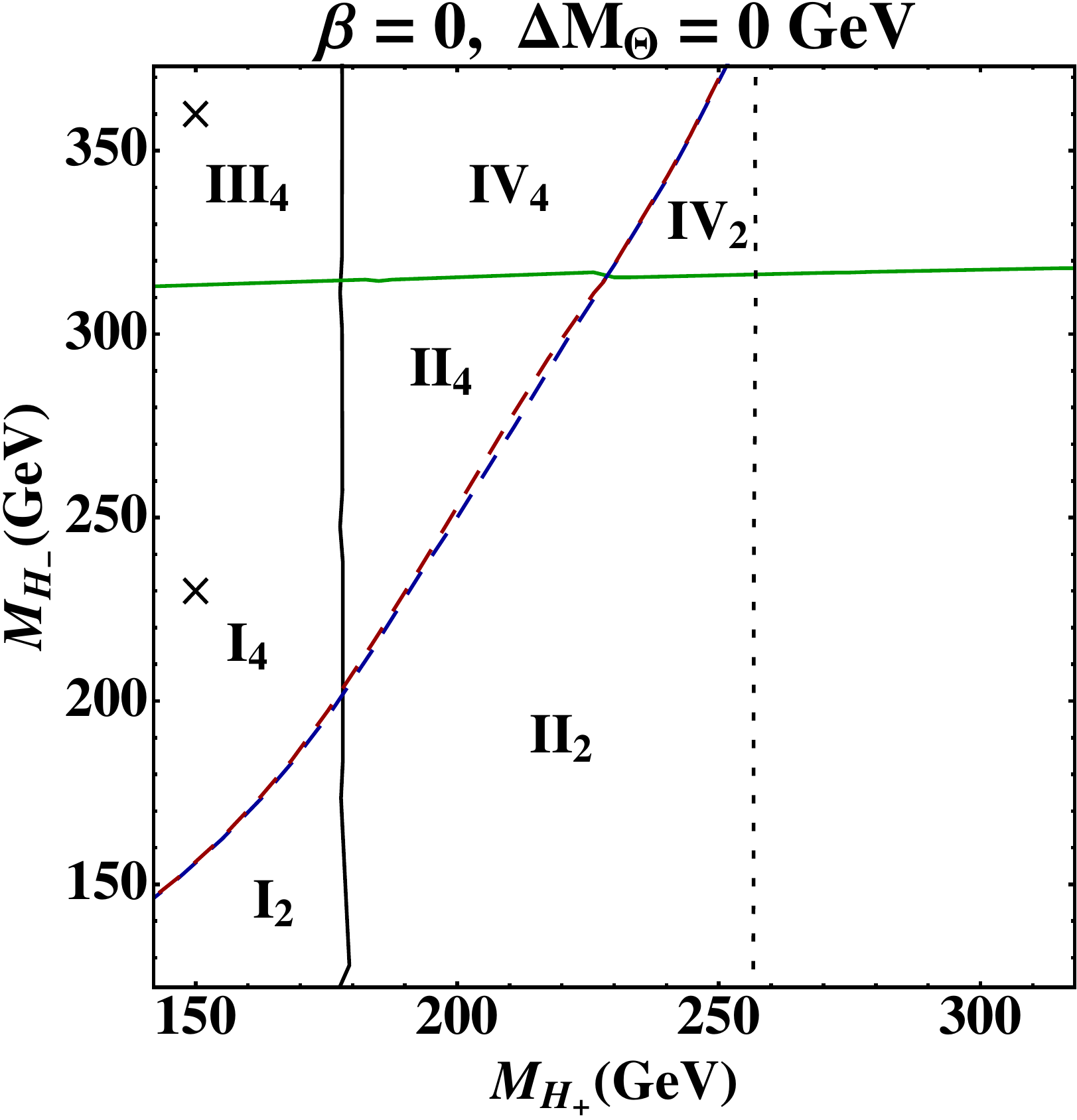}\hspace{0.4cm}\includegraphics[height=5.5cm,width=5.2cm]{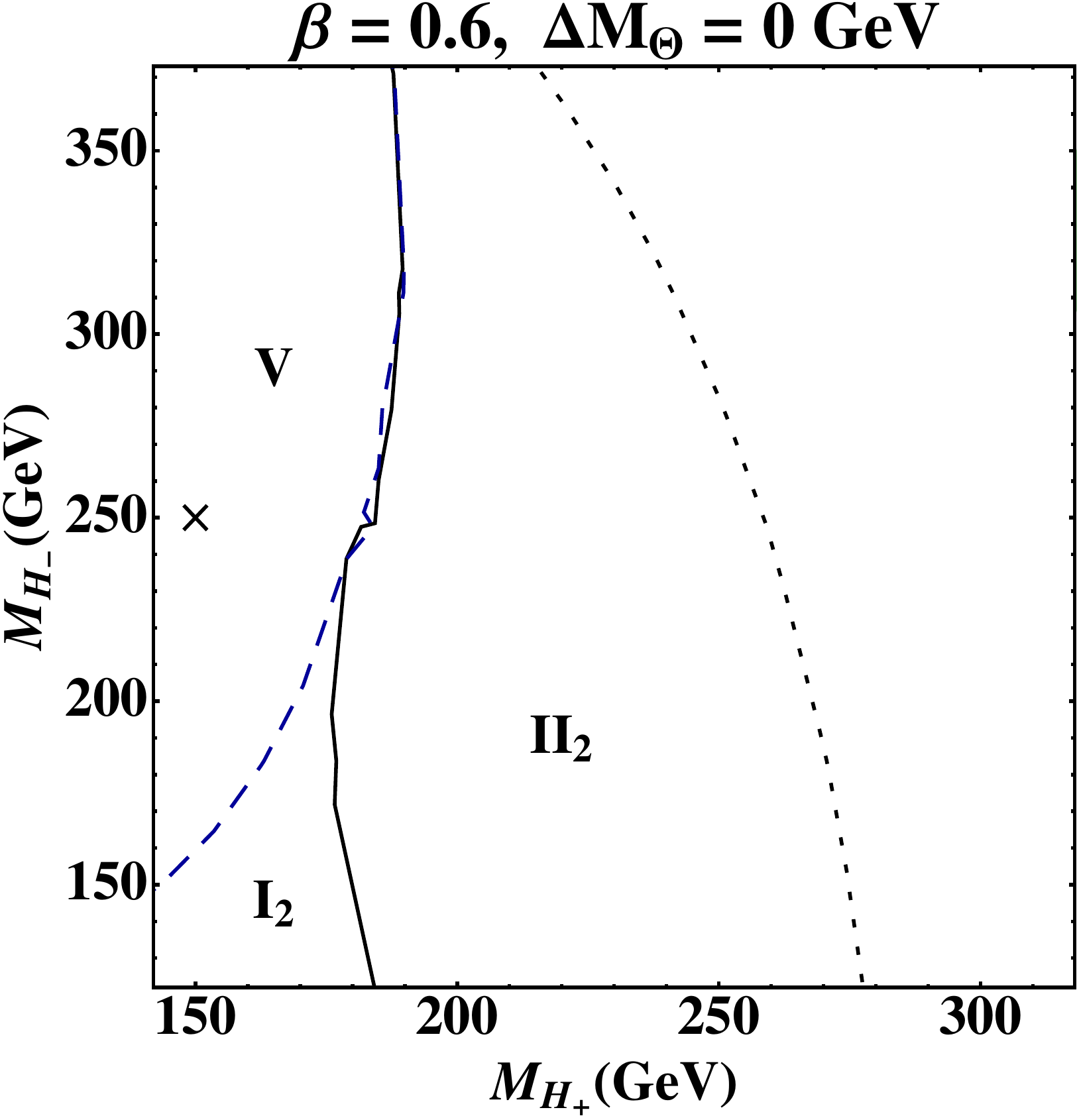}


 \includegraphics[height=5.5cm,width=5.2cm]{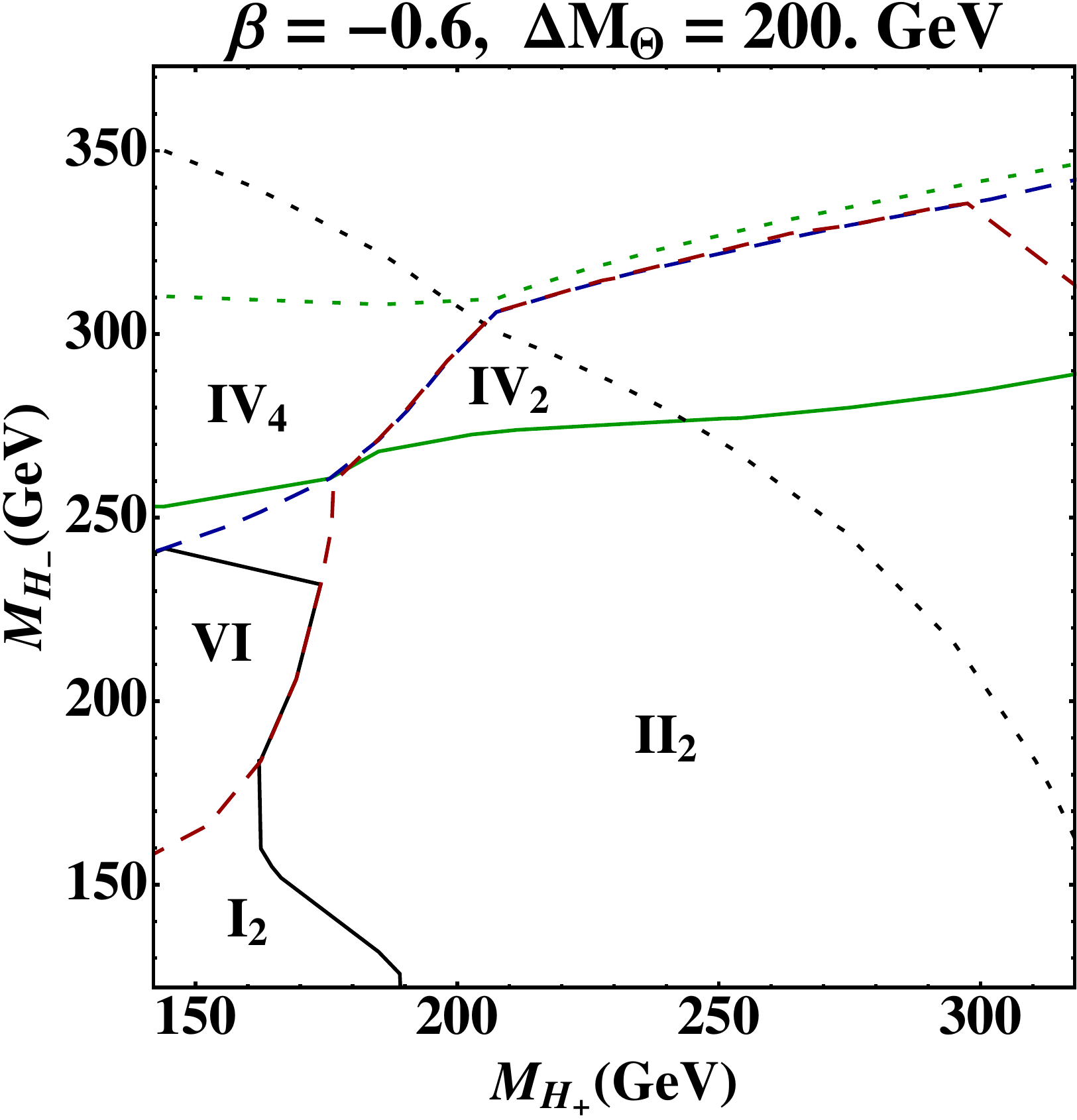}\hspace{0.4cm}\includegraphics[height=5.5cm,width=5.2cm]{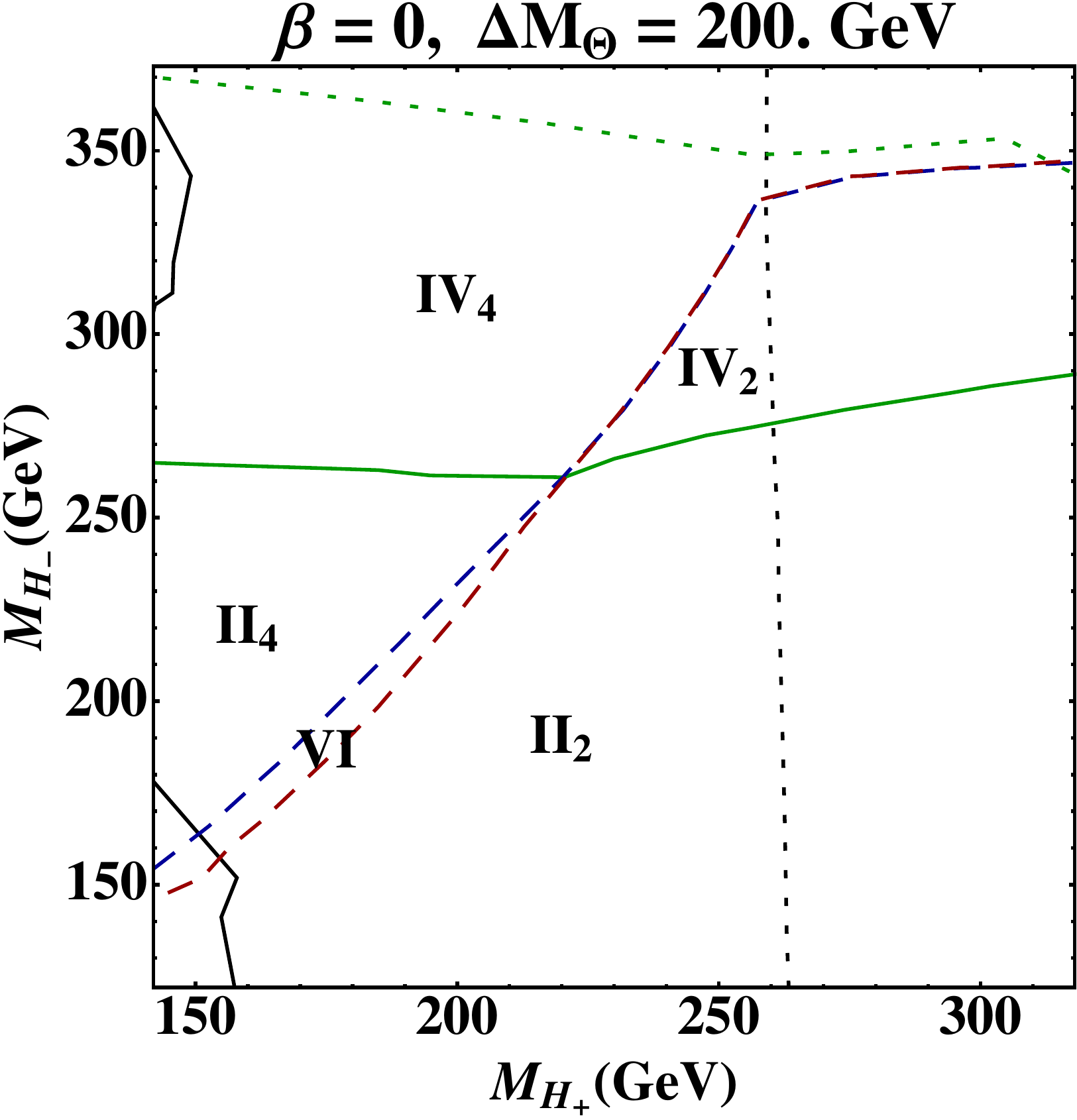}\hspace{0.4cm}\includegraphics[height=5.5cm,width=5.2cm]{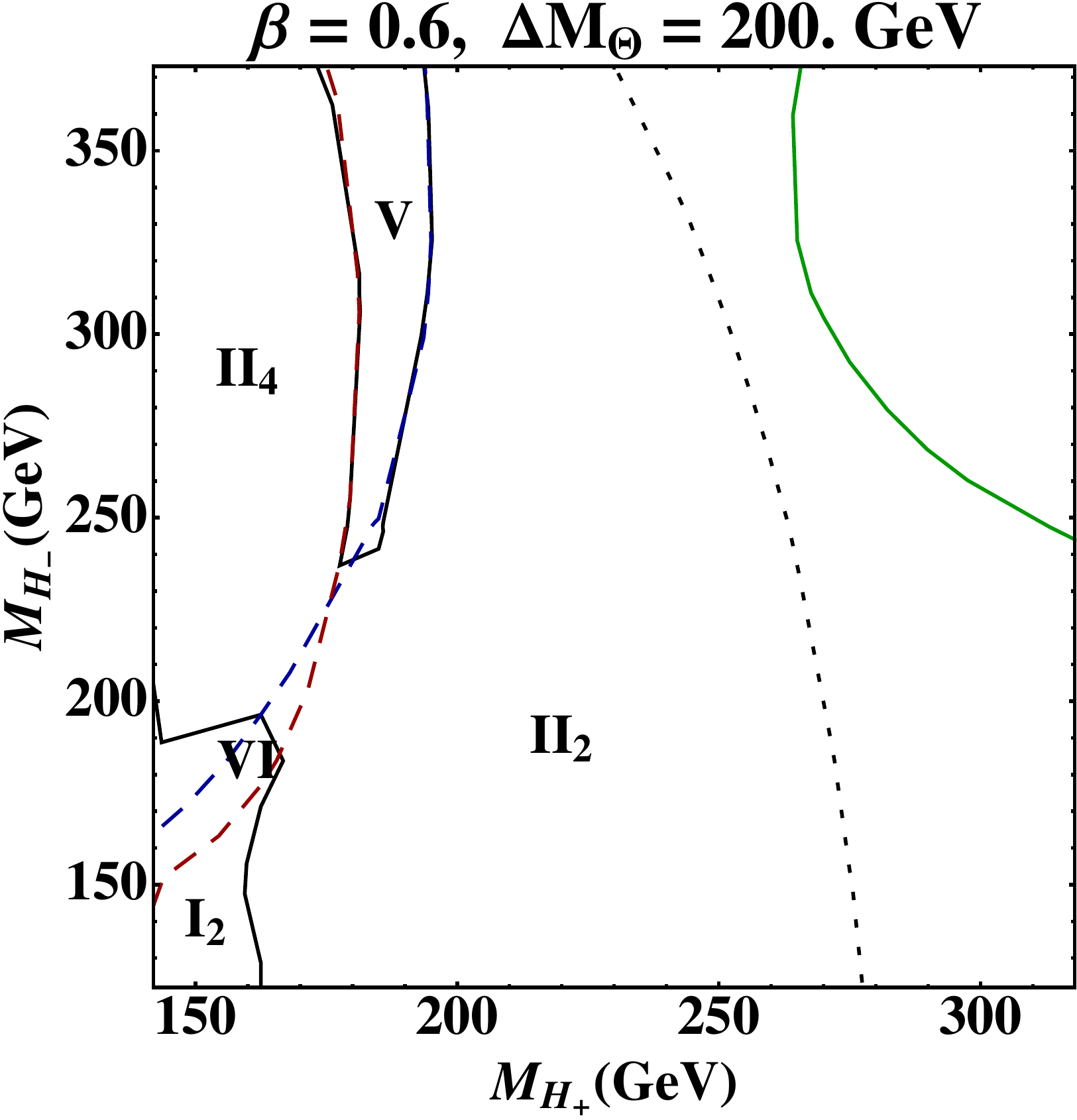}
}
\caption{Maps of the phase transition behavior when the Higgs masses are varied in the ``without EW'' scanario for various values of $\beta$ and $\Delta M_\Theta$ as indicated in the labels. We fixed $\Delta M_{\Pi_4} = 250$~GeV, $\Delta M_{\Pi_2} = 500$~GeV, and $v_2=300$~GeV. The black (green) joined and dotted curves denote where the $\sigma_4$ ($\sigma_2$) transition changes from first to second order and where the transition disappears, respectively. The dashed curves indicate a change in the order of the critical temperatures. For the meaning of labels see text and Fig.~\ref{fig:condT}. }\label{fig:phased}
\end{figure}

\begin{figure}
{
 \includegraphics[height=2.4cm,width=5.2cm]{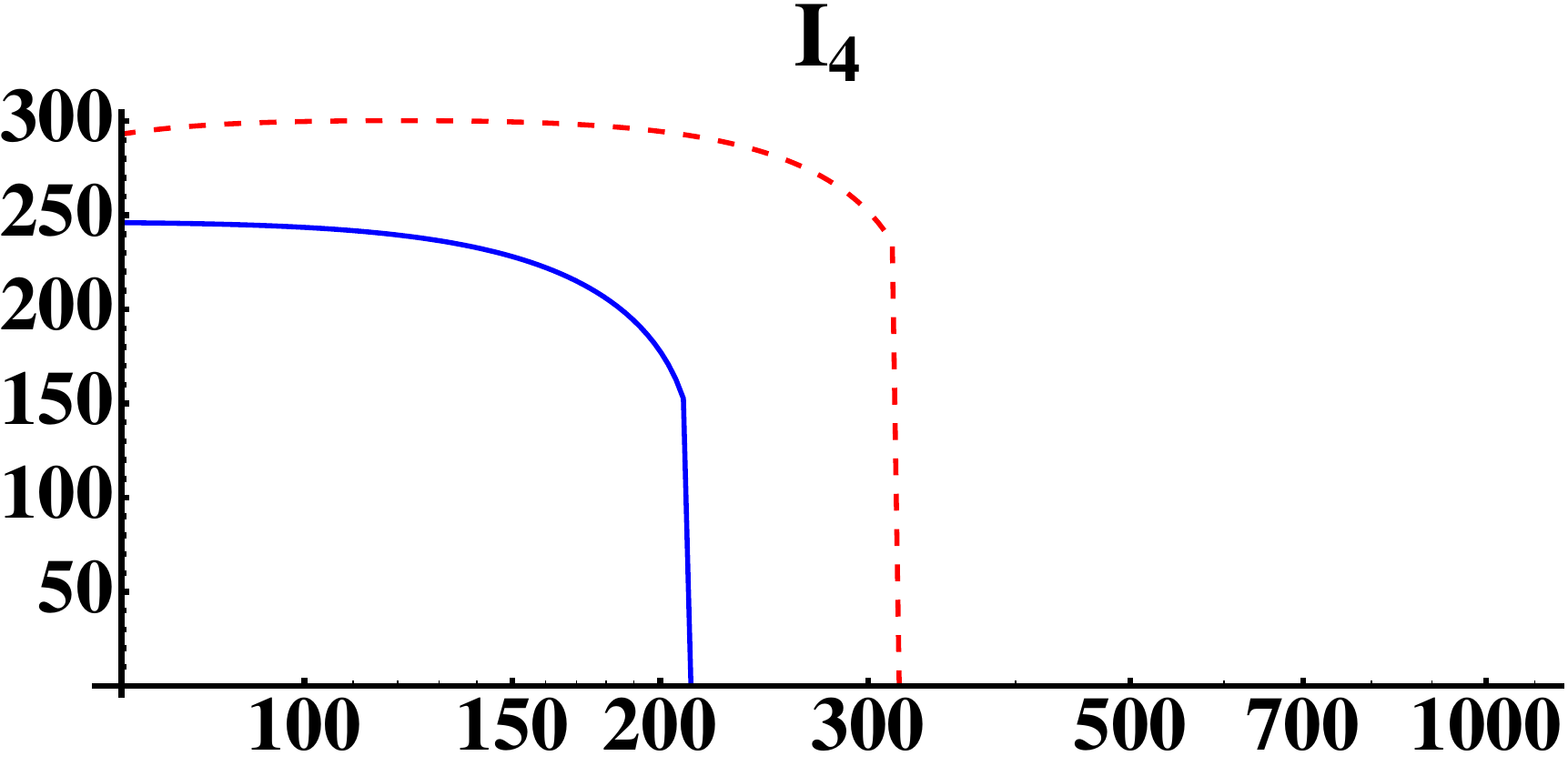}\hspace{0.4cm}\includegraphics[height=2.4cm,width=5.2cm]{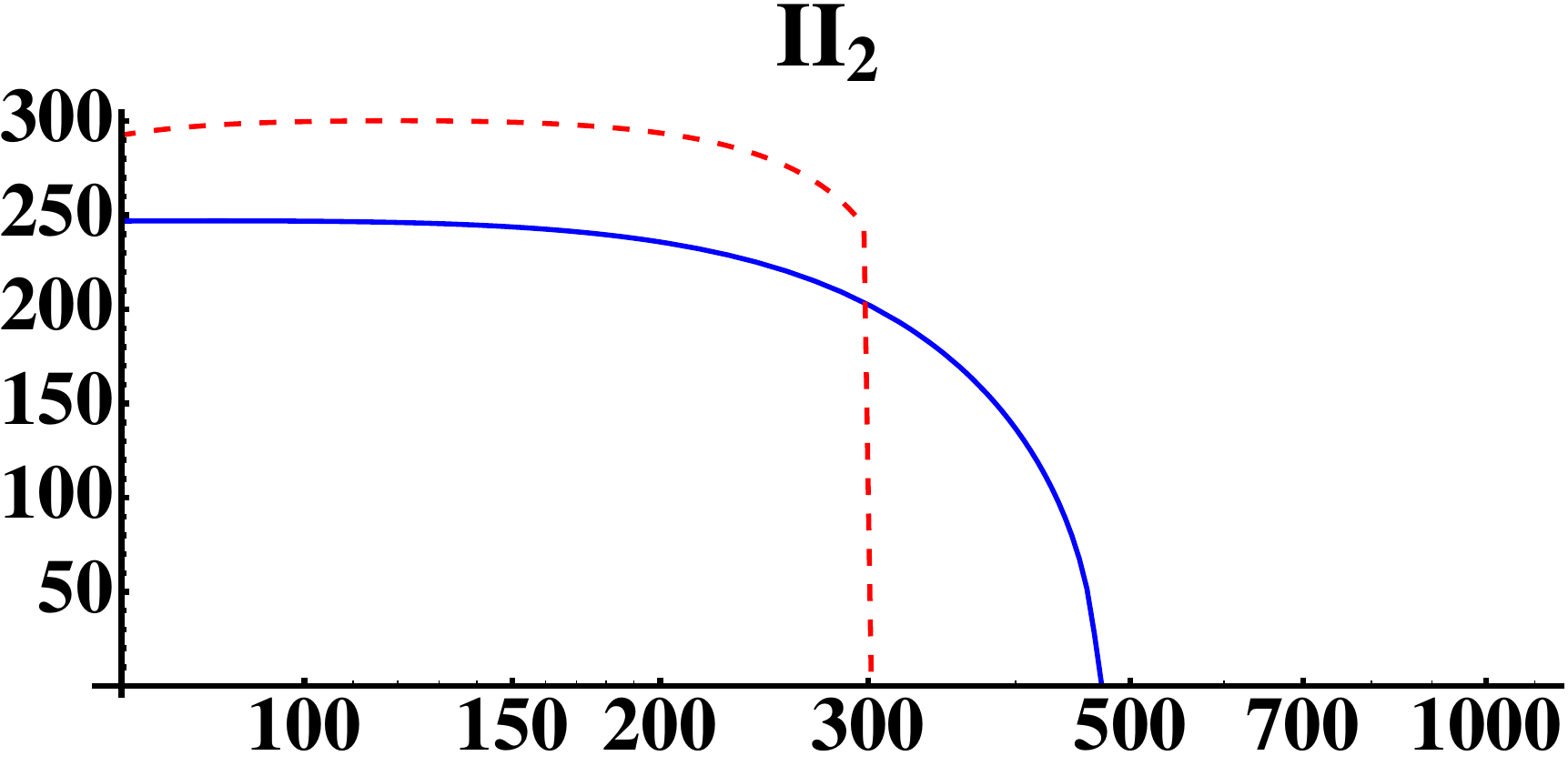}\hspace{0.4cm}\includegraphics[height=2.4cm,width=5.2cm]{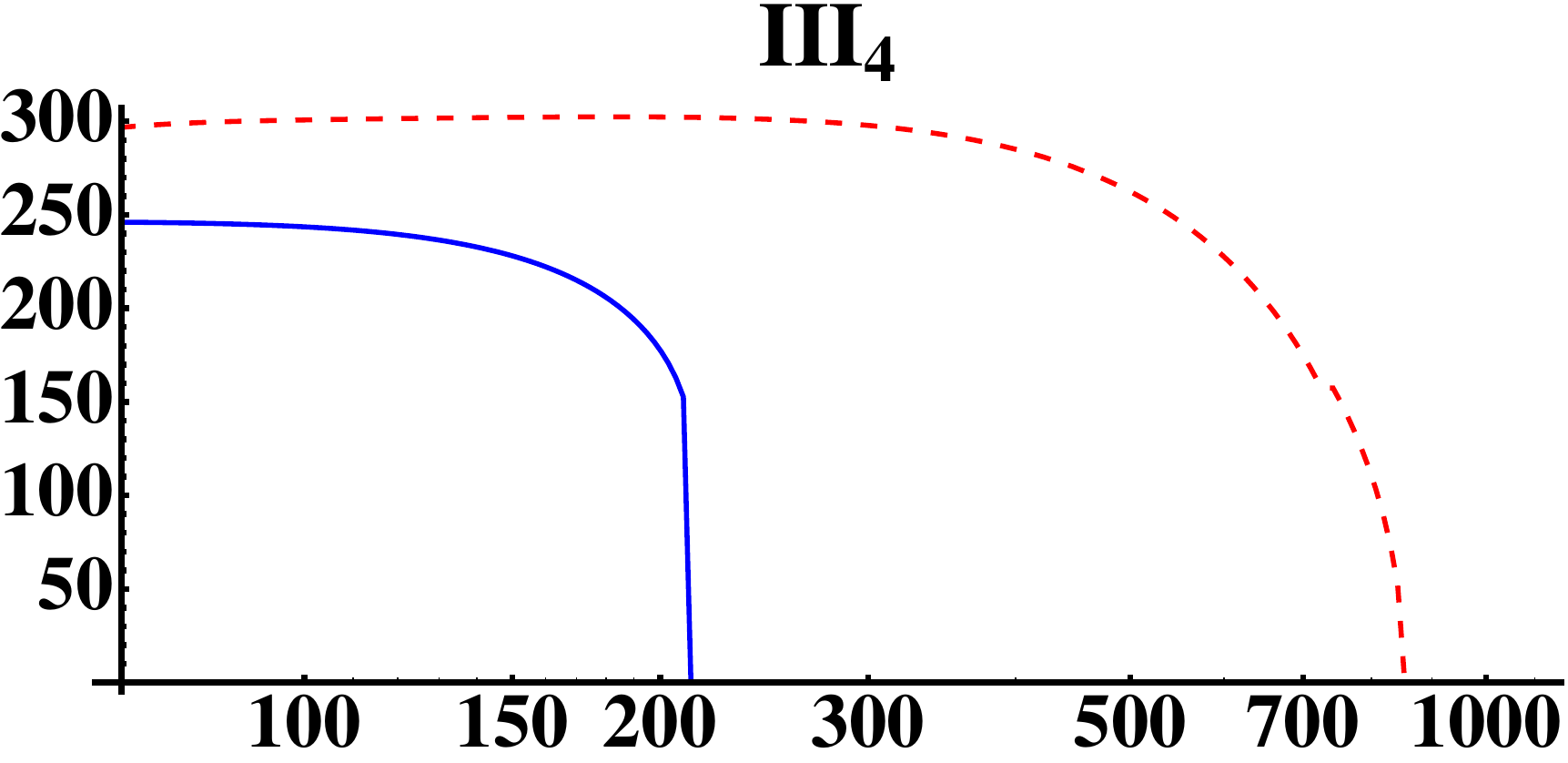}


 \includegraphics[height=2.4cm,width=5.2cm]{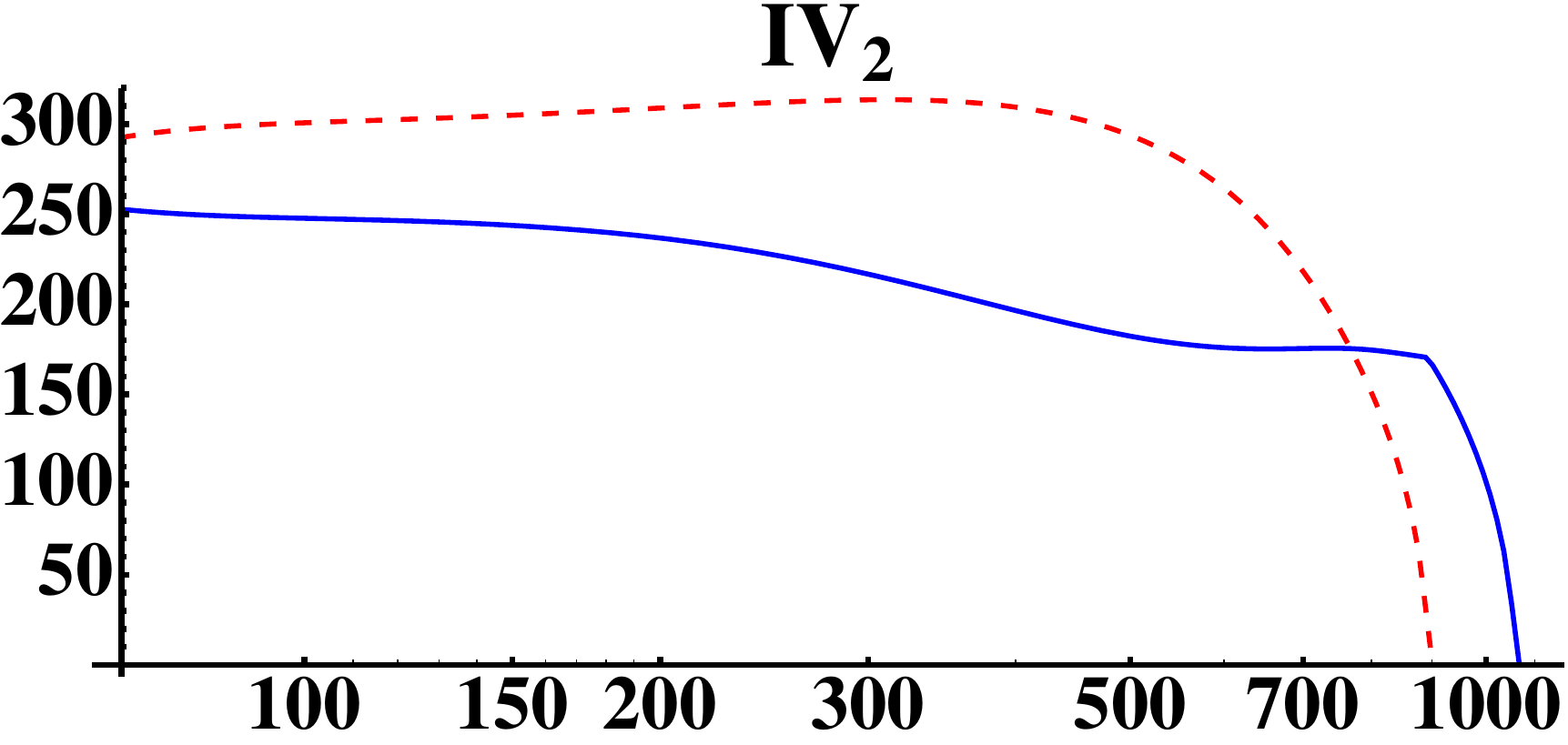}\hspace{0.4cm}\includegraphics[height=2.4cm,width=5.2cm]{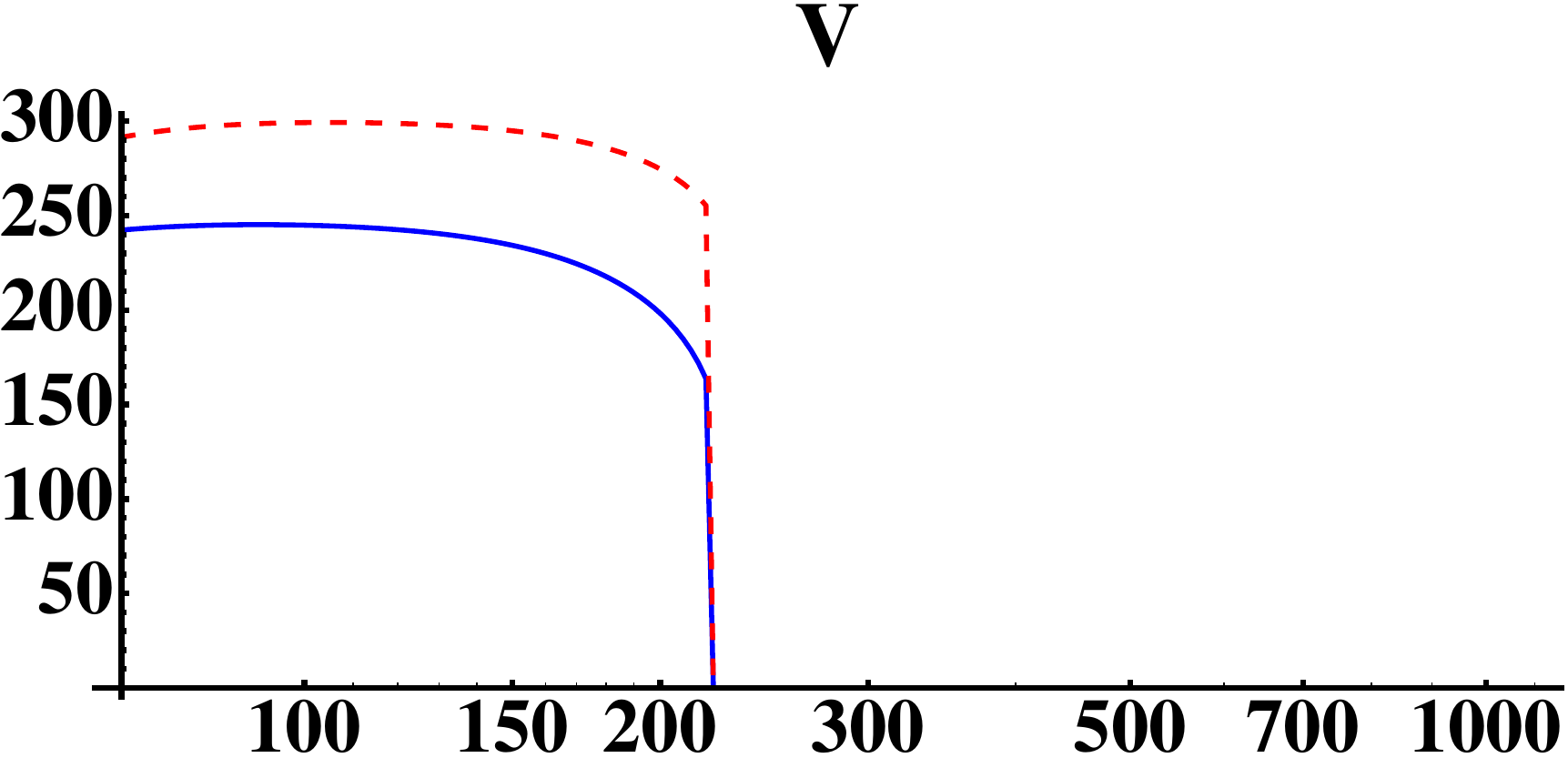}\hspace{0.4cm}\includegraphics[height=2.4cm,width=5.2cm]{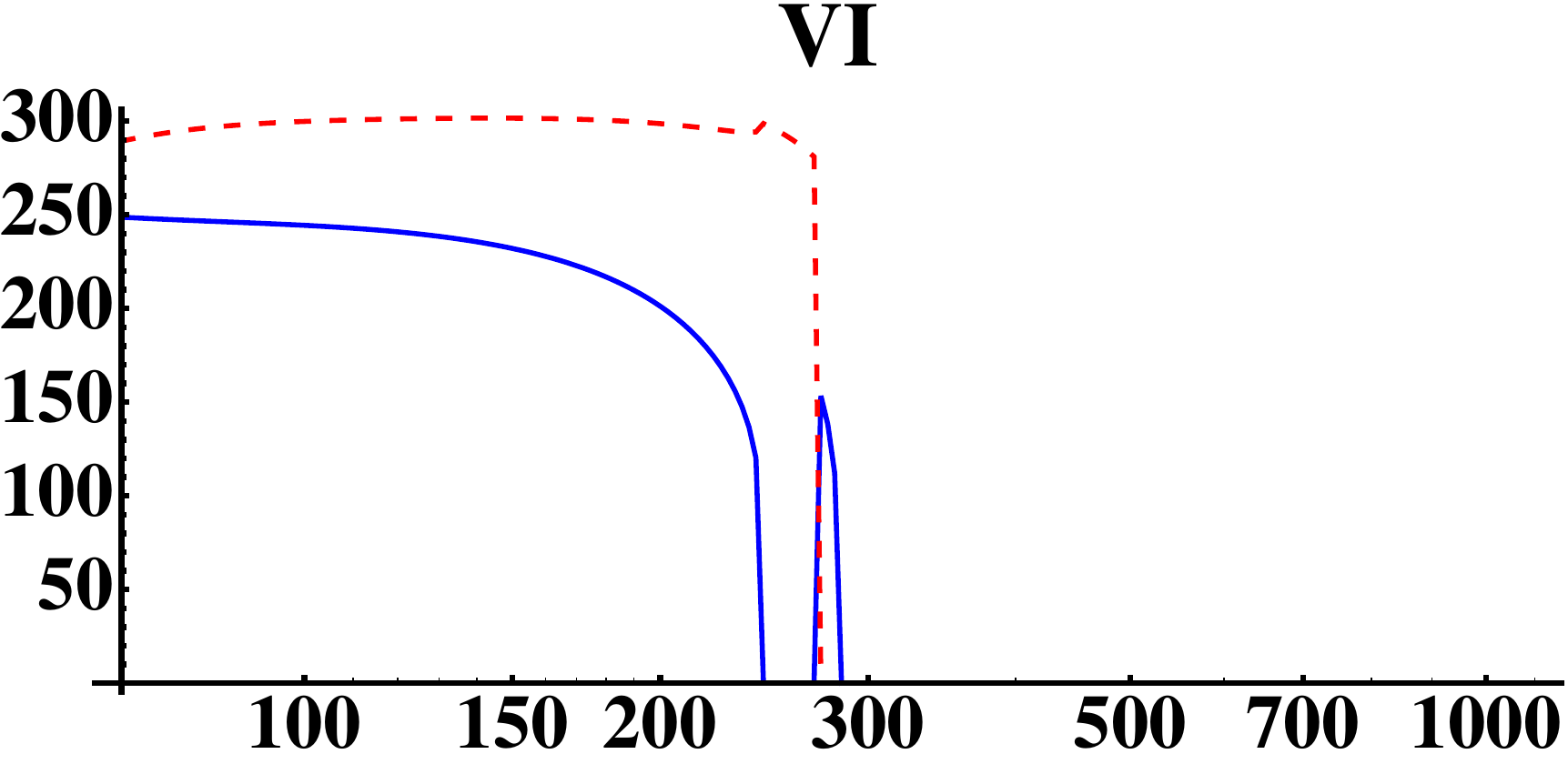}
}
\caption{The behavior of the condensates as a function of temperature for a set of parameter values which are maked with crosses in the maps of Fig.\ref{fig:phased}. The temperature (note logarithmic scale) and both the condensates $\sigma_4$ (joined blue line) and $\sigma_2$ (dashed red line) are given in GeV.
 } \label{fig:condT}
\end{figure}


Let us then go on with the more interesting and complicated case of coupling two first order transitions. We choose  reference values $\Delta M_{\Pi_4} =250$~GeV, $\Delta M_{\Pi_2} =500$~GeV, and, $v_2=300$~GeV, for which both phase transitions are first order for a wide range of Higgs masses (see Figs.~\ref{fig:decoupled4} and ~\ref{fig:decoupled2}). We start by mapping the qualitative behavior of the transitions for different values of $\beta$ and $\Delta M_\Theta$ as the Higgs masses are varied in Fig.~\ref{fig:phased}. 
The mostly vertical joined (black) curves indicate the change of the $\sigma_4$ transition from first to second order (with increasing $M_{H_+}$). Also very weak $\sigma_{4,c}/T_{4,c}<0.1$ first order transitions are counted as second order ones in the plots. Similarly the mostly horizontal joined (green) curves indicate where the same change occurs for the $\sigma_2$ transition. The dotted vertical black (horizontal green) curves show where the $\sigma_4$ ($\sigma_2$) transition ceases to exist, with increasing $M_{H_+}$. The additional dashed (red and blue) curves represent a change in the order of the critical temperatures. The meaning of the labels I-VI is as follows:
\begin{itemize}
 \item[I] both transitions first order
 \item[II] a second order $\sigma_4$ transition and a first order $\sigma_2$ transition
 \item[III] a first order $\sigma_4$ transition and a second order $\sigma_2$ transition
 \item[IV] both transitions second order
 \item[V] simultaneous (first order) transition of both condensates
 \item[VI] a ``bounce back'' of the  $\sigma_4$ condensate at the  (first order)  $\sigma_2$ transition
\end{itemize}
The lower index in the labels, $4$ or $2$, refers to the transition that has a lower critical temperature. Examples are shown in Fig.~\ref{fig:condT}. We did not label any regions which are small or where either of the phase transitions is absent.
Notice that in the regions V and VI the order of the critical temperatures is not defined. Thus they are surrounded by the red and blue joined curves, which are the borders of the regions where $T_{4,c}>T_{2,c}$ and $T_{4,c}<T_{2,c}$, respectively, and the order is well defined. 
These cases will be discussed in detail below. Notice that in the $\beta=0=\Delta M_\Theta$ plots the lines which denote changes in the behavior of the $\sigma_4$ ($\sigma_2$) transition are exactly vertical (horizontal). This is a sign of decoupling of the transitions. From (\ref{Hmm}) we also see that the Higgs mass matrix is independent of $\beta$ if $M_{H_+}=M_{H_-}$. This is reflected in Fig.~\ref{fig:phased} as the maps for different values of $\beta$ coincide on the  $M_{H_+}=M_{H_-}$ line.

\begin{figure}
{
 \includegraphics[height=4.5cm,width=3.8cm]{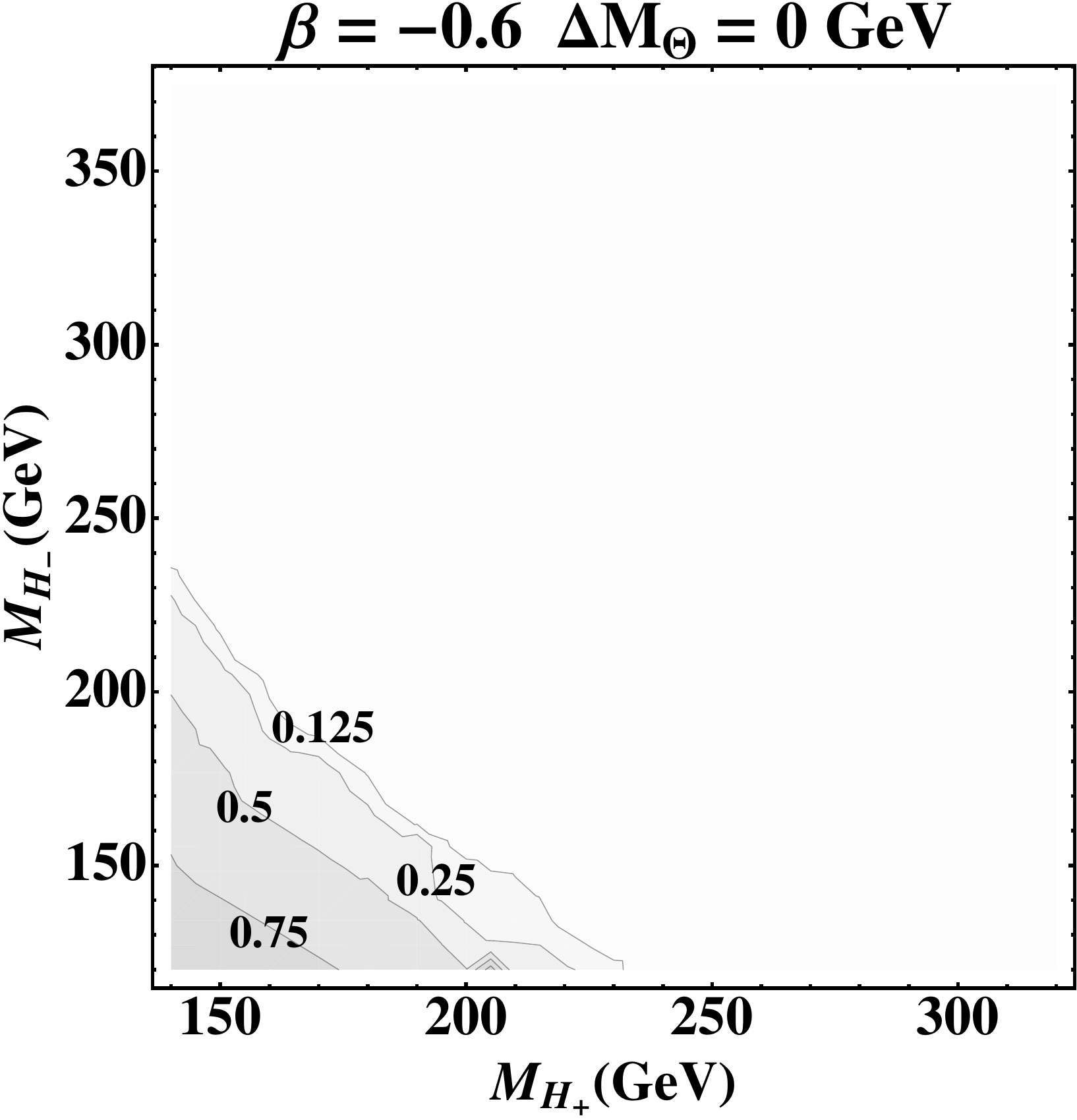}\hspace{0.4cm}\includegraphics[height=4.5cm,width=3.8cm]{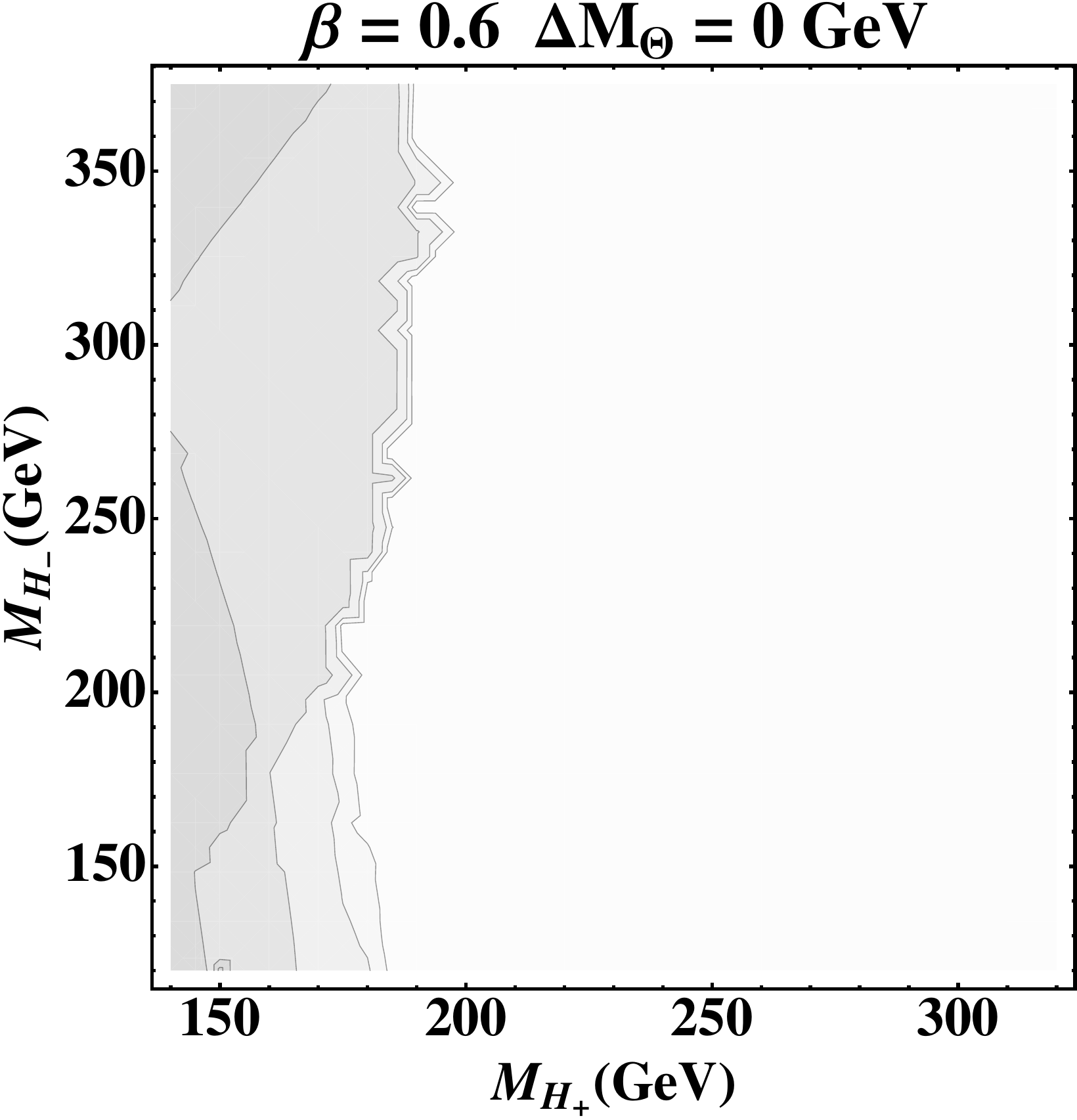}\hspace{0.4cm}\includegraphics[height=4.5cm,width=3.8cm]{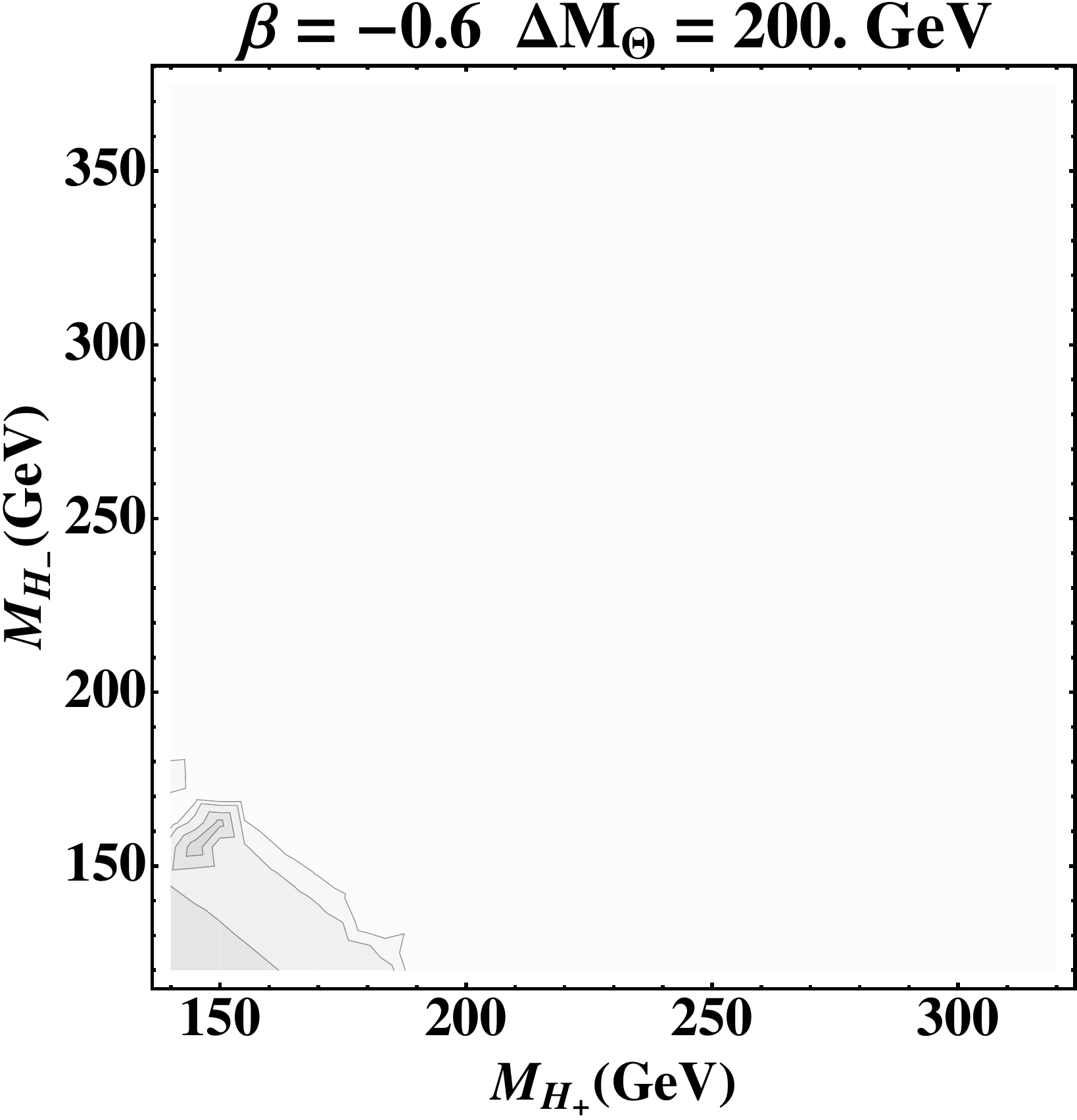}\hspace{0.4cm}\includegraphics[height=4.5cm,width=3.8cm]{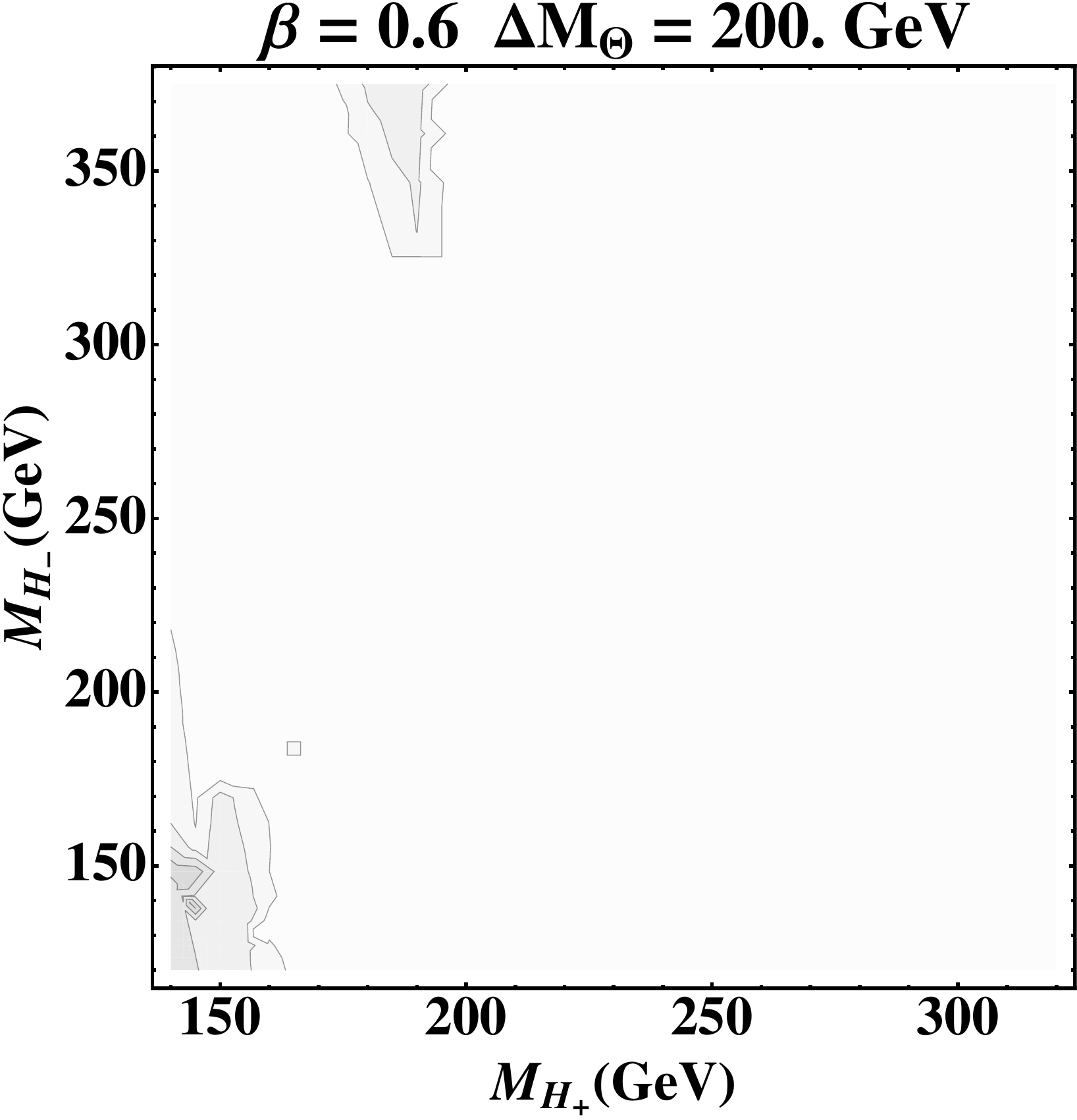}
 \includegraphics[height=4.5cm,width=3.8cm]{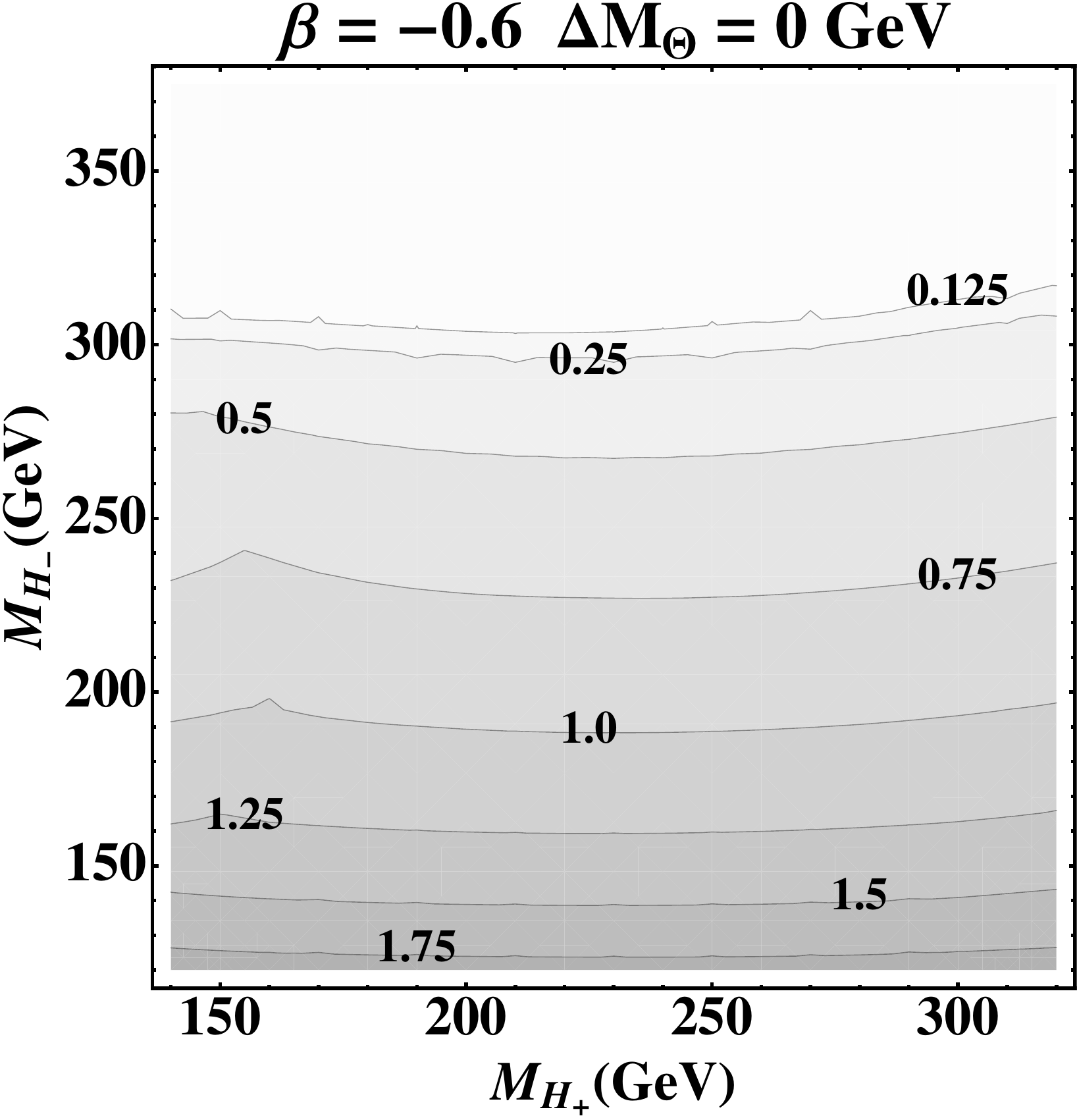}\hspace{0.4cm}\includegraphics[height=4.5cm,width=3.8cm]{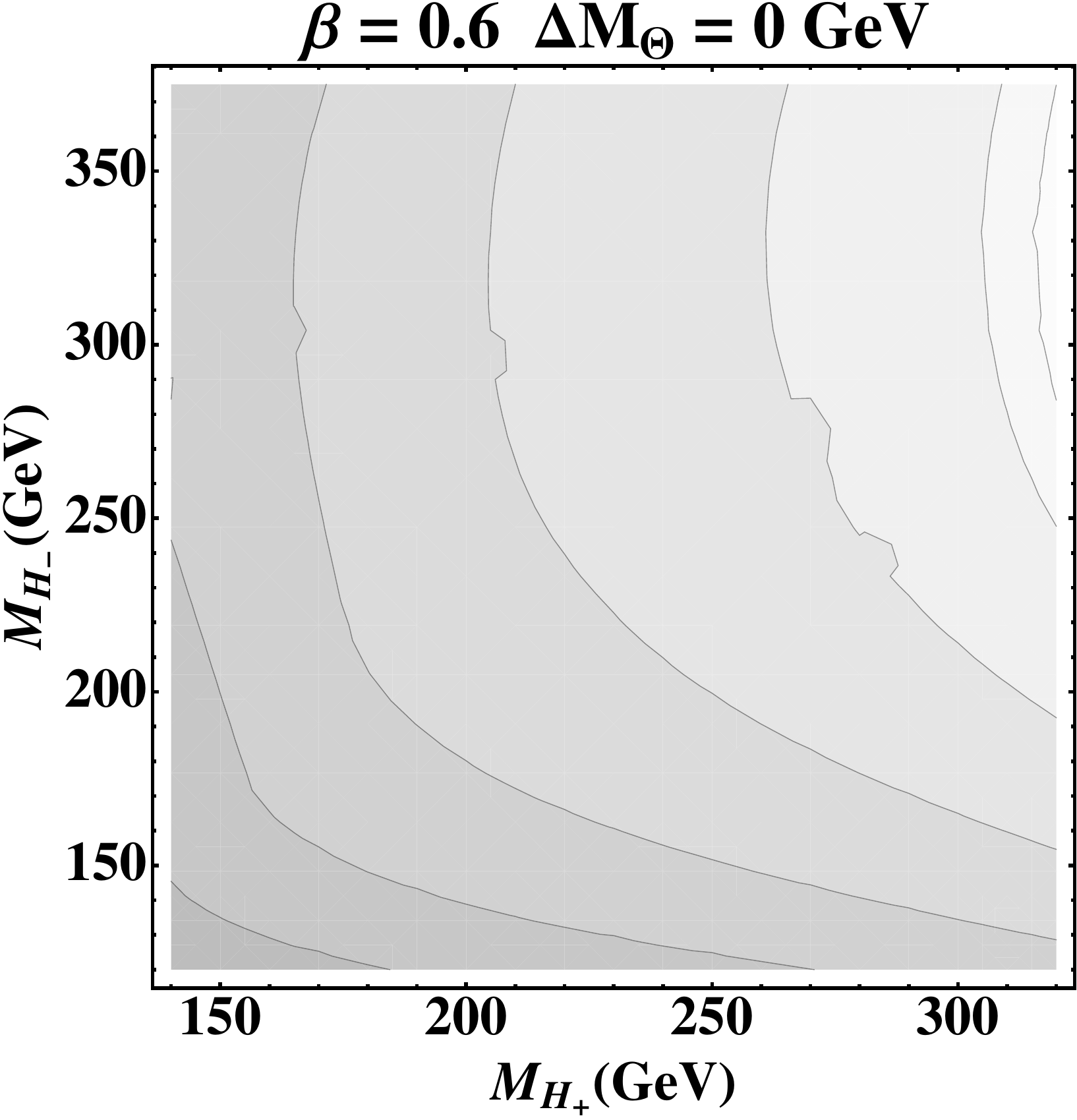}\hspace{0.4cm}\includegraphics[height=4.5cm,width=3.8cm]{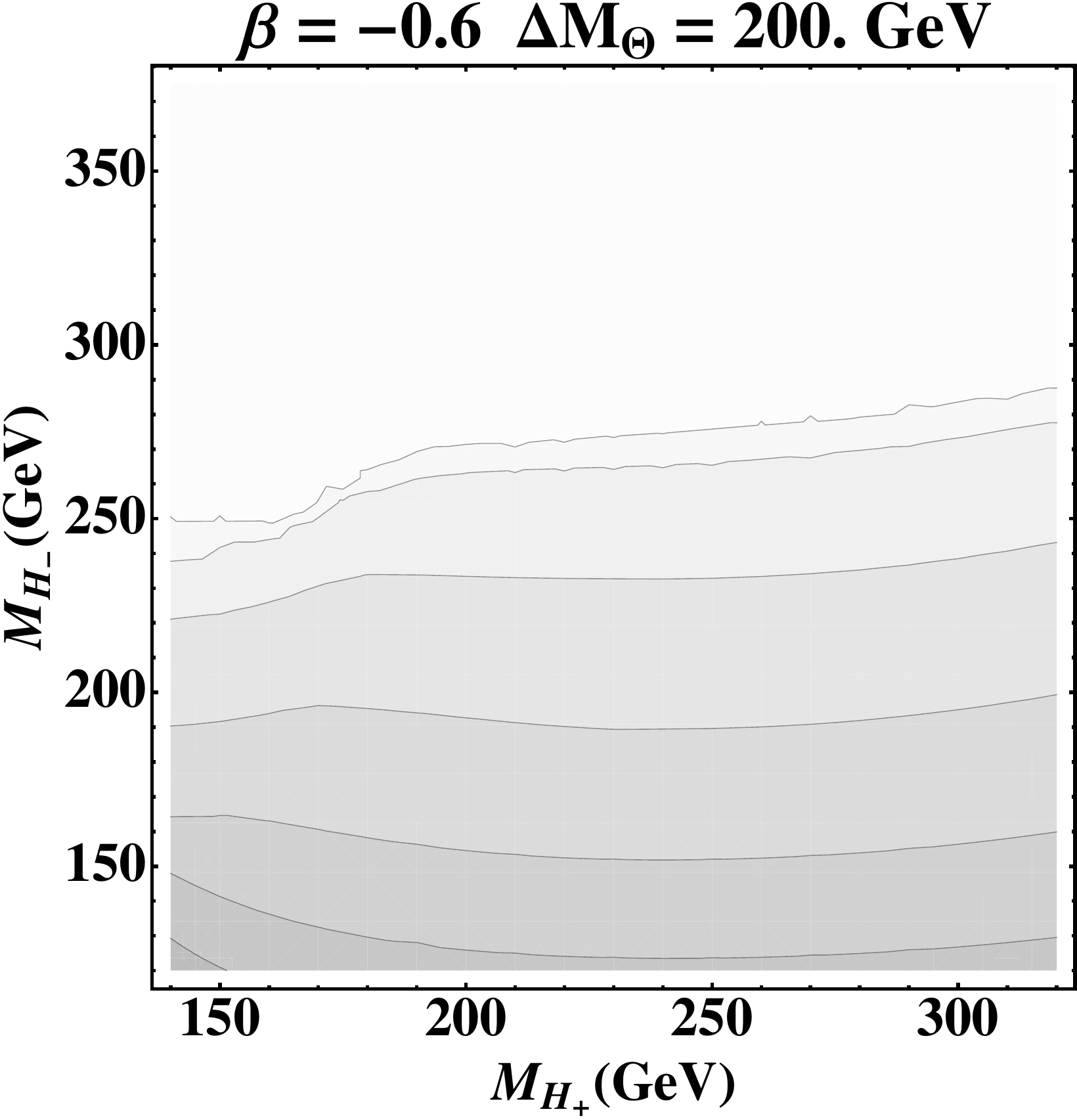}\hspace{0.4cm}\includegraphics[height=4.5cm,width=3.8cm]{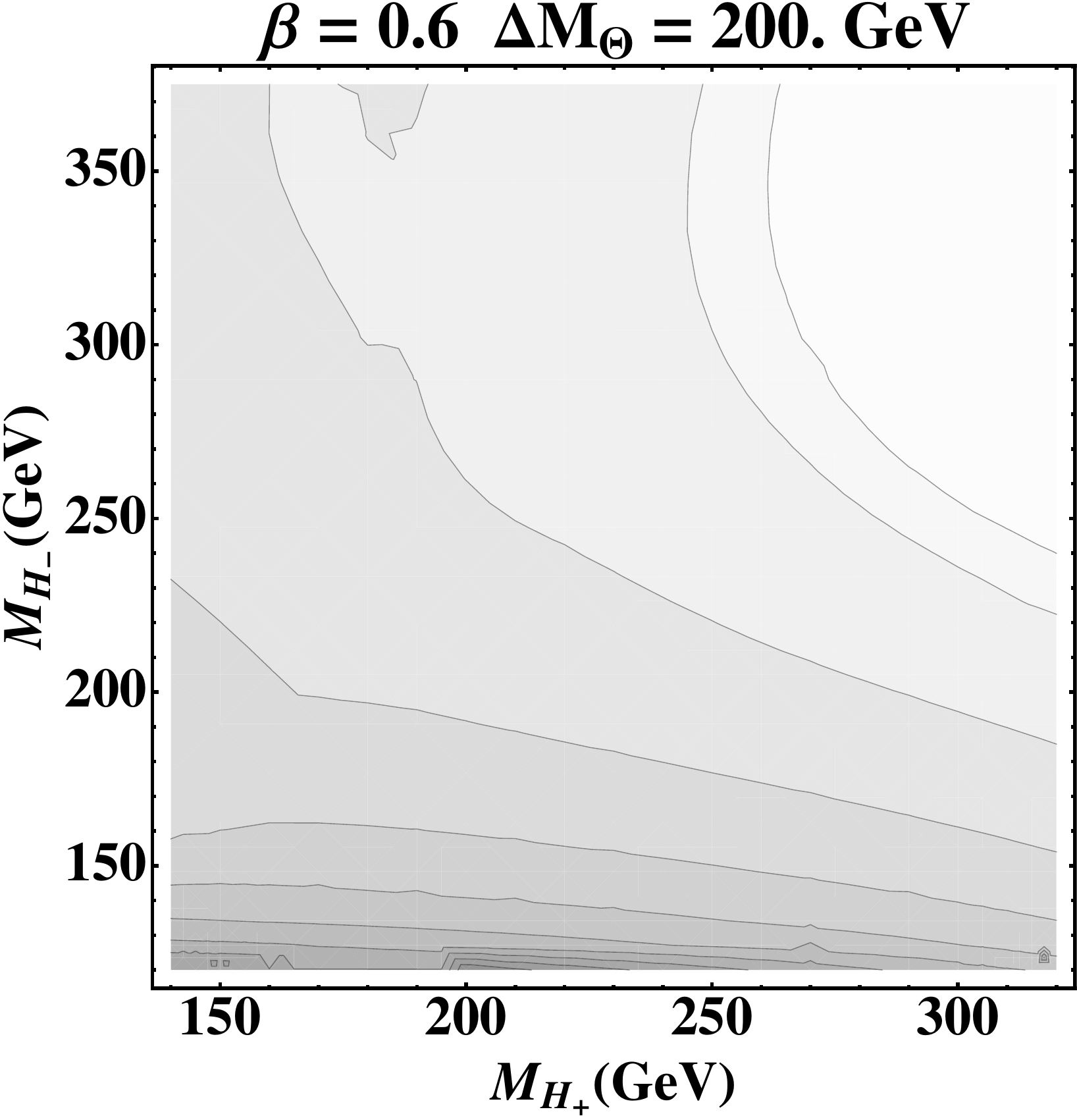}
}
\caption{The ratios $\sigma_{4,c}/T_{4,c}$ (top row) and  $\sigma_{2,c}/T_{2,c}$ (bottom row) as functions of the Higgs masses in the ``without EW'' scenario. We use $\Delta M_\Theta = 0$, 200~GeV  and $\beta= -0.6$,  0.6 as indicated in the labels and otherwise the parameters of Fig.~\ref{fig:phased} ($\Delta M_{\Pi_4} = 250$~GeV, $\Delta M_{\Pi_2} = 500$~GeV, and $v_2=300$~GeV).}\label{fig:phioverTwithout}
\end{figure}

As seen from  Fig.~\ref{fig:condT}, the critical temperatures are larger or approximately equal to the electroweak scale. Thus the ratios $\sigma/T$ are at most of the order one, and very strong transitions are not expected. This is seen explicitly from the plots of Fig.~\ref{fig:phioverTwithout}, which show the strengths $\sigma/T$ for the parameter values of Fig.~\ref{fig:phased}. We chose the parameter values such that the plots are in one to one correspondence with Fig.~\ref{fig:phasedEW}, except that we excluded the $\beta=0$ plots where coupling effects are small.
In particular, $\sigma_{4,c}/T_{4,c}$ remains small even when the first order transitions are coupled. The values of $\sigma/T$ slightly increase if larger values of $\Delta M_{ \Pi_2}$ and  $\Delta M_{\Pi_4}$ are chosen.

\begin{figure}
{
 \includegraphics[height=5.5cm,width=5.2cm]{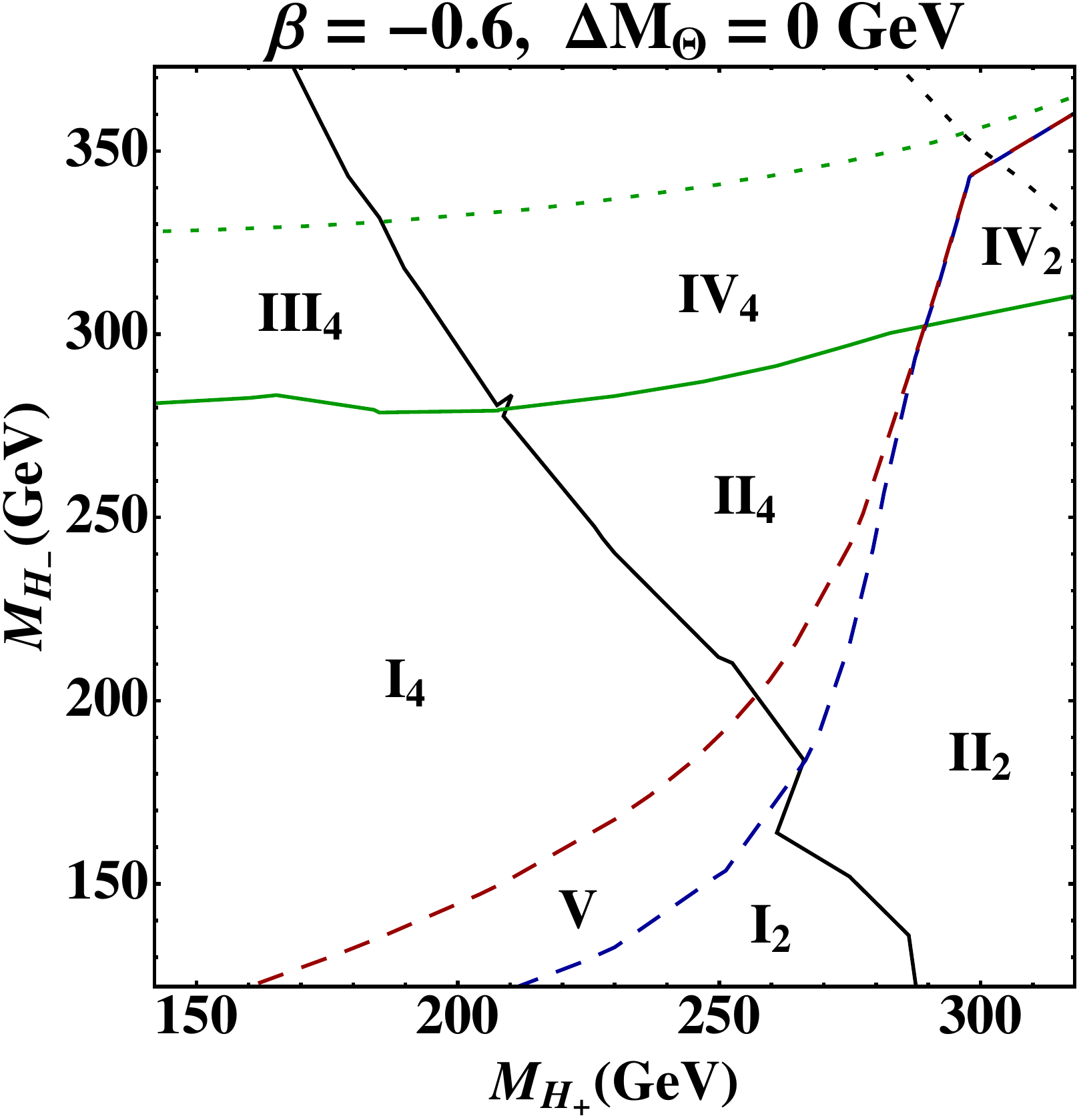}\hspace{0.4cm}\includegraphics[height=5.5cm,width=5.2cm]{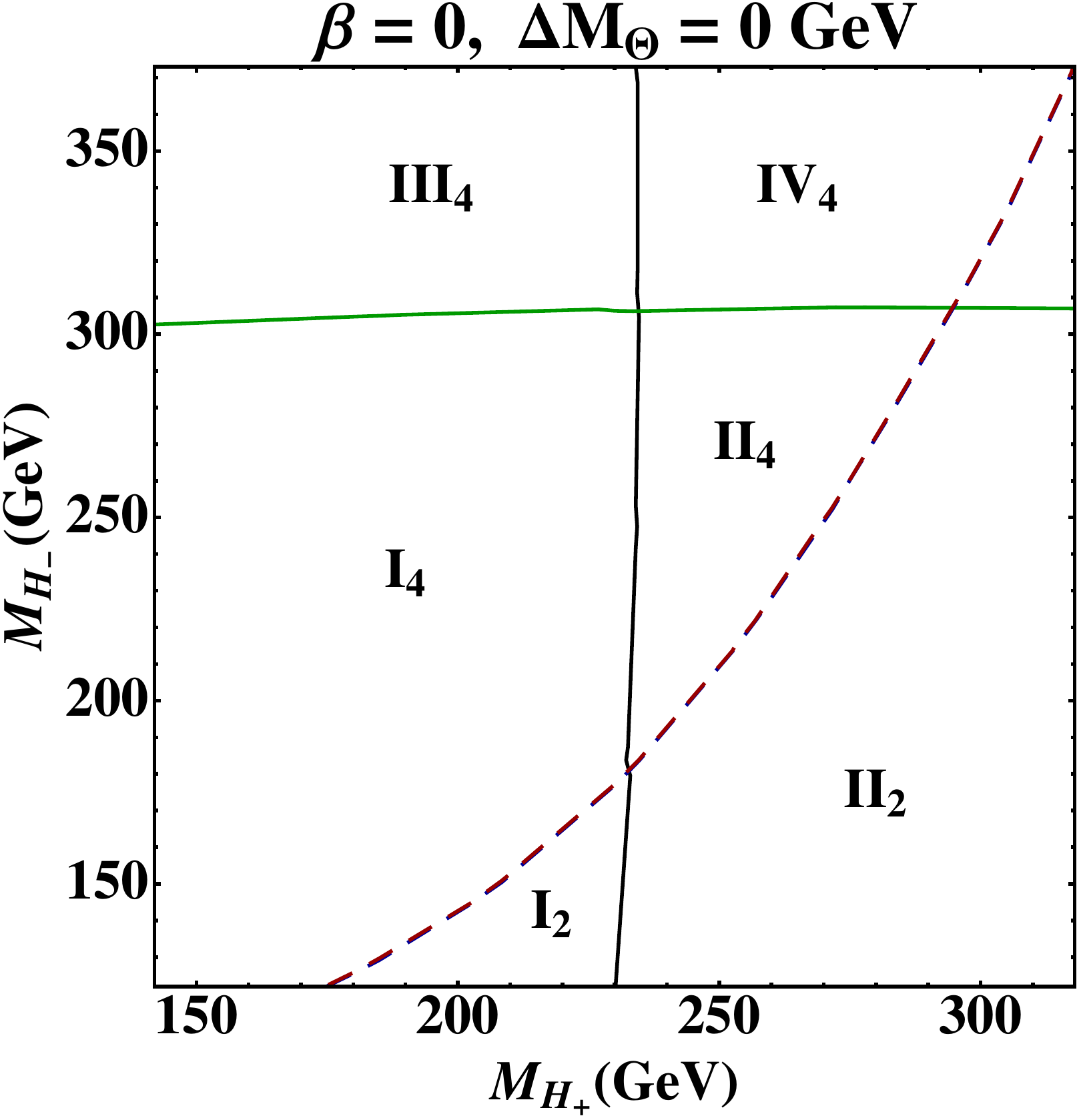}\hspace{0.4cm}\includegraphics[height=5.5cm,width=5.2cm]{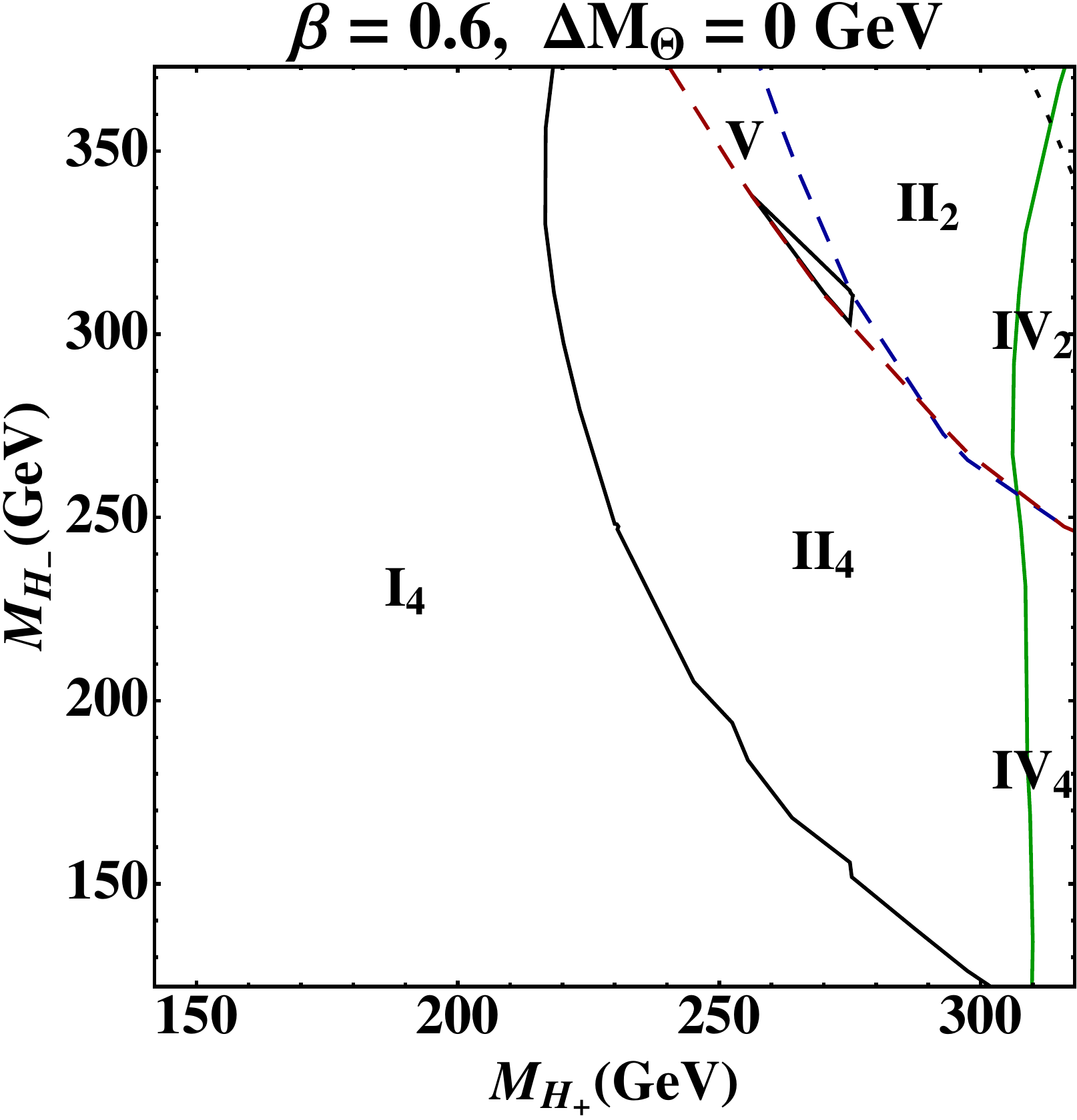}


 \includegraphics[height=5.5cm,width=5.2cm]{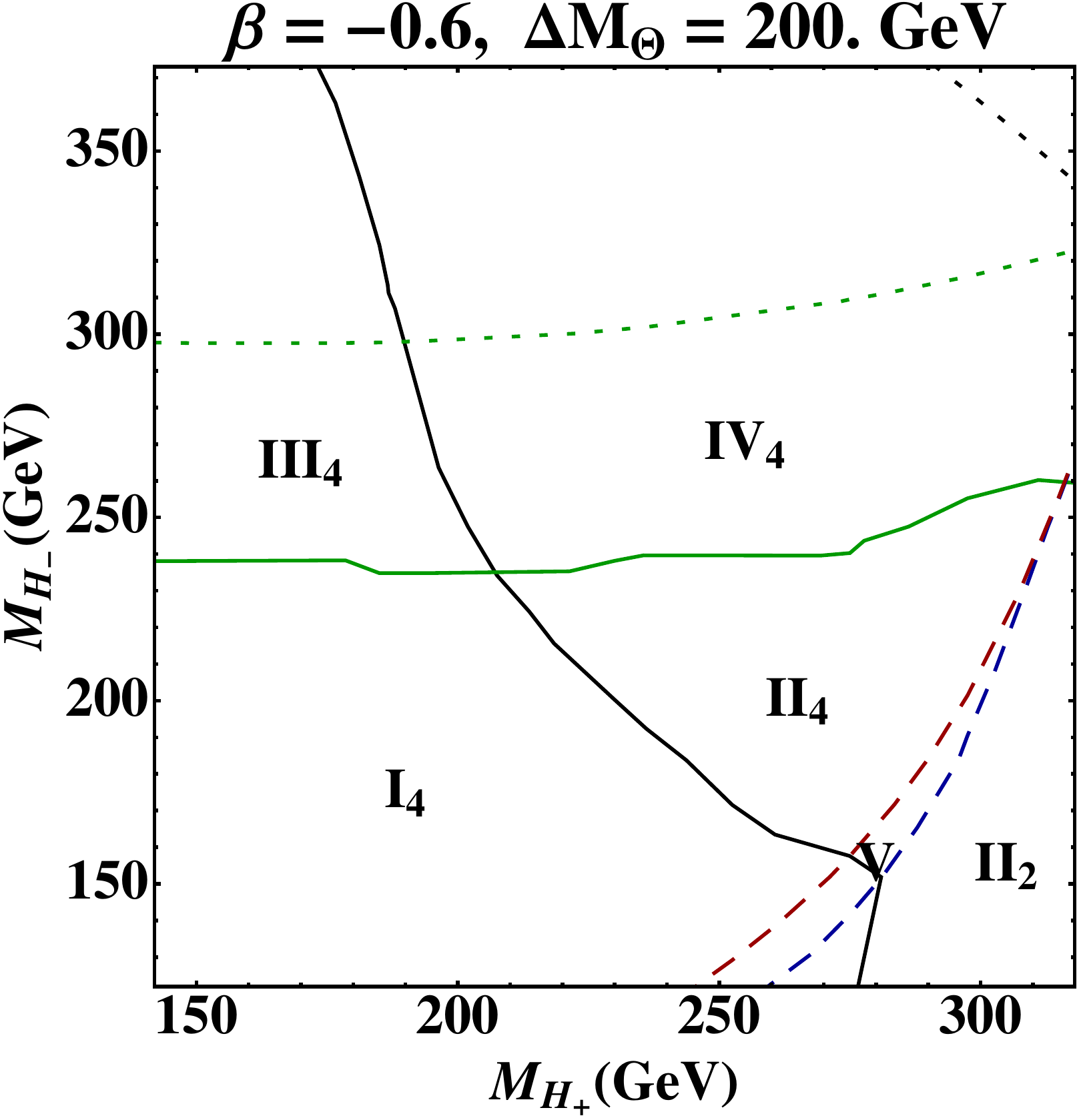}\hspace{0.4cm}\includegraphics[height=5.5cm,width=5.2cm]{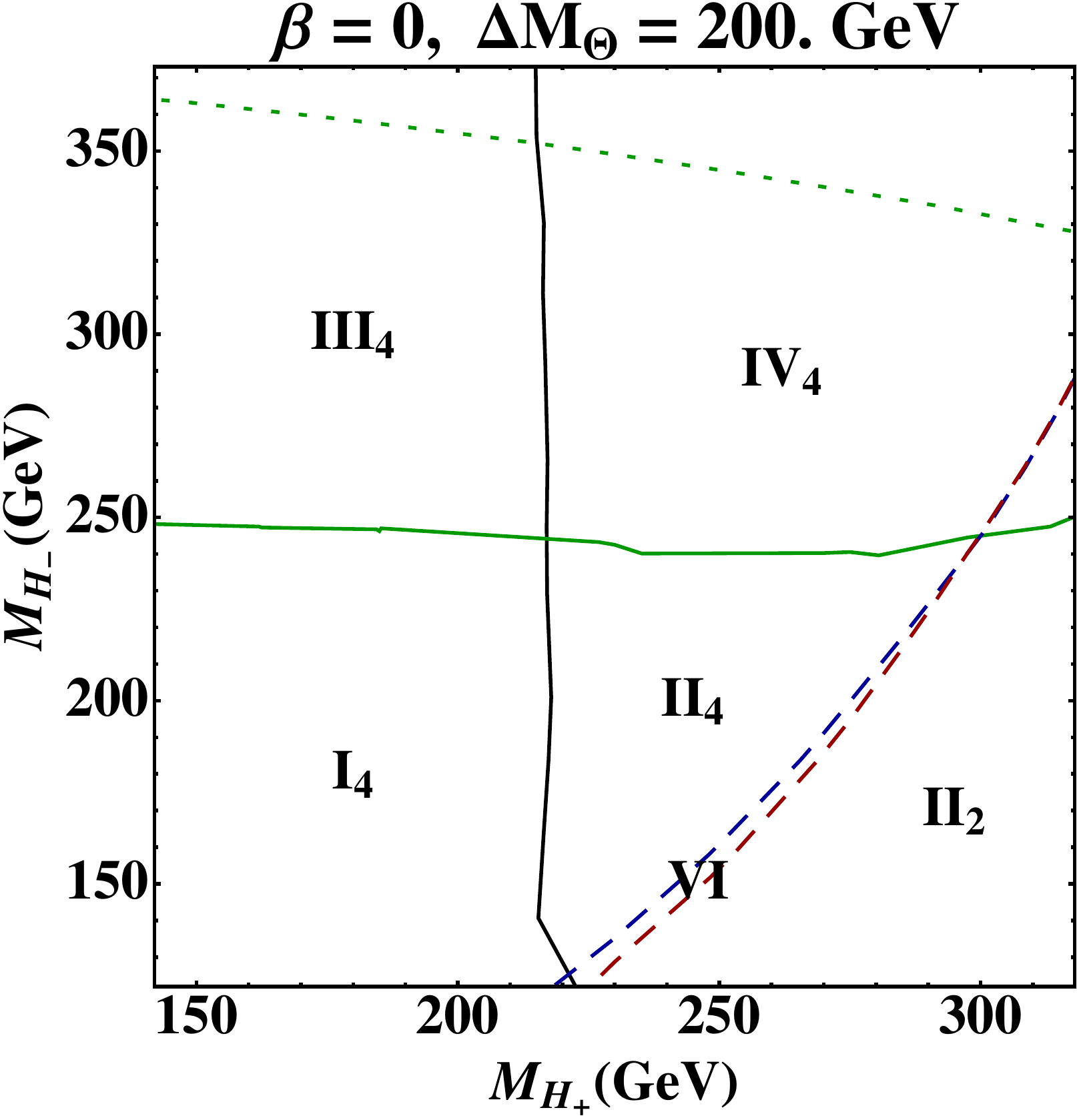}\hspace{0.4cm}\includegraphics[height=5.5cm,width=5.2cm]{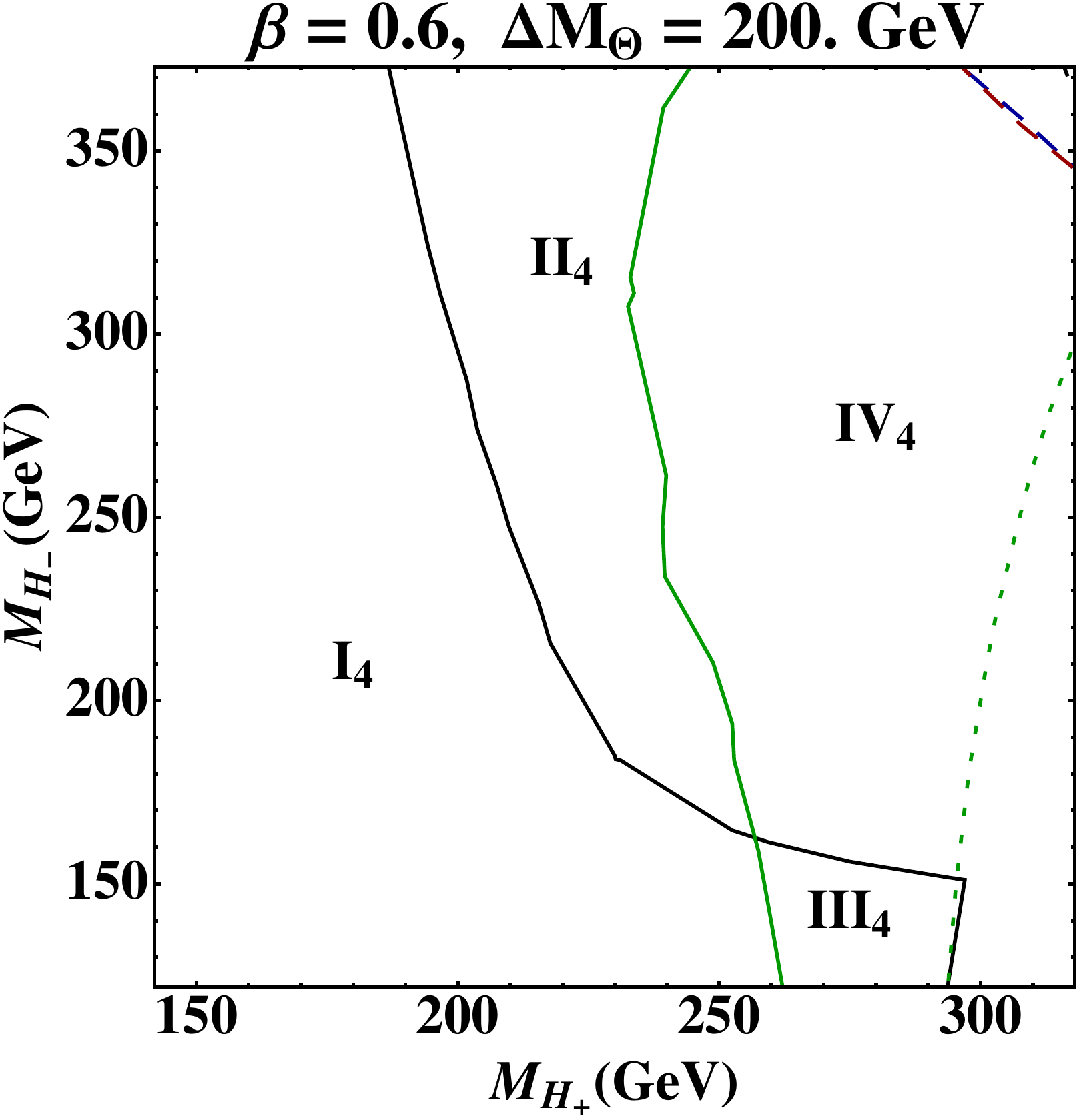}
}
\caption{The same as Fig.~\ref{fig:phased} ($\Delta M_{\Pi_4} = 250$~GeV, $\Delta M_{\Pi_2} = 500$~GeV, and $v_2=300$~GeV) but with EW interactions switched on. }\label{fig:phasedEW}
\end{figure}

\subsection{With EW}

Let us then discuss the effects of electroweak physics. First, when either of the transitions is second order, we see similar behavior as in the ``without EW'' scenario: mixing typically weakens the transitions. That is why we start directly by studying the coupling of two first order transitions.
Fig.~\ref{fig:phasedEW} presents the maps of the phase transition behavior exactly for the same parameter values as Fig.~\ref{fig:phased}, but now with EW interactions switched on. As mentioned above, the difference between Figs.~\ref{fig:phased} and~\ref{fig:phasedEW} is dominated by the inclusion of the top quark. The temperature dependent one-loop term is significantly enhanced when the top is included reducing the critical temperature and allowing for a heavier composite Higgs. Thus the interesting mixing effects (regions V and VI) take place at much higher Higgs mass $M_{H_+}$ in the ``with EW'' scenario.

\begin{figure}
{
 \includegraphics[height=5.5cm,width=5.2cm]{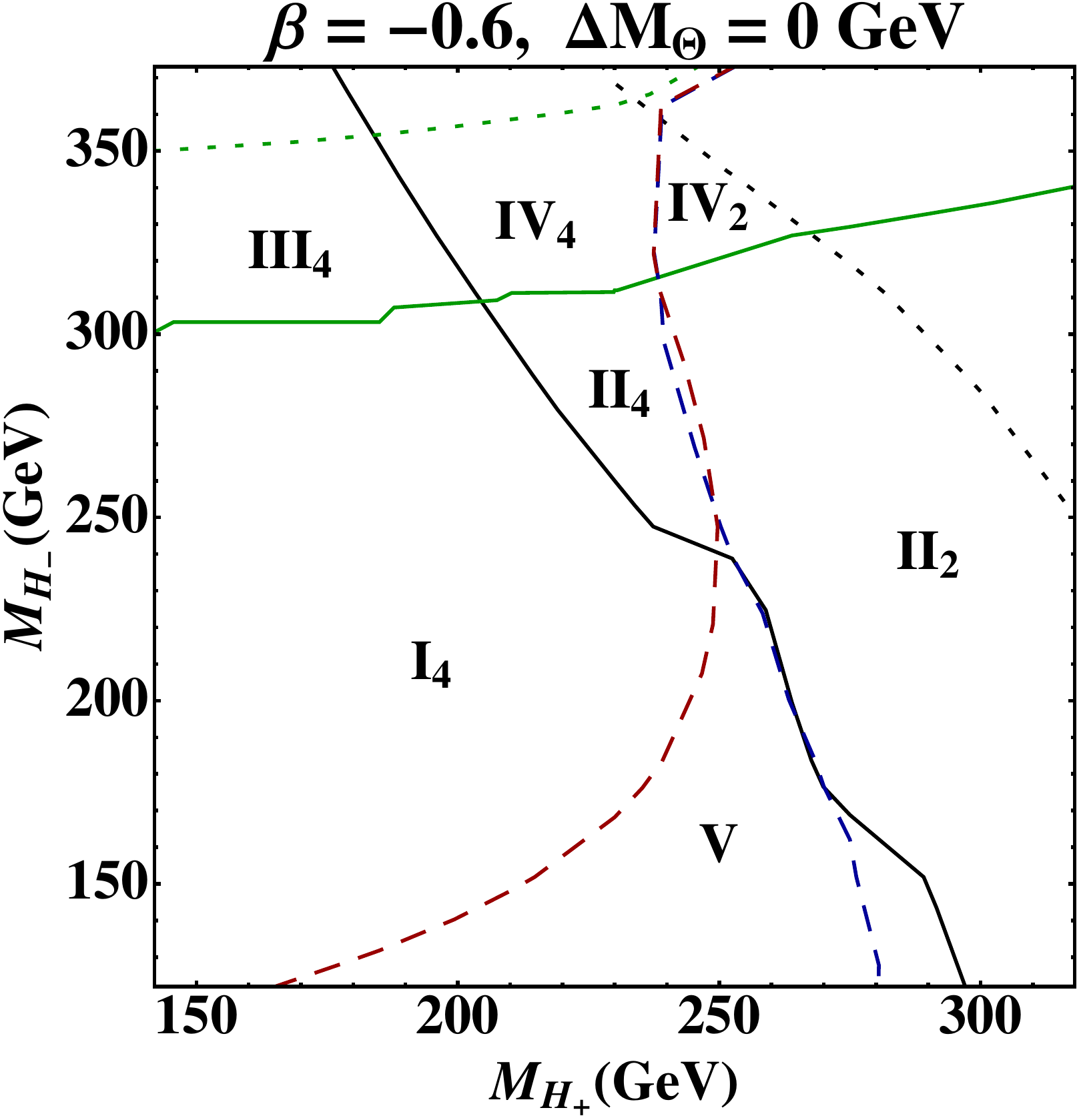}\hspace{0.4cm}\includegraphics[height=5.5cm,width=5.2cm]{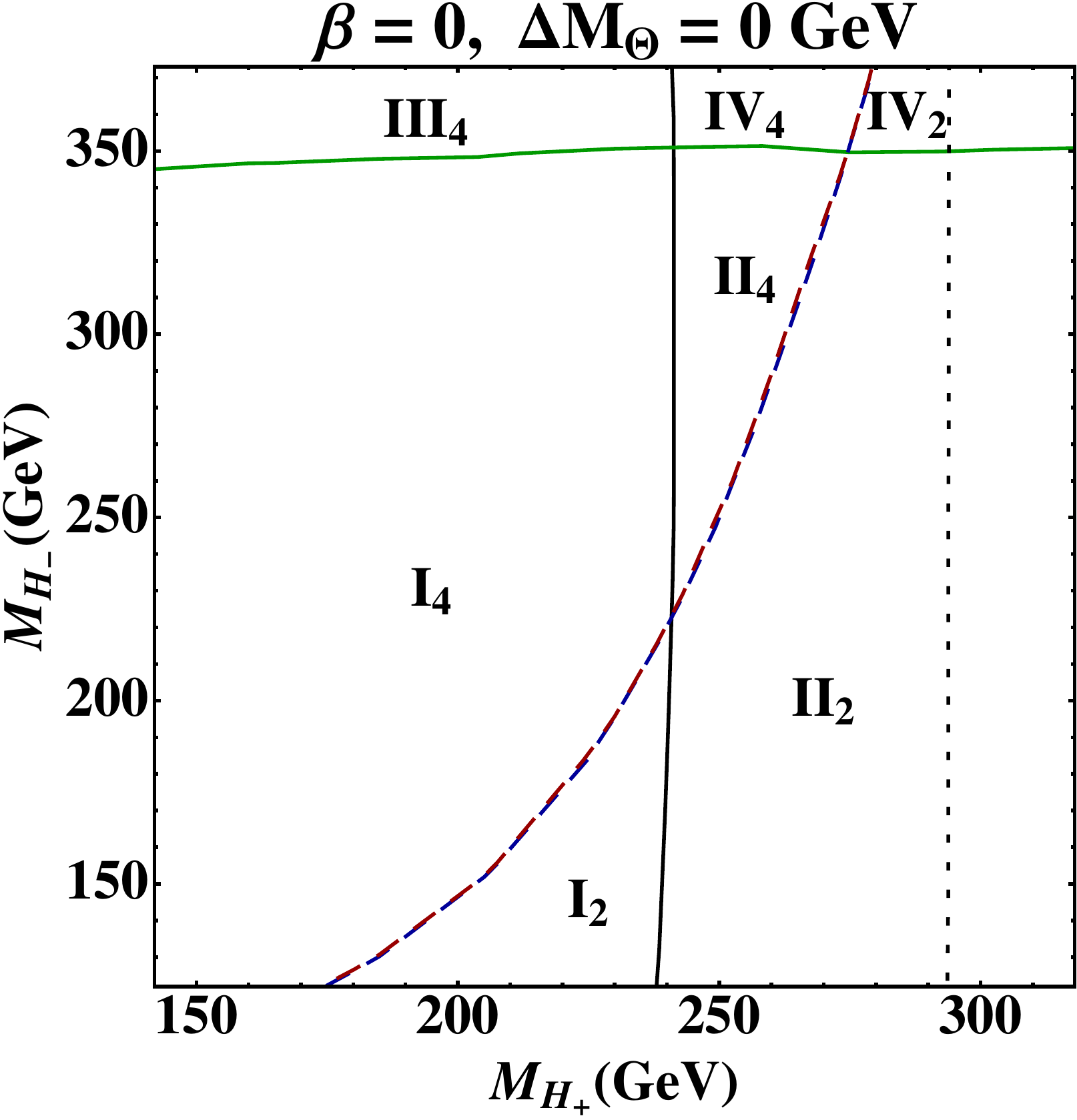}\hspace{0.4cm}\includegraphics[height=5.5cm,width=5.2cm]{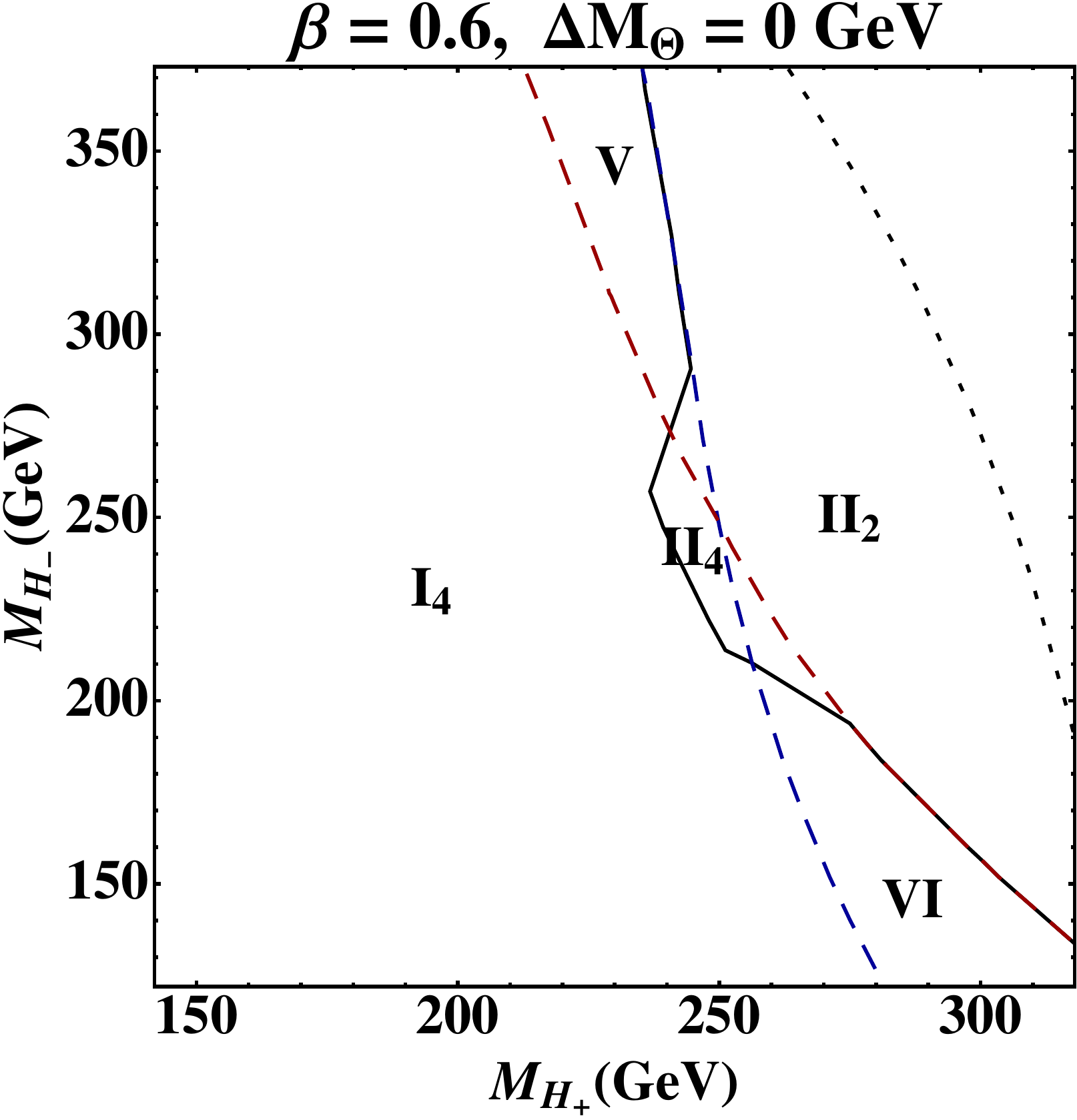}


 \includegraphics[height=5.5cm,width=5.2cm]{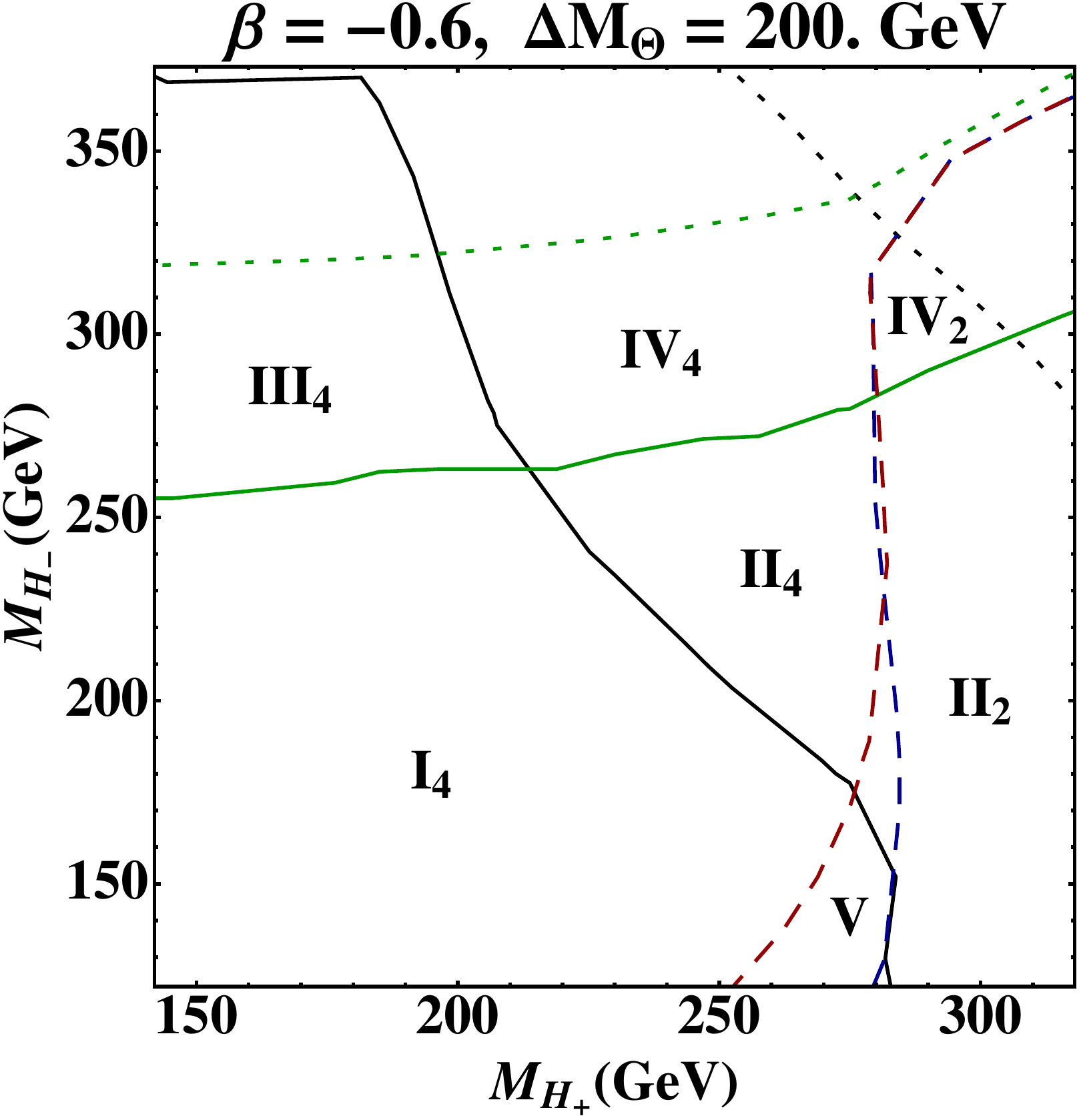}\hspace{0.4cm}\includegraphics[height=5.5cm,width=5.2cm]{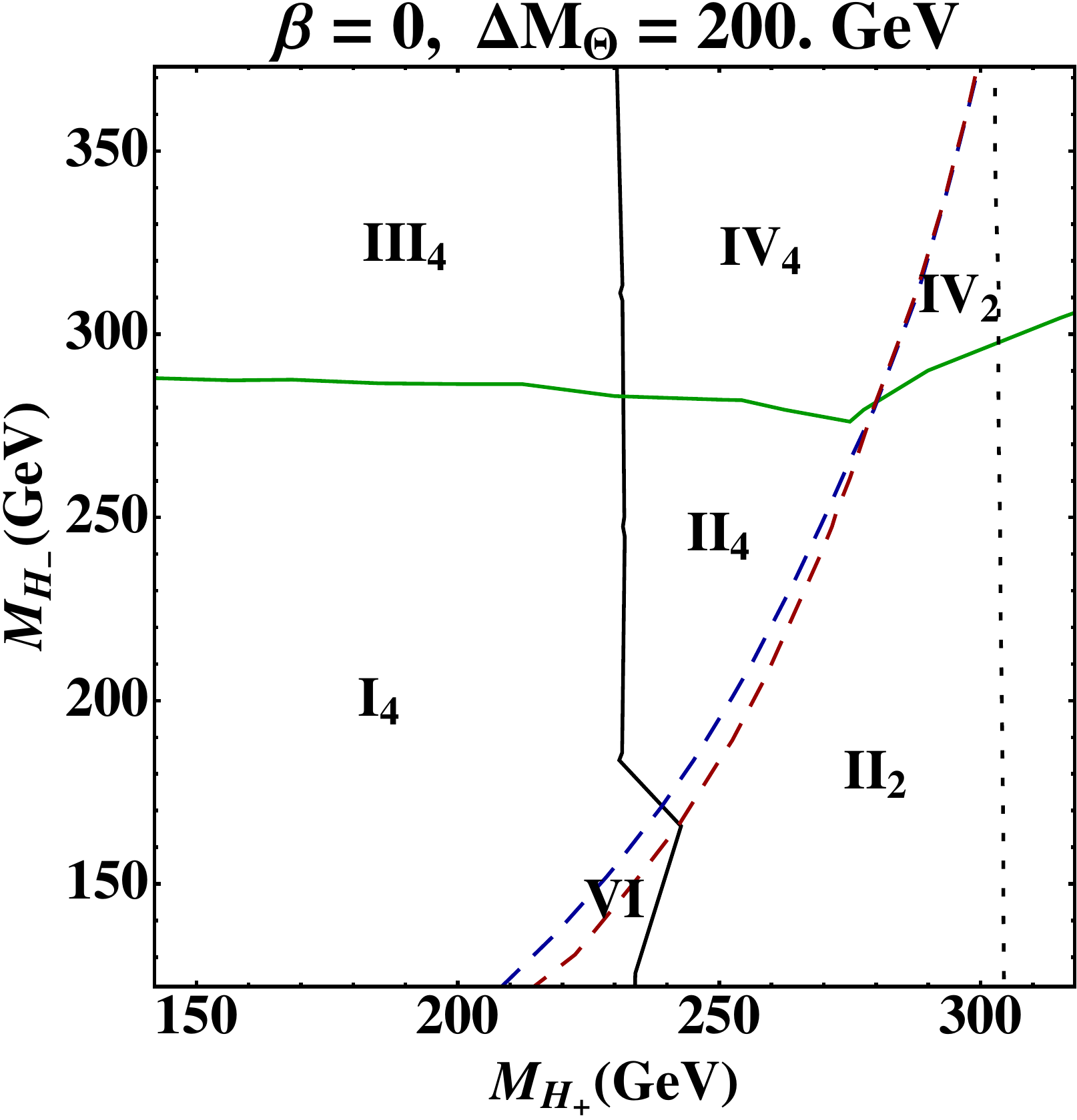}\hspace{0.4cm}\includegraphics[height=5.5cm,width=5.2cm]{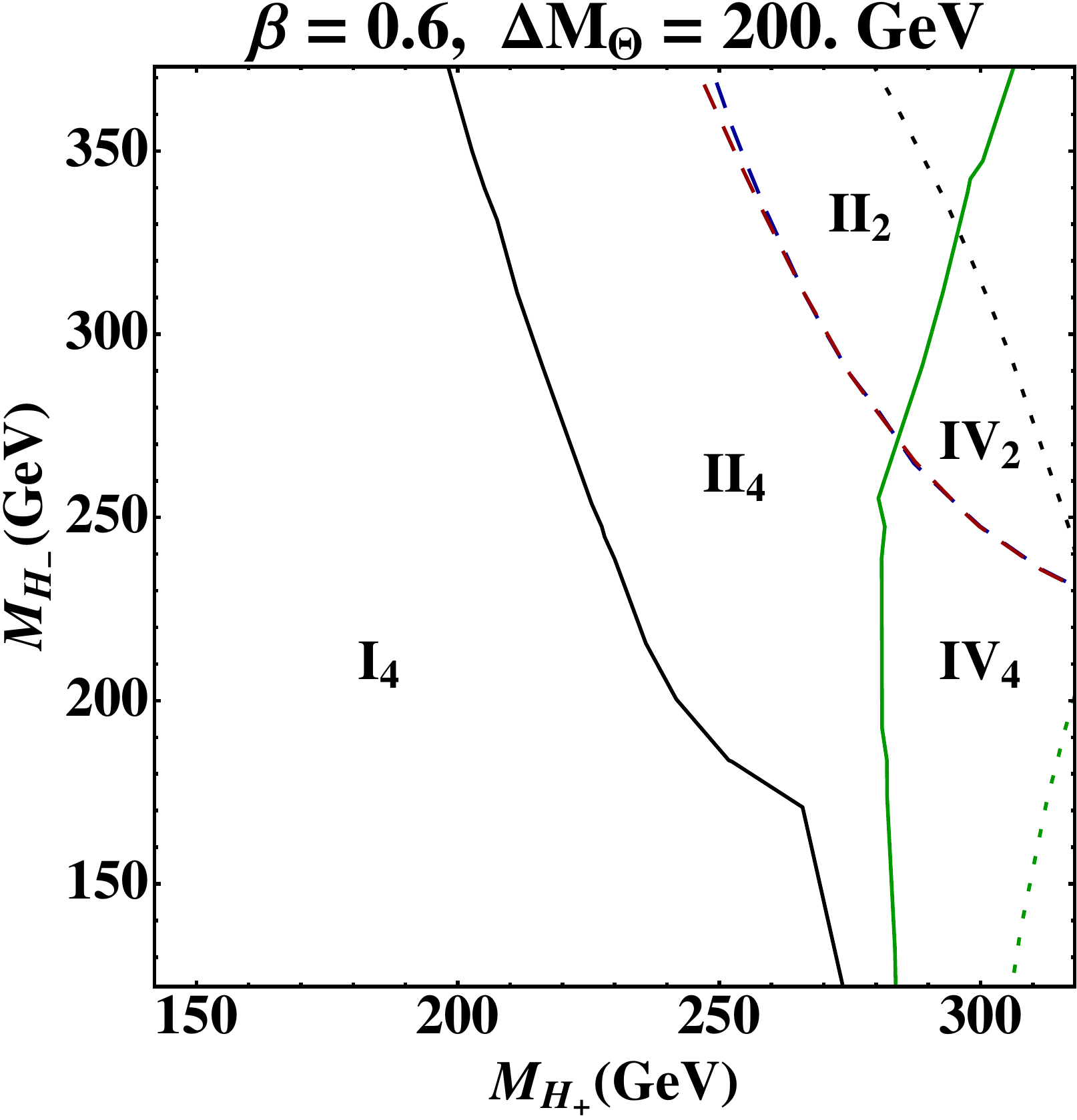}
}
\caption{The same as Figs.~\ref{fig:phased}, and ~\ref{fig:phasedEW} but for different parameter values. We use the ``with EW'' scenario with $\Delta M_{\Pi_4}=300$~GeV, $\Delta M_{\Pi_2} = 550$~GeV and $v_2=350$~GeV. }\label{fig:phasedEW2}
\end{figure}

In particular, the parameters of Figs.~\ref{fig:phased} and~\ref{fig:phasedEW} were chosen to have strong transitions and coupling effects in the ``without EW'' scenario. This is why we also present the maps with EW interactions but for a different parameter setting in Fig.~\ref{fig:phasedEW2}.

\subsection{Simultaneous and extra phase transitions}

Let us discuss the physics related to the regions V and VI. Recall that in the region V the two phase transitions take place simultaneously whereas in VI there is a bounce back of the condensate $\sigma_4$ at the $\sigma_2$ transition. It is also possible to have the condensate  $\sigma_2$ bouncing back at the $\sigma_4$ transition, but this does not take place in the parameter space of Figs.~\ref{fig:phased}, \ref{fig:phasedEW} and \ref{fig:phasedEW2}.

As pointed out in \cite{Jarvinen:2009wr}, regions V and VI appear naturally when two first order transitions are coupled. The main requirement is that the critical temperatures of the transitions lie close to each other and that they coincide for some values of the input parameters when the transitions are decoupled. When the coupling between the two Higgses is turned on one is bound to see either a simultaneous transition or several phase transitions as a function of temperature, as in region VI. In the case of UMT one can specify quantitatively which parts of the parameter space include regions V and which include VI. First, notice from Figs.~\ref{fig:phased}, \ref{fig:phasedEW} and \ref{fig:phasedEW2} that these phenomenona are mostly driven by the nonzero values of $\beta$: in the $\beta=0$, $\Delta M_\Theta=200$~GeV plots coupling effects are small. 
The relevant quantity is the nondiagonal element in the Higgs mass matrix (\ref{Hmm}) which is
given by $1/2 \sin(2\beta)(M_{H+}^2-M_{H_-}^2)$ and thus
independent of $\Delta M_\Theta$. Notice that it is directly related to the shape of the tree level potential (\ref{Vtree}) by
\be \label{ndelem}
 \frac{1}{2}\sin(2\beta)(M_{H+}^2-M_{H_-}^2) = 2 \left(\delta -2 v_2^2 \delta^\prime\right) \ .
\ee
When the nondiagonal element is positive (negative), region VI (V) appears as a result of mixing. Hence the VI (V) regions lie 
above (below) the $M_{H_+}=M_{H_-}$ line in
the $\beta=-0.6$ plots and vice versa for $\beta=+0.6$. This statement is exact for $\Delta M_\Theta=0$ but also appears to be a fairly good approximation for $\Delta M_\Theta=200$~GeV.
The fact that (\ref{ndelem}) vanshes for $\beta=0$ and for $M_{H_+}=M_{H_-}$ is in agreement with the earlier observation that (for $\Delta M_\Theta=0$) the $\beta\ne 0$ plots coincide with the $\beta=0$ plot on the line of equal Higgs masses where the phase transitions are thus decoupled.

\begin{figure}
{
 \includegraphics[height=4.5cm,width=3.8cm]{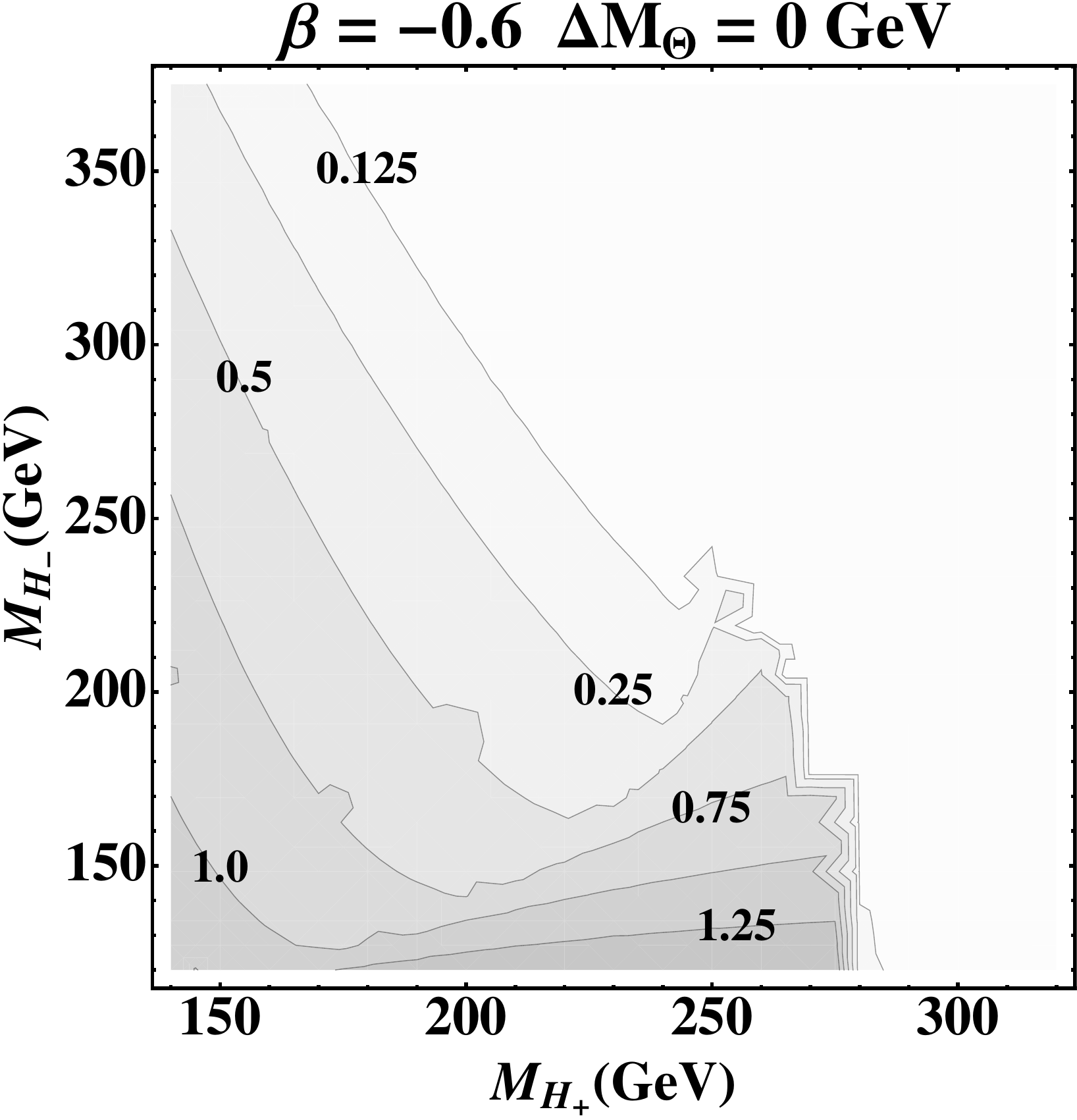}\hspace{0.4cm}\includegraphics[height=4.5cm,width=3.8cm]{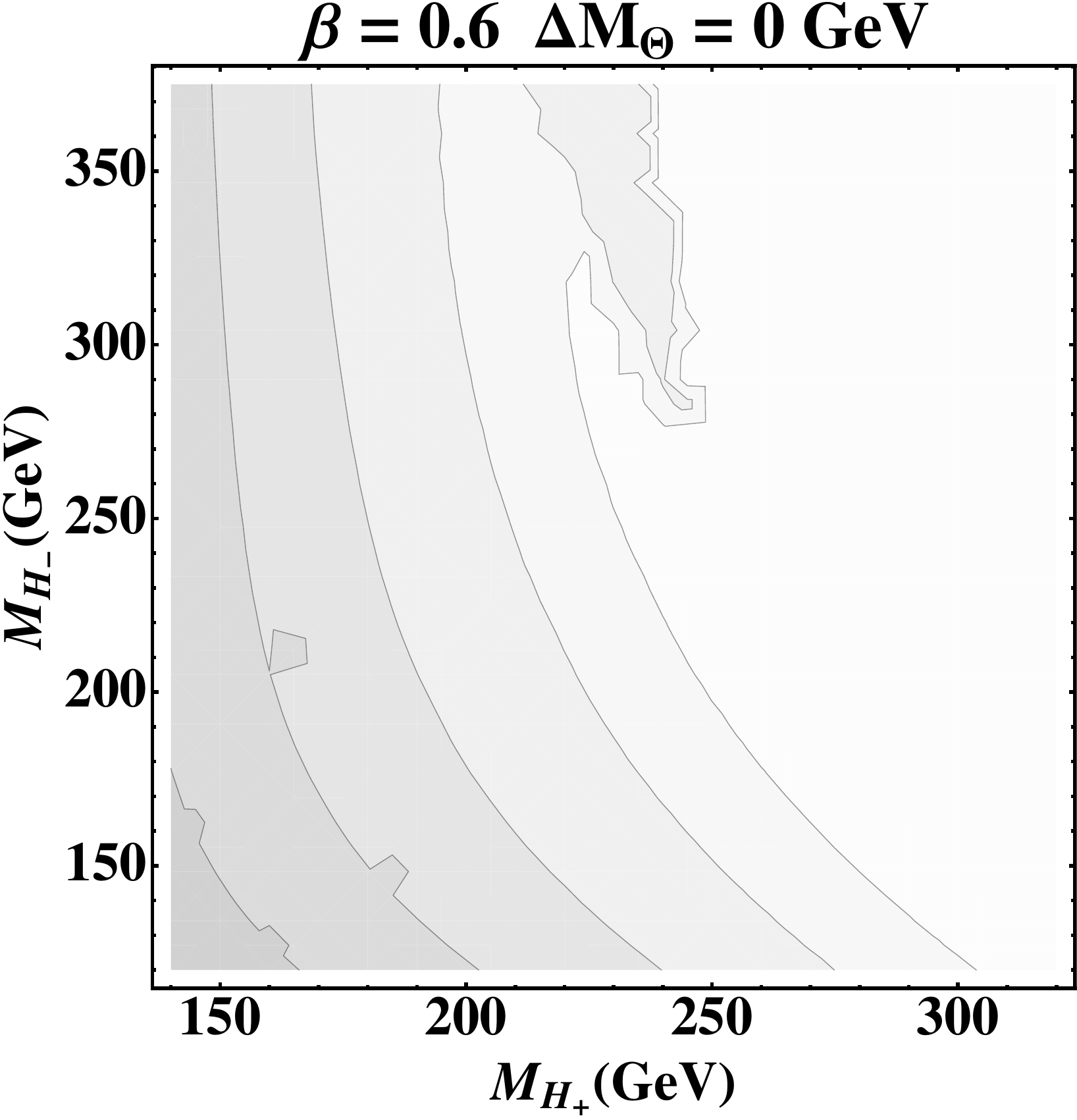}\hspace{0.4cm}\includegraphics[height=4.5cm,width=3.8cm]{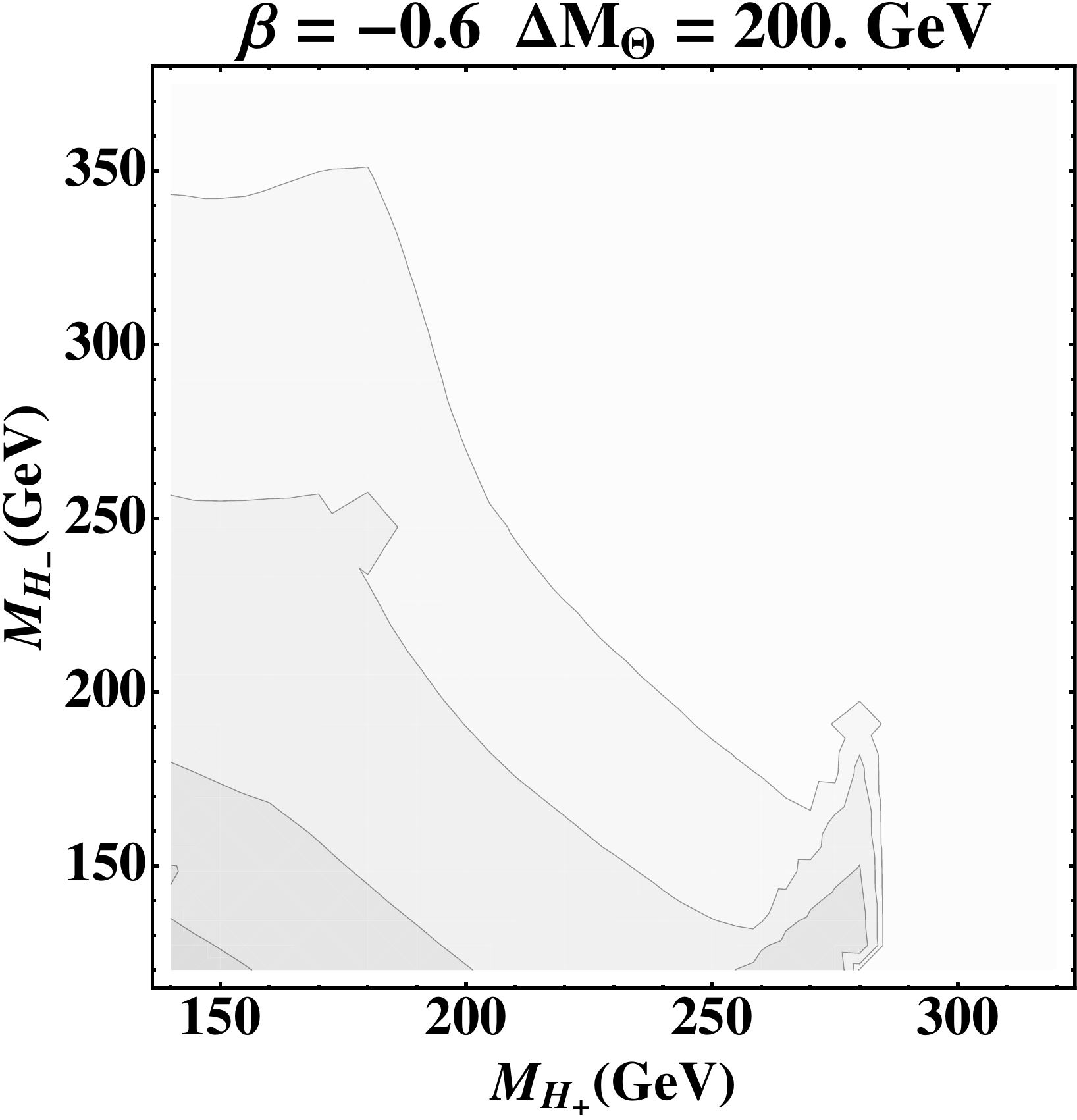}\hspace{0.4cm}\includegraphics[height=4.5cm,width=3.8cm]{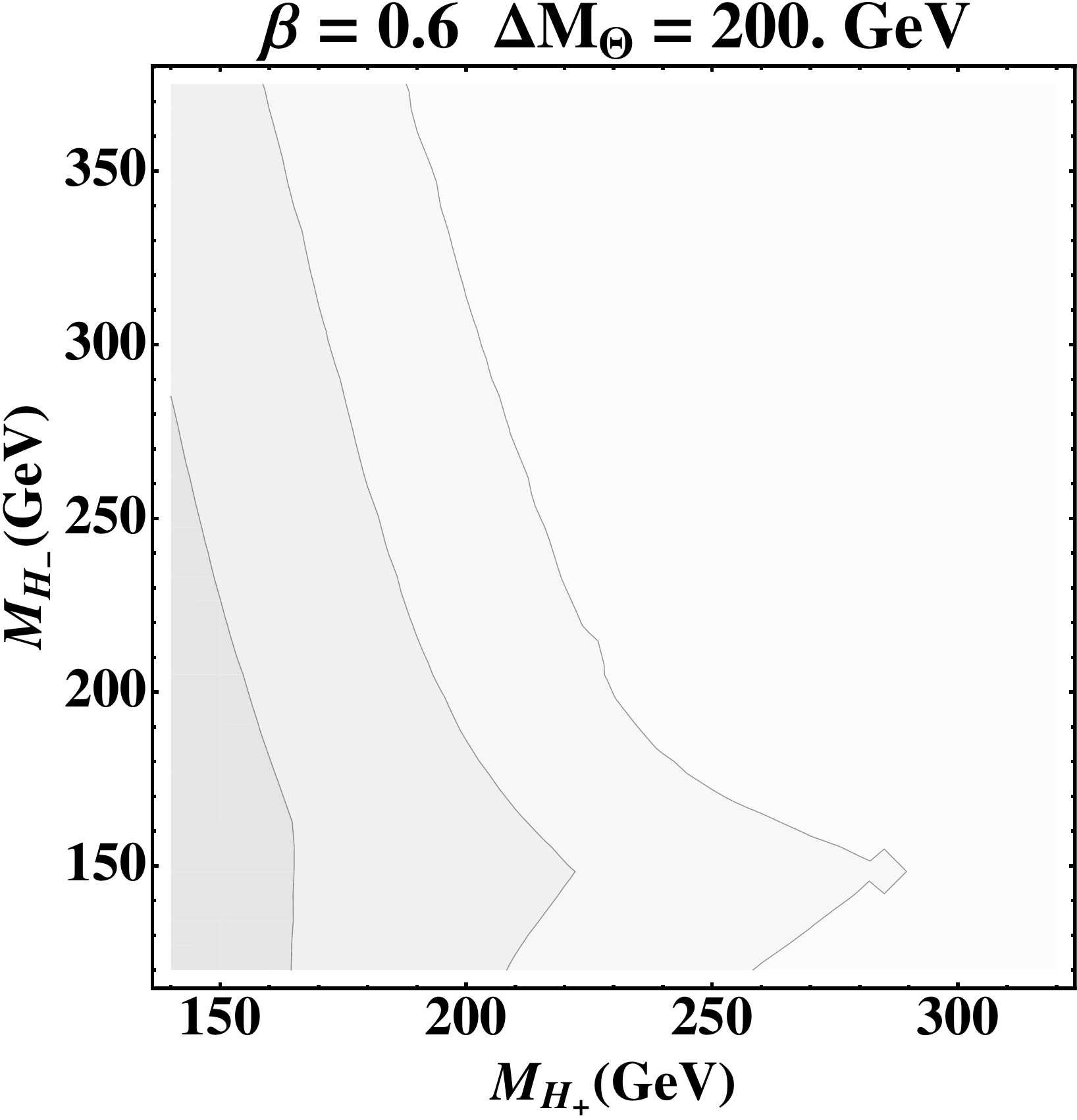}
 \includegraphics[height=4.5cm,width=3.8cm]{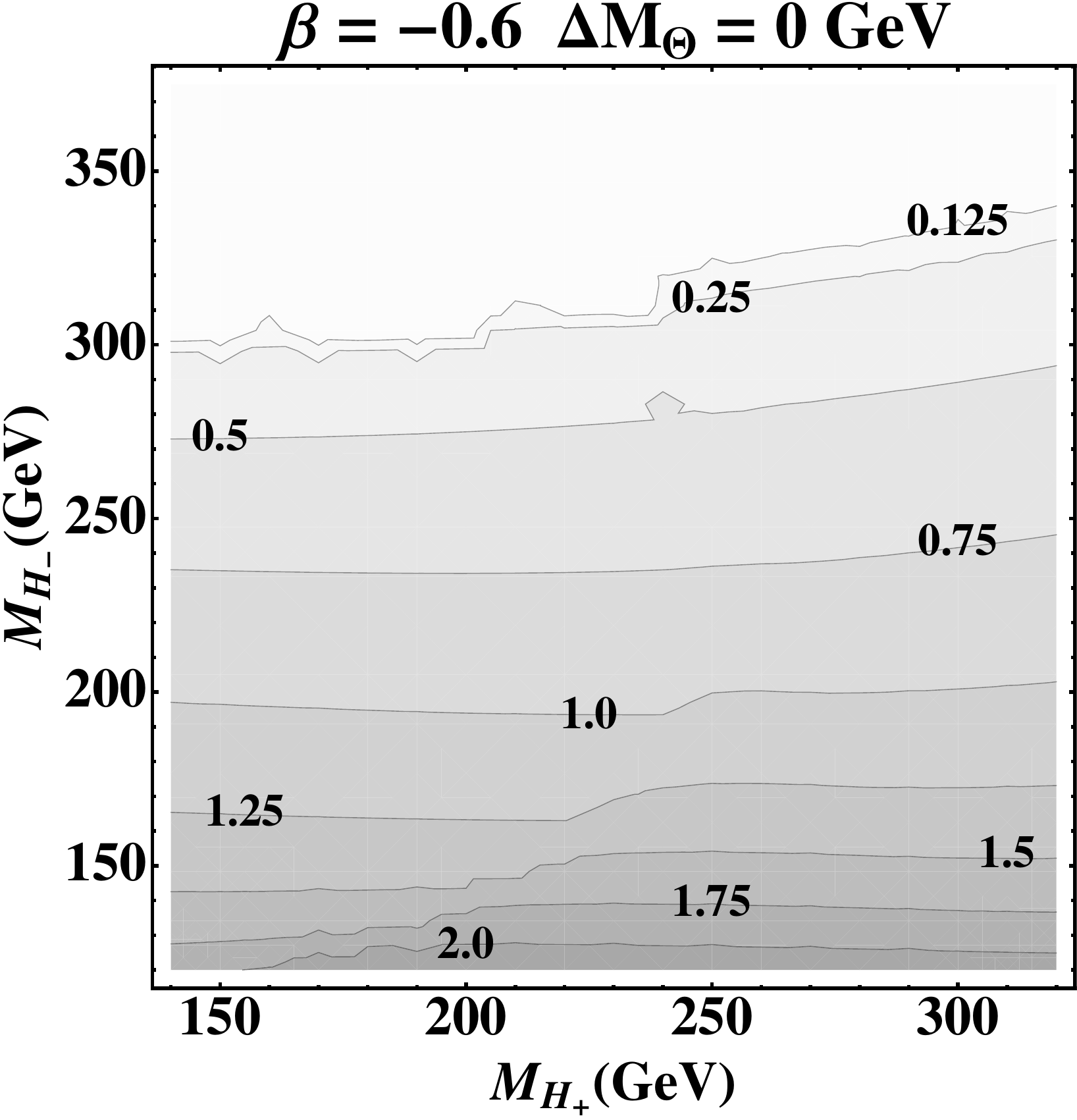}\hspace{0.4cm}\includegraphics[height=4.5cm,width=3.8cm]{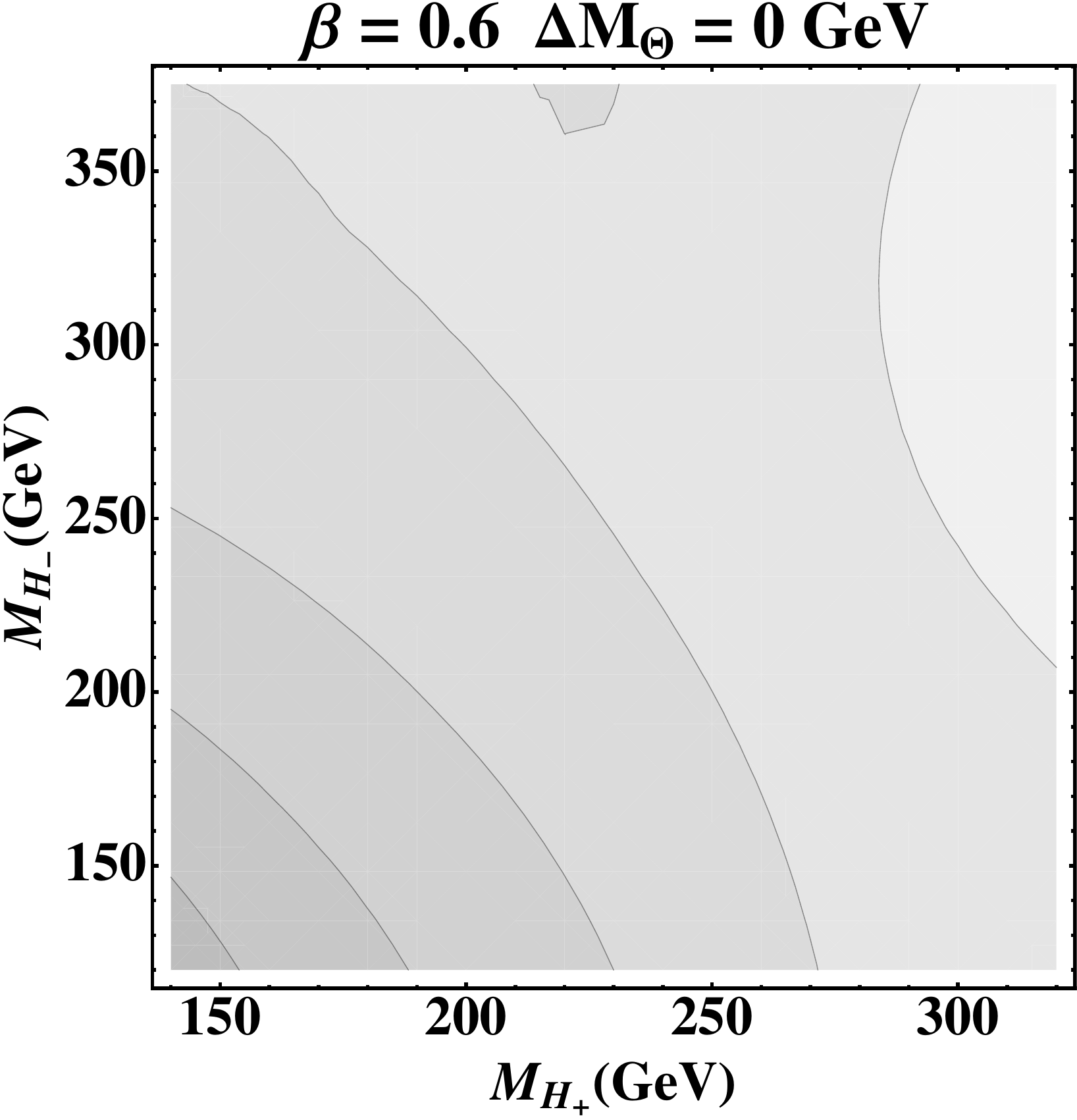}\hspace{0.4cm}\includegraphics[height=4.5cm,width=3.8cm]{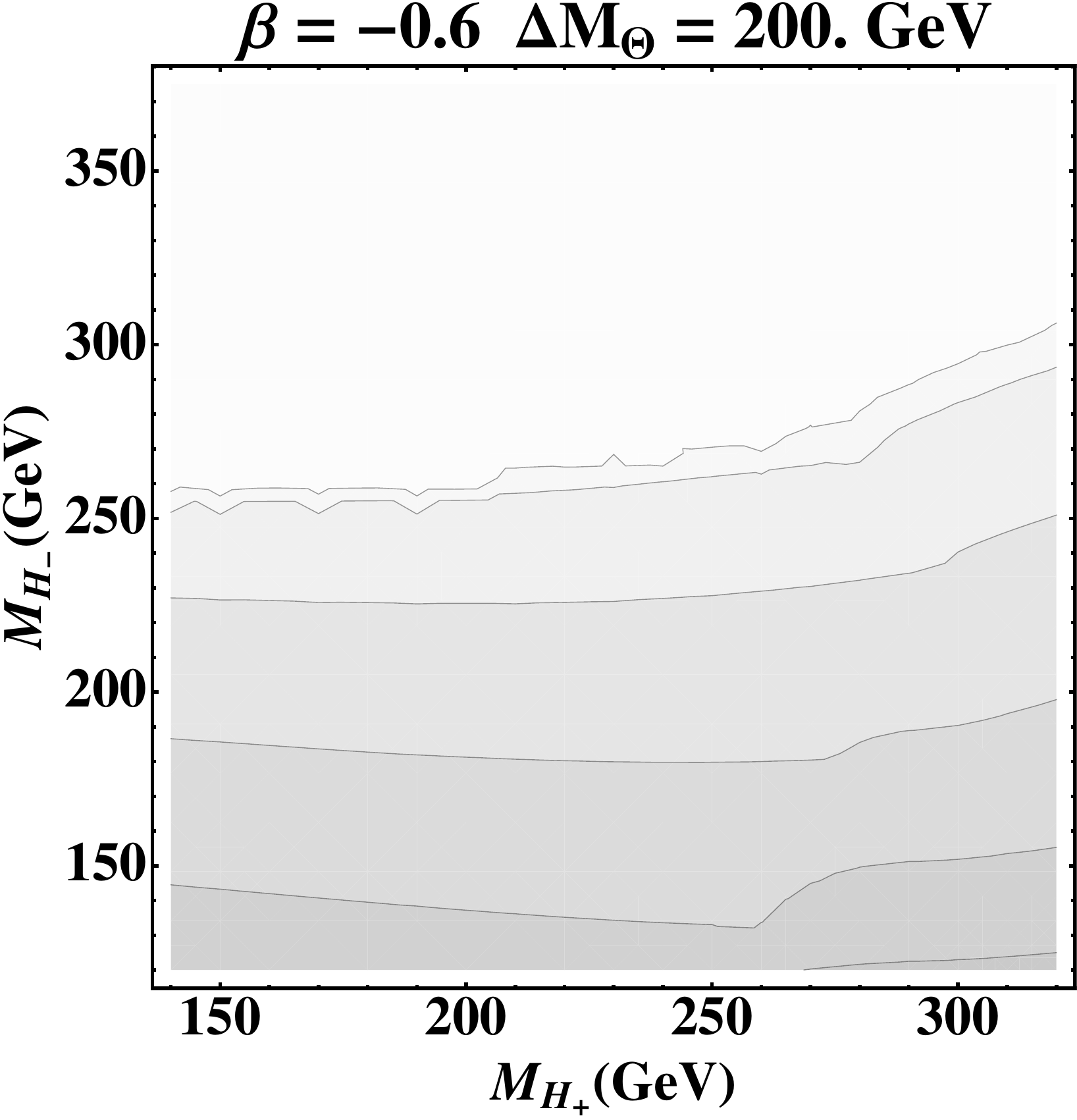}\hspace{0.4cm}\includegraphics[height=4.5cm,width=3.8cm]{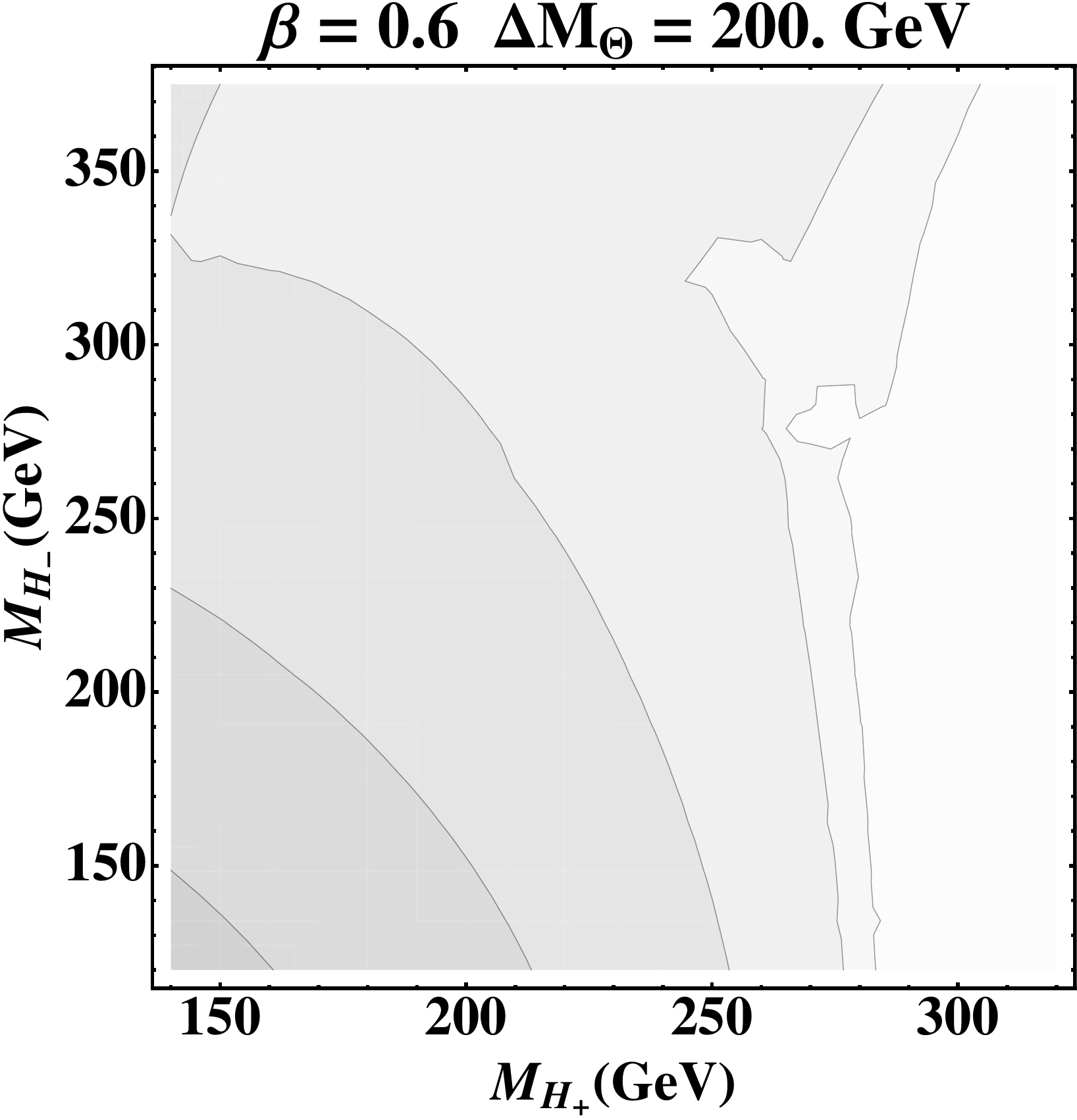}
}
\caption{The ratios $\sigma_{4,c}/T_{4,c}$ (top row) and  $\sigma_{2,c}/T_{2,c}$ (bottom row) as functions of the Higgs masses in the ``with EW'' scenario. We use $\Delta M_\Theta = 0$, 200~GeV  and $\beta= -0.6$,  0.6 as indicated in the labels and otherwise the parameters of Fig.~\ref{fig:phasedEW2} ($\Delta M_{\Pi_4}=300$~GeV, $\Delta M_{\Pi_2} = 550$~GeV and $v_2=350$~GeV).}\label{fig:phioverTwith}
\end{figure}

In Fig.~\ref{fig:phioverTwith} we present the behavior of the ratios $\sigma_c/T_c$ for the $\beta \ne 0$ plots of  Fig.~\ref{fig:phasedEW2}.
In general,  $\sigma_{4,c}/T_{4,c}$ shows similar behavior as above in the low $\Delta M_{\Pi_2}$ region of Fig.~\ref{fig:coupl1}: it is mainly reduced for nonzero $\beta$.
However, in the region of simultaneous transitions (V in Fig.~\ref{fig:phasedEW2})  $\sigma_{4,c}/T_{4,c}$ is significantly and  $\sigma_{2,c}/T_{2,c}$ somewhat enhanced. Similar behavior is in fact seen also in the ``without EW'' plots of Fig.~\ref{fig:phioverTwithout}.
Notice that the actual region V may be larger than in the plots if one of the phase transitions triggers the other.
We also see that the $\sigma_4$ transition is weak in the regions VI.

\section{Conclusion}
We uncovered the phase diagram as function of the temperature associated with the low energy effective theory possessing the same global symmetries of the UMT extension of the SM.  We discovered a very rich structure ranging from the occurrence of extra electroweak phase transitions \cite{Jarvinen:2009wr} appearing at different values of the temperature to the possibility of having simultaneous phase transitions. We find a region of parameters of the effective Lagrangian which allows for a sufficiently strong electroweak first order phase transition of interest for the electroweak baryogenesis problem. We note that having different sectors the model exhibits several phase transitions. Given that gravity couples to all the sectors we expect that all the sufficiently strong phase transitions, also the ones not directly related to the EWPT, are potential sources of gravitational waves.  If discovered, these gravitational waves could hint to a UMT-like structure behind a dynamical breaking of the electroweak symmetry.

\newpage
\appendix

\section{SU(4) and SU(2) Generators}
\label{generators}

Here we construct the explicit realization of the generators of $SU(4)$ and $SU(2)$. We denote the fifteen generators of $SU(4)$ by $S_4^a$ and $X_4^i$ with $a=1,\ldots,10$ and $i=1,\ldots,5$. They can be represented as:
\begin{eqnarray}
S_4^a = \left( \begin{array}{cc}
\textbf{A} & \textbf{B} \\
\textbf{B}^{\dagger} & -\textbf{A}^T
\end{array} \right) \ , \qquad X_4^i = \left( \begin{array}{cc}
\textbf{C} & \textbf{D} \\
\textbf{D}^{\dagger} & \textbf{C}^T
\end{array} \right) \ ,
\end{eqnarray}
where $A$ is Hermitian, $C$ is Hermitian and traceless, $B$ is symmetric and $D$ is antisymmetric. The $S_4^a$ obey the relation $(S_4^a)^TE + ES_4^a = 0$ and are a representation of $Sp(4)$. They are explicitly given by:
\begin{eqnarray}
S_4^a &=& \frac{1}{2\sqrt{2}} \left( \begin{array}{cc}
\tau^a & \textbf{0} \\
\textbf{0} & -\tau^{aT}
\end{array} \right) \ , \qquad a=1,\ldots,4 \\
S_4^a &=& \frac{1}{2\sqrt{2}} \left( \begin{array}{cc}
\textbf{0} & \textbf{B}^a \\
\textbf{B}^{a\dagger} & \textbf{0}
\end{array} \right) \ , \qquad a=5,\ldots,10
\end{eqnarray}
where $\tau^{1,2,3}$ are the usual Pauli matrices, $\tau^4=\textbf{1}$ and:
\begin{eqnarray}
\begin{array}{ccc}
B^5 = \textbf{1} \ , & B^7=\tau^3 \ , & B^9=\tau^1 \ , \\
B^6 = i\textbf{1}\ , & B^8 = i\tau^3\ , & B^{10} = i\tau^1\ .
\end{array}
\end{eqnarray}

The remaining five generators are explicitly given by:
\begin{eqnarray}
X_4^i &=& \frac{1}{2\sqrt{2}} \left( \begin{array}{cc}
\tau^i & \textbf{0} \\
\textbf{0} & \tau^{iT}
\end{array} \right)\ , \qquad i=1,\ldots,3 \\
X_4^i &=& \frac{1}{2\sqrt{2}} \left( \begin{array}{cc}
\textbf{0} & \textbf{D}^i \\
\textbf{D}^{i\dagger} & \textbf{0}
\end{array} \right) \ , \qquad i=4,5
\end{eqnarray}
with:
\begin{eqnarray}
D^4= \tau^2 \ , \qquad D^5 = i\tau^2 \ .
\end{eqnarray}

The generators are normalized according to:

\begin{eqnarray}
\text{Tr} \left[ S_4^aS_4^b \right] = \frac{1}{2} \delta^{ab} \ , \qquad \text{Tr} \left[ X_4^iX_4^j \right] = \frac{1}{2} \delta^{ij} \ , \qquad \text{Tr}  \left[ S_4^aX_4^i \right] = 0 \ .
\end{eqnarray}

The generators of $SU(2)$ are similarly divided into the two that are broken $X_2^i =\frac{\tau^i}{2},\ i=1,2$ and the one that leaves the vacuum invariant $S_2^1 = \frac{\tau^3}{2}$.

\section{Zero-Temperature Background-Dependent Scalar Masses} \label{AppA}

In terms of the underlying degrees of freedom the composite states transform as
\begin{equation}
\begin{array}{rclcrcl}\label{mesons2}
 v_4+H_4  & \equiv & \sigma_4 \sim \overline{U}U + \overline{D}D & ,~~~~ & \Theta_4 & \sim & i \left( \overline{U}\gamma^5 U + \overline{D} \gamma^5 D \right) \ , \\
\Pi^0 & \equiv & \Pi_4^3 \sim i \left( \overline{U} \gamma^5 U - \overline{D} \gamma^5 D \right) & ,~~~~ & \tilde{\Pi}^0 & \equiv & \tilde{\Pi}_4^3  \sim \overline{U}U - \overline{D}D \ , \\
\Pi^+ & \equiv & {\displaystyle \frac{\Pi_4^1 - i \Pi_4^2}{\sqrt{2}} \sim i \overline{D} \gamma^5 U} & ,~~~~ & \tilde{\Pi}^+ & \equiv & \frac{\tilde{\Pi}_4^1 - i \tilde{\Pi}_4^2}{\sqrt{2}} \sim \overline{D} U\ , \\
\Pi^- & \equiv & \frac{\Pi_4^1 + i \Pi_4^2}{\sqrt{2}} \sim i\overline{U} \gamma^5 D & ,~~~~ & \tilde{\Pi}^- & \equiv & \frac{\tilde{\Pi}_4^1 + i \tilde{\Pi}_4^2}{\sqrt{2}} \sim \overline{U}D \ , \\
\Pi_{UD} & \equiv & \frac{\Pi_4^4 + i \Pi_4^5}{\sqrt{2}} \sim U^T C D & ,~~~~ & \tilde{\Pi}_{UD} & \equiv & \frac{\tilde{\Pi}_4^4 + i \tilde{\Pi}_4^5}{\sqrt{2}} \sim iU^T C \gamma^5 D \ , \\
\Pi_{\overline{U}\overline{D}} & \equiv & \frac{\Pi_4^4 - i \Pi_4^5}{\sqrt{2}} \sim \overline{U} C \overline{D}^T & ,~~~~ & \tilde{\Pi}_{\overline{U}\overline{D}} & \equiv & \frac{\tilde{\Pi}_4^4 - i \tilde{\Pi}_4^5}{\sqrt{2}} \sim i \overline{U} C \gamma^5  \overline{D}^T \ , \\
\end{array}
\end{equation}
and
\begin{equation}
\begin{array}{rclcrcl}\label{mesons3}
 v_2 +H_2  & \equiv & \sigma_2 \sim \overline{\lambda}_D \lambda_D & ,~~~~ & \Theta_2 & \sim & i\overline{\lambda}_D \gamma^5 \lambda_D \ , \\
\Pi_{\lambda \lambda} & \equiv & \frac{\Pi_2^1-i\Pi_2^2}{\sqrt{2}} \sim \lambda_D^T C \lambda_D & ,~~~~ & \tilde{\Pi}_{\lambda \lambda} & \equiv & \frac{\tilde{\Pi}_2^1-i\tilde{\Pi}_2^2}{\sqrt{2}}  \sim i \lambda_D^T C \gamma_5 \lambda_D\ , \\
\Pi_{\overline{\lambda}\overline{\lambda}} & \equiv & \frac{\Pi_2^1+i\Pi_2^2}{\sqrt{2}} \sim \overline{\lambda}_D C \overline{\lambda}_D^T & ,~~~~ & \tilde{\Pi}_{\overline{\lambda}\overline{\lambda}} & \equiv & \frac{\tilde{\Pi}_2^1+i\tilde{\Pi}_2^2}{\sqrt{2}}  \sim i \overline{\lambda}_D C \gamma_5 \overline{\lambda}_D^T \ , \\
\end{array}
\end{equation}

Here $U=(U_L,U_R)^T$, $D=(D_L,D_R)^T$ and $\lambda_D=(\lambda^1, - i\sigma^2 {\lambda^2}^{\ast})^T$. In terms of the charge eigenstates the two matrices $M_4$ and $M_2$ can be written as
\begin{eqnarray}
M_4 =
\begin{pmatrix}
0 & -\frac{\Pi_{UD}-i\tilde{\Pi}_{UD}}{\sqrt{2}} & \frac{\sigma_4 + i\Theta_4 + i\Pi_0 + \tilde{\Pi}_0}{2} & \frac{i\Pi_+ + \tilde{\Pi}_+}{\sqrt{2}} \\
\frac{\Pi_{UD}-i\tilde{\Pi}_{UD}}{\sqrt{2}} & 0 & \frac{i\Pi_-+\tilde{\Pi}_-}{\sqrt{2}} & \frac{\sigma_4 + i\Theta_4 - i\Pi_0 - \tilde{\Pi}_0}{2} \\
-\frac{\sigma_4 + i\Theta_4 + i\Pi_0 + \tilde{\Pi}_0}{2} & -\frac{i\Pi_-+\tilde{\Pi}_-}{\sqrt{2}} & 0 & \frac{\Pi_{\bar{U}\bar{D}}-i \tilde{\Pi}_{\bar{U}\bar{D}}}{\sqrt{2}} \\
- \frac{i\Pi_+ + \tilde{\Pi}_+}{\sqrt{2}} & - \frac{\sigma_4 + i\Theta_4 - i\Pi_0 - \tilde{\Pi}_0}{2} & - \frac{\Pi_{\bar{U}\bar{D}}-i \tilde{\Pi}_{\bar{U}\bar{D}}}{\sqrt{2}} & 0
\end{pmatrix} \ ,
\end{eqnarray}
\begin{eqnarray}
M_2 =
\begin{pmatrix}
i\Pi_{\lambda\lambda} + \tilde{\Pi}_{\lambda\lambda} & \frac{\sigma_2 + i \Theta_2}{\sqrt{2}} \\
\frac{\sigma_2 + i \Theta_2}{\sqrt{2}} & i\Pi_{\bar{\lambda}\bar{\lambda}} + \tilde{\Pi}_{\bar{\lambda}\bar{\lambda}}
\end{pmatrix} \ .
\end{eqnarray}
The complete Lagrangian of the new Higgs sector including the scalar potential and the ETC mass terms reads
\begin{eqnarray}
\mathcal{L} &=& \frac{1}{2}\text{Tr}\left[ D_{\mu}M_4 D^{\mu}M_4^{\dagger} \right] + \frac{1}{2}\text{Tr}\left[ \partial_{\mu}M_2 \partial^{\mu}M_2^{\dagger} \right]  \nonumber \\
&& -\frac{m_4^2}{2}\text{Tr}\left[ M_4M_4^{\dagger} \right] + \frac{\lambda_4}{4}\text{Tr}\left[ M_4M_4^{\dagger} \right]^2 + \lambda_4'\text{Tr}\left[ M_4M_4^{\dagger}M_4M_4^{\dagger} \right] -\frac{m_2^2}{2}\text{Tr}\left[ M_2M_2^{\dagger} \right] \nonumber \\
&&  + \frac{\lambda_2}{4}\text{Tr}\left[ M_2M_2^{\dagger} \right]^2 + \lambda_2'\text{Tr}\left[ M_2M_2^{\dagger}M_2M_2^{\dagger} \right] +\frac{\delta}{2} \text{Tr}\left[ M_4M_4^{\dagger} \right]\text{Tr}\left[ M_2M_2^{\dagger} \right]  \\
&&  + 4 \delta' \left[ \left( \det M_2 \right)^2\text{Pf}\ M_4 + \text{h.c.} \right] + \frac{m_{4,ETC}^2}{4}\text{Tr} \left[ \tilde{M}_4B_4\tilde{M}_4^{\dagger}B_4 + \tilde{M}_4\tilde{M}_4^{\dagger} \right] \nonumber \\
&& + \frac{m_{2,ETC}^2}{4}\text{Tr} \left[ \tilde{M}_2B_2\tilde{M}_2^{\dagger}B_2 + \tilde{M}_2\tilde{M}_2^{\dagger} \right] + m_{1,ETC}^2 \left[ \text{Pf}\ P_4 + \text{h.c.} \right] - \frac{m_{1,ETC}^2}{2} \left(\det P_2 + \text{h.c.} \right) \nonumber
\end{eqnarray}

Due to the presence of the determinant/Pfaffian terms the Higgs particles and their associated partners mix respectively. We find the following scalar mass squared eigenstates
\begin{eqnarray}
M^2_{H_4 / H_2}(\sigma_4,\sigma_2) &=& \frac{1}{2} \left[ -m_4^2 - m_2^2 + \left( \delta + 3\lambda_4 + 3\lambda_4' \right) \sigma_4^2 + \left( \delta + 3\lambda_2 + 6 \lambda_2' \right) \sigma_2^2 \right.\nonumber \\
&&
  - \delta' \left( \sigma_2^2 + 6 \sigma_4^2 \right) \sigma_2^2 \pm \sqrt{\left(m_4^2 -m_2^2 + \left( \delta -3 \lambda_4 - 3 \lambda_4' \right) \sigma_4^2 \right.} \nonumber \\
  &&
\left. \overline{\left. + \left( -\delta +3 \lambda_2 + 6 \lambda_2' \right) \sigma_2^2 + \delta' \left( \sigma_2^2 -6 \sigma_4^2 \right) \sigma_2^2 \right)^2 + 16 \sigma_2^2\sigma_4^2\left( \delta - 2\delta' \sigma_2^2 \right)^2}\right] \nonumber \\
M^2_{\Theta_4 / \Theta_2}(\sigma_4,\sigma_2) &=& \frac{1}{2} \left[ 2m_{1,ETC}^2 -m_4^2 - m_2^2 + \left( \delta + \lambda_4 + \lambda_4' \right) \sigma_4^2 + \left( \delta + \lambda_2  +2 \lambda_2' \right)\right. \sigma_2^2 \nonumber \\
&&
 + \delta' \left( \sigma_2^2 + 6\sigma_4^2 \right) \sigma_2^2 \pm \sqrt{\left( m_4^2 - m_2^2 + \left( -\delta + \lambda_2 +2\lambda_2' \right) \sigma_2^2 \right.} \nonumber \\
&&
\left. \overline{\left. + \left( \delta - \lambda_4 - \lambda_4' \right) \sigma_4^2 - \delta' \left( \sigma_2^2 - 6 \sigma_4^2 \right)\sigma_2^2 \right)^2+64 \delta' \sigma_4^2 \sigma_2^6} \right] \nonumber \\
M^2_{\Pi^{\pm}_{0}} (\sigma_4,\sigma_2) &=& -m_4^2 + \left( \lambda_4 + \lambda_4' \right) \sigma_4^2 + \left( \delta - \delta' \sigma_2^2 \right) \sigma_2^2 \nonumber \\
M^2_{\Pi_{UD}} (\sigma_4,\sigma_2) &=&  -m_4^2 + \left(\lambda_4 + \lambda_4' \right) \sigma_4^2 + \left( \delta - \delta'\sigma_2^2 \right) \sigma_2^2 + m_{4,ETC}^2 \nonumber \\
M^2_{\tilde{\Pi}^{\pm}_{0}}(\sigma_4,\sigma_2) &=& M^2_{\tilde{\Pi}_{UD}}(\sigma_4,\sigma_2) =  - m_4^2 + \left( \lambda_4 + 3\lambda_4' \right) \sigma_4^2 + \left( \delta + \delta' \sigma_2^2 \right) \sigma_2^2 +m_{1,ETC}^2 \nonumber \\
M^2_{\Pi_{\lambda\lambda}}(\sigma_4,\sigma_2) &=&  - m_2^2 + \left( \lambda_2 + 2\lambda_2' \right) \sigma_2^2 + \delta \sigma_4^2 -2\delta' \sigma_4^2 \sigma_2^2 + m_{2,ETC}^2 \nonumber \\
M^2_{\tilde{\Pi}_{\lambda\lambda}}(\sigma_4,\sigma_2) &=&  -m_2^2 + \left( \lambda_2 + 6\lambda_2' \right) \sigma_2^2 + \delta \sigma_4^2 + 2\delta'\sigma_4^2 \sigma_2^2 + m_{1,ETC}^2 \ .
\end{eqnarray}
The notation is such that $M_{A/B}$ denotes the mass of the state which is a linear combination of $A$ and $B$.

\section{Temperature and Background Dependent Scalar Masses} \label{AppB}

The temperature-dependent scalar masses are found by first computing the one-loop thermal correction $V_T^{(1)}$ to the potential \cite{Arnold:1992rz}. This should be done as a function of all the scalar fields. Then by adding to the above zero-temperature scalar mass matrix the thermal corrections
\begin{eqnarray}
\frac{\partial^2}{\partial v_i \partial v_j}V_T^{(1)} \ ,
\end{eqnarray}
we obtain the effective thermal mass matrix. Here $v_i,\ i=1,\ldots,18$ denotes all the scalar fields. Due to the presence of the top quark the thermal corrections mix the two Higgs particles with $\tilde{\Pi}_0$ and similarly for their associated partners. Summarizing, the temperature and background dependent scalar masses squared are
\begin{eqnarray}
M^2_{H_4/H_2/\tilde{\Pi}_0}(\sigma_4, \sigma_2,T) &=&
\begin{pmatrix}
  k_1& 2\sigma_2 \sigma_4 \left( \delta -2 \delta' \sigma_2^2 \right) & \frac{m_{\rm top}^2 T^2}{2v_4^2} \\
2\sigma_2 \sigma_4 \left( \delta -2 \delta' \sigma_2^2 \right) & k_2 & 0 \\
\frac{m_{\rm top}^2 T^2}{2v_4^2} & 0 & k_3
\end{pmatrix} \nonumber \\
M^2_{\Theta_4/ \Theta_2/ \Pi_0}(\sigma_4, \sigma_2,T) &=&
\begin{pmatrix}
\tilde{k}_1 & 4 \delta' \sigma_4 \sigma_2^3 & \frac{m_{\rm top}^2T^2}{2v_4^2} \\
4 \delta' \sigma_4 \sigma_2^3   & \tilde{k}_2 & 0 \\
 \frac{m_{\rm top}^2T^2}{2v_4^2}  & 0  &  \tilde{k}_3
\end{pmatrix} \nonumber \\
M^2_{\Pi^{\pm}}(\sigma_4, \sigma_2,T) &=& M^2_{\Pi^{\pm}}(T=0) + \frac{T^2}{48} \left( 24 \delta + 56 \lambda_4 + 96 \lambda_4' + 3g'^2 + 9 g^2 \right) \nonumber \\
M^2_{\Pi_{UD}}(\sigma_4, \sigma_2,T) &=& M^2_{\Pi_{UD}}(T=0) + \frac{T^2}{6} \left( 3 \delta  + 7 \lambda_4 + 12\lambda_4' \right) \nonumber \\
M^2_{\tilde{\Pi}^{\pm}}(\sigma_4, \sigma_2,T) &=& M^2_{\tilde{\Pi}^{\pm}}(T=0) + \frac{T^2}{48} \left( 24 \delta + 56 \lambda_4 + 96 \lambda_4' + 3g'^2 + 9 g^2 \right) \nonumber \\
M^2_{\tilde{\Pi}_{UD}}(\sigma_4, \sigma_2,T)  &=& M^2_{\tilde{\Pi}_{UD}} (T=0) + \frac{T^2}{6} \left( 3 \delta  + 7 \lambda_4 + 12\lambda_4' \right) \nonumber \\
M^2_{\Pi_{\lambda\lambda}}(\sigma_4, \sigma_2,T) &=& M^2_{\Pi_{\lambda\lambda}} (T=0) + \frac{T^2}{3} \left( 3\delta + 2 \lambda_2 + 6 \lambda_2' \right) \nonumber \\
M^2_{\tilde{\Pi}_{\lambda\lambda}}(\sigma_4, \sigma_2,T) &=& M^2_{\tilde{\Pi}_{\lambda\lambda}} (T=0) + \frac{T^2}{3} \left( 3\delta + 2 \lambda_2 + 6 \lambda_2' \right)
\end{eqnarray}
where it is understood that the masses of the first $3+3$ states are the eigenvalues of the corresponding matrices. Also
\begin{eqnarray}
k_1 &=& -m_4^2 +  \left( \delta - \delta' \sigma_2^2 \right) \sigma_2^2 + 3 \left( \lambda_4 + \lambda_4' \right) \sigma_4^2 \nonumber \\
&& +\frac{T^2}{48}\left( 3g'^2 + 9 g^2 + 8 \left( 7\lambda_4 + 3 \left( \delta + \frac{m_{\rm top}^2}{v_4^2} + 4 \lambda_4' \right) \right) \right)  \nonumber \\
k_2 &=& -m_2^2 + \left( \delta - 6 \delta' \sigma_2^2 \right) \sigma_4^2 + 3 \left( \lambda_2 +2 \lambda_2' \right) \sigma_2^2 + \frac{T^2}{3} \left(3 \delta + 2 \lambda_2 + 6\lambda_2' \right) \nonumber \\
k_3 &=& M^2_{\tilde{\Pi}_0}(T=0 ) +\frac{T^2}{48}\left( 3g'^2 + 9 g^2 + 8 \left( 7\lambda_4 + 3 \left( \delta + \frac{m_{\rm top}^2}{v_4^2} + 4 \lambda_4' \right) \right) \right) \nonumber \\
\tilde{k}_1 &=& m_{1,ETC}^2 -m_4^2  +  \left( \delta + \delta' \sigma_2^2 \right) \sigma_2^2 + \left( \lambda_4 + \lambda_4' \right) \sigma_4^2 \nonumber \\
&& +\frac{T^2}{48}\left( 3g'^2 + 9 g^2 + 8 \left( 7\lambda_4 + 3 \left( \delta + \frac{m_{\rm top}^2}{v_4^2} + 4 \lambda_4' \right) \right) \right)  \nonumber \\
\tilde{k}_2 &=& m_{1,ETC}^2 - m_2^2 + \left( \delta + 6 \delta' \sigma_2^2 \right) \sigma_4^2 +  \left( \lambda_2 +2 \lambda_2' \right) \sigma_2^2 + \frac{T^2}{3} \left(3 \delta + 2 \lambda_2 + 6\lambda_2' \right) \nonumber \\
\tilde{k}_3 &=& M^2_{\Pi_0}(T=0) +\frac{T^2}{48}\left( 3g'^2 + 9 g^2 + 8 \left( 7\lambda_4 + 3 \left( \delta + \frac{m_{\rm top}^2}{v_4^2} + 4 \lambda_4' \right) \right) \right) \ . \nonumber
\end{eqnarray}

\section{Gauge Boson Mass Matrices} \label{AppC}

The background dependent transverse and longitudinal gauge boson mass matrices are
\begin{eqnarray}
M_T^2(\sigma_4)=\frac{\sigma_4^2}{4}
\left( \begin{array}{cccc}
g^2 & 0& 0 & 0 \\
0 & g^2 & 0 & 0 \\
0 & 0 & g^2 & -g'g \\
0 & 0 & -g'g & g'^2
\end{array} \right) \ , \qquad \qquad M_L^2(\sigma_4) = M_T^2(\sigma_4) + \Pi_L \ ,
\end{eqnarray}
where
\begin{eqnarray}
\Pi_L =
\left( \begin{array}{cccc}
2g^2T^2 & 0& 0 & 0 \\
0 & 2g^2T^2 & 0 & 0 \\
0 & 0 & 2g^2T^2 & 0 \\
0 & 0 & 0 & 2g'^2T^2
\end{array} \right) \ .
\end{eqnarray}

The longitudinal mass matrix receives finite temperature contributions both from the SM and also the technicolor sector. From the technicolor sector only the Higgs particle and its partner associated to $SU(4)$ contribute. The remaining composites are neutral under the electroweak symmetry. The contributions are computed as follows
\begin{eqnarray}
U(1): && \qquad \Pi_L^S = \frac{g'^2T^2}{3} \sum_S Y^2_S \ , \qquad \Pi_L^F = \frac{g'^2T^2}{6} \sum_F Y^2_F \ ,  \\
SU(N): && \qquad \Pi_L^S = \frac{g^2T^2}{3} \sum_S T(r_S) \ , \qquad \Pi_L^F = \frac{g^2T^2}{6} \sum_F T(r_F) \ , \qquad \Pi_L^V = \frac{N}{3}g^2T^2 \ \ \ \ \
\end{eqnarray}
The sums are over complex scalars and Weyl fermions respectively. The last contribution in the non-abelian case is due to the gauge bosons self interaction. Also $T(r)$ is the trace normalization factor $\text{Tr}\left[T^a_rT^b_r\right] = T(r) \delta^{ab}$.


\end{document}